\documentclass[12pt,a4paper]{article}
\pdfoutput=1
\usepackage{jheppub}
\usepackage{amsfonts,amsthm}
\usepackage[english]{babel}
\usepackage[utf8]{inputenc}
\usepackage{indentfirst}
\usepackage{slashed}
\usepackage{mathrsfs}
\usepackage{amssymb}
\usepackage{color}
\hypersetup{unicode}
\usepackage{natbib}
\usepackage{mathrsfs}
\usepackage{array}

\newcommand{\ma}[1]{\mbox{$\mathcal{#1}$}}

\newcommand{\mrm}[1]{\mbox{$\mathrm{#1}$}}

\newcommand{\AdS}{\mathrm{AdS}}
\newcommand{\vol}{\mathrm{vol}}

\newcommand{\ie}{\emph{i.e.}}
\newcommand{\eg}{\emph{e.g.}}
\newcommand{\cf}{\emph{cf.}}

\DeclareMathOperator{\tr}{tr}
\DeclareMathOperator{\arccot}{arccot}

\newcommand{\RR}{\mathbb{R}}

\newcommand{\U}[1]{\mathrm{U}(#1)}
\newcommand{\SU}[1]{\mathrm{SU}(#1)}
\newcommand{\SO}[1]{\mathrm{SO}(#1)}
\newcommand{\ISO}[1]{\mathrm{ISO}(#1)}

\newcommand{\dd}{d}
\newcommand{\ee}{e}

\title{Holographic reconstruction for defect CFTs from $\mathrm{AdS}_p \times S^q$ spacetimes}

\author[a]{Federico Faedo}
\author[b]{Nicol\`o Petri}
\author[a]{Alessia Segati}

\affiliation[a]{International Centre for Theoretical Physics Asia–Pacific,
University of Chinese Academy of Sciences, 100190 Beijing, China.}
\affiliation[b]{
INFN, Sezione di Milano, Via Celoria 16, 20133 Milano, Italy.}

\emailAdd{federico.faedo@ucas.ac.cn}
\emailAdd{nicolo.petri@mi.infn.it}
\emailAdd{alessia.segati@ucas.ac.cn}

\abstract{
We study defects in superconformal field theories using holography, focusing on the precise derivation of the defect observables from supergravity. We consider $\mathrm{AdS}_p \times S^q$ spacetimes fibered over an interval and coupled to higher-form gauge fields as well as scalar fields. We determine the coordinate system in which the defect geometry admits an asymptotically flat boundary and, in this setup, we systematically apply holographic renormalization to compute the fundamental observables of the defect theory.
In particular, we derive the one-point correlators of the bulk fields, the holographic stress tensor, and its Ward identities. We implement explicitly this procedure for line and surface defects in five- and six-dimensional Romans supergravity. The relevant geometries are $\mathrm{AdS}_2\times S^2$, $\mathrm{AdS}_2\times S^3$ and $\mathrm{AdS}_3\times S^2$ backgrounds warped over an interval, preserving four supercharges and asymptotically $\mathrm{AdS}_5$ and $\mathrm{AdS}_6$. In each case, we discuss the implications of our results and compare them with the standard literature on defects in conformal field theory.
}

\begin{document}
\maketitle
\flushbottom

\section{Introduction}

Defects provide a powerful probe of strongly coupled quantum field theories and their non-perturbative effects. They arise in a wide variety of contexts, ranging from condensed matter systems to string theory constructions. In supersymmetric settings, they often admit a geometric realization in terms of intersecting D-branes.

In this work we focus on configurations of this type admitting a holographic description in terms of warped AdS solutions in supergravity. By means of holography, these geometries provide computational access to strongly coupled observables in the dual field theory. In this context, our main goal is to develop a systematic framework allowing to compute precisely the holographic observables of the defect, directly from the asymptotic structure of these AdS backgrounds.

Among holographic defects, a prominent role is played by conformal ones. These are described by bulk geometries preserving a lower-dimensional conformal subgroup of the full conformal group that characterizes the ambient theory. Early realizations of this idea were formulated in terms of probe branes embedded inside AdS vacua \cite{Karch:2000gx,DeWolfe:2001pq,Aharony:2003qf,Bachas:2001vj}. In this framework, the defect degrees of freedom are localized on the branes worldvolume. Once the backreaction of the defect is fully included, the dual spacetime is described by warped AdS solutions in (super)gravity.

These solutions are typically characterized by the coexistence of two different AdS structures. In fact, on the one hand, their metric has the form of a warped product containing an AdS$_p$ factor, associated with the defect worldvolume, while on the other hand they asymptote to a higher-dimensional AdS$_D$ vacuum, describing the ambient conformal field theory. The transverse direction interpolating between these two regions encodes the bulk realization of the defect profile. Typically, this geometric structure is realized by large classes of Janus and interface solutions known in the literature (see, \eg, \cite{Bak:2003jk,Clark:2004sb,Clark:2005te,DHoker:2006qeo,DHoker:2006vfr,DHoker:2007zhm,DHoker:2009lky,Aharony:2011yc,Bobev:2020fon} for a selected list of references on the topic).

From the field theory viewpoint, the presence of the defect breaks the original conformal symmetry of the ambient theory down to the subgroup compatible with the lower-dimensional subspace where the defect degrees of freedom lie. The appearance of lower-dimensional AdS factors in the bulk geometry is the direct gravitational manifestation of this symmetry breaking pattern. In particular, the defect conformal symmetry is geometrically realized by the AdS$_p$ slicing, while the asymptotic AdS$_D$ region retains the holographic information associated with the ambient theory. As a consequence, warped defect geometries simultaneously contain information about both the defect and the higher-dimensional ambient theory in which the defect is embedded.

The interest in these configurations stems from their microscopic brane interpretation in string theory. In some cases, warped AdS geometries can be reproduced as near-horizon limits of suitable intersecting brane systems. Establishing this correspondence is often highly non-trivial, especially when the number of supersymmetries preserved by the system is low. Nevertheless, the {\itshape brane picture} plays a central role in clarifying the physical origin of the defect. This becomes particularly relevant for supersymmetric AdS$_3$ and AdS$_2$ solutions, which frequently arise as near-horizon geometries of extremal black holes.\footnote{The literature on defect CFTs from string theory is vast. For some relevant references in defect CFT, holography and AdS spaces, we refer to \cite{Erdmenger:2002ex,Constable:2002xt,Kapustin:2005py,Gaiotto:2008sa,Gaiotto:2008sd,Jensen:2015swa,deLeeuw:2015hxa,Billo:2016cpy,Estes:2014hka,Andrei:2018die,Chalabi:2021jud,Bianchi:2018zpb,Bianchi:2019sxz,Ghodsi:2023pej,Bianchi:2026sax}. For a non-exhaustive list of references on AdS$_2$ and AdS$_3$ solutions with holographic interpretation as conformal defects and relevant for this paper, we refer to \cite{DHoker:2007mci,Chiodaroli:2009yw,Chiodaroli:2011fn,Bobev:2013yra,Dibitetto:2017tve,Dibitetto:2017klx,Gutperle:2017nwo,Dibitetto:2018iar,Dibitetto:2018gtk,Gutperle:2018fea,Dibitetto:2019nyz,Chen:2019qib,Gutperle:2019dqf,Chen:2020mtv,Faedo:2020nol,Faedo:2020lyw,Chen:2020efh,Dibitetto:2020bsh,Lozano:2021fkk,Lozano:2022ouq,Anabalon:2022fti,Lozano:2022swp,Lozano:2022vsv,Lozano:2024idt,Conti:2024qgx,Gutperle:2024yiz}.}

In the last few years, a large class of defect solutions containing AdS$_3$ and AdS$_2$ factors has been constructed both in Type II and eleven-dimensional supergravity, as well as in lower-dimensional gauged supergravities obtained through consistent truncations.  These solutions preserve $\mathcal N=(0,4)$ supersymmetry in $p=3$ and $\mathcal N=4$ supersymmetry in $p=2$, and share the important property of admitting a clear microscopic brane interpretation. In particular, the corresponding geometries arise as near-horizon limits of intersecting brane configurations supporting line or surface defects inside higher-dimensional conformal field theories.

In this paper we focus on three specific classes of such defect geometries, obtained through consistent truncations to lower-dimensional supergravities. We are interested in solutions that are completely regular and supported by a two-form gauge potential. The first two examples are solutions of six-dimensional $\mathrm{F}(4)$ Romans supergravity. These backgrounds are warped AdS$_3\times S^2$ over an interval and they are studied in \cite{Dibitetto:2018iar,Lozano:2019emq,Lozano:2019jza,Faedo:2020nol,Faedo:2020lyw,Couzens:2021veb,Lozano:2022ouq} and AdS$_2\times S^3$ foliations, studied in \cite{Dibitetto:2018gtk, Chen:2019qib,Lozano:2020txg,Lozano:2020sae, Lozano:2022swp}. Both these solutions asymptotically approach the AdS$_6$ vacuum. In the AdS$_3$ case, we present a new particular solution whose geometry smoothly approaches AdS$_3\times \mathbb R^3$ near the defect. Similarly, we will consider AdS$_2\times S^3$ solutions that close off smoothly, describing an AdS$_2\times \mathbb R^4$ geometry in the near-defect region~\cite{Chen:2019qib}.
The third class consists of solutions of five-dimensional minimal $\SU{2} \times \U{1}$ gauged supergravity, obtained in \cite{Lozano:2021fkk,Lozano:2022vsv,Faedo:2025kjf}. These backgrounds are described by AdS$_2\times S^2$ foliations over an interval. They asymptotically approach an AdS$_5$ geometry in one limit and close off as AdS$_2\times \mathbb R^3$ in the opposite one \cite{Faedo:2025kjf}.

From the holographic viewpoint, these solutions describe line and surface defects inside higher-dimensional superconformal field theories. More precisely, the six-dimensional backgrounds are dual to defects within five-dimensional $\mathcal N=2$ SCFTs. On the other hand, the AdS$_2\times S^2$ solutions describe line defects either inside $\mathcal N=4$ SYM \cite{Lozano:2021fkk,Faedo:2025kjf} or within Gaiotto--Maldacena theories \cite{Lozano:2021fkk,Lozano:2022vsv}, depending on whether one considers their higher-dimensional Type IIB or Type IIA interpretation.
A common feature of these solutions is that they are typically supported by two-form gauge potentials along the lower-dimensional AdS factors and/or the internal spherical directions. These higher-form fields are intrinsically tied to the presence of the defect and, in some sense, play a role analogous to the vector fields appearing in the so-called {\itshape monodromy defects}, examples of which have been studied in \cite{Arav:2024exg,Arav:2024wyg,Conti:2025wwf,Conti:2025wyj,Couzens:2026qne}.

Despite the substantial progress in constructing explicit solutions, the systematic computation of defect observables from string theory backgrounds remains conceptually and technically subtle. The main difficulty originates from the coexistence of the two AdS structures described above. Indeed, the coordinates naturally adapted to the warped AdS$_p$ slicing are not the coordinates in which the flat conformal frame associated with the asymptotic AdS$_D$ boundary is manifest. Consequently, the holographic radial direction associated with the ultraviolet region of the ambient theory is not directly visible in the natural coordinates of the defect geometry. Important examples applying holographic renormalization techniques to defects solutions are \cite{Papadimitriou:2004rz,Chiodaroli:2011fn,Chen:2019qib,Gutperle:2019dqf,Chen:2020mtv}.

This point has important consequences for holographic renormalization. Since the AdS$_p$-sliced metric does not admit a flat boundary, but only a \textit{conformally} flat one, the standard holographic dictionary cannot be implemented straightforwardly. As a consequence, extracting the one-point functions and deriving the associated Ward identities requires a careful reconstruction of the flat asymptotic geometry. To this end, it is of utmost importance to identify a suitable set of Fefferman--Graham coordinates.

In this work we develop a systematic framework for the holographic analysis of warped defect geometries supported by higher-form fields. Starting from the three classes mentioned above, we determine the coordinate transformation bringing the metric in the appropriate asymptotic Fefferman--Graham form. Using this construction, we identify the holographic sources and expectation values associated with the defect configuration, and compute the holographic stress tensor together with its Ward identities.

The paper is organized as follows.
In Section \ref{ref:GeneralSection} we introduce the general strategy adopted in this paper to compute the holographic observables from defect AdS solutions. In particular, we discuss the construction of the suitable asymptotic Fefferman--Graham frame that enables us to obtain the flat boundary required for the holographic analysis.
In Section \ref{sec:Defects6D} we consider line and surface defects in five-dimensional SCFTs described by warped AdS$_2\times S^3 \times I$ and AdS$_3\times S^2 \times I$ solutions of six-dimensional $\mathrm{F}(4)$ Romans supergravity.
In Section \ref{sec:Defects5D} we turn to line defects in four-dimensional SCFTs represented by warped AdS$_2\times S^2 \times I$ solutions of five-dimensional $\SU{2} \times \U{1}$ gauged supergravity. For all these backgrounds, we explicitly compute the holographic one-point functions, and the stress tensor.
We close with three appendices, containing the conventions, in Appendix~\ref{app:conventions}, and technical details of the holographic renormalization procedure, including the counterterms and the renormalized on-shell actions in $D=6$ (\ref{app:counterterms-6d}) and $D=5$ (\ref{app:counterterms-5d}).

\section{The holographic description of a defect CFT}
\label{ref:GeneralSection}

In this section we present the general framework that will be employed in the holographic analysis of defect solutions throughout this work. The discussion is formulated in a model-independent way and applies uniformly to the supergravity solutions considered later. Our goal is to isolate the structural ingredients that are common to warped $\AdS_p \times S^q\times I$ backgrounds describing superconformal defects and to clarify the logic underlying the extraction of holographic observables.

The main focus of this paper will be solutions arising in gauged supergravity theories in $D=p+q+1$ dimensions, whose metric takes the schematic form
\begin{equation} \label{generalJanus}
d s^2_D = e^{2U(\theta)}\, d s^2_{\AdS_p} + e^{2W(\theta)}\, d s^2_{S^q} + e^{2V(\theta)}\, d\theta^2 \, ,
\end{equation}
where the warp factors depend on a single transverse coordinate $\theta$ parametrizing an interval. The $\AdS_p$ term captures the (super)conformal symmetry preserved by the defect, while the sphere $S^q$ accounts for the transverse rotational symmetry, as will become clear shortly. The solutions we consider are supported by scalar fields and higher-form gauge potentials whose profiles vary non-trivially along $\theta$.
The higher-form gauge fields are typically of the form
\begin{equation}
 B_p = b(\theta)\,\vol_{\AdS_p} \qquad\quad \text{and/or} \qquad\quad B_q = b(\theta)\,\vol_{S^q}\,,
\end{equation}
and their presence is directly tied to the defect in the bulk. As shown in various examples (see for instance \cite{Dibitetto:2017klx,Dibitetto:2018iar,Dibitetto:2018gtk,Faedo:2020nol,Faedo:2020lyw,Lozano:2021fkk,Lozano:2022swp,Lozano:2022vsv,Faedo:2025kjf}), after uplifting to ten or eleven dimensions these fields are directly related to D-brane fluxes.

Since our interest lies in conformal defects embedded in superconformal theories, the dual geometries must be asymptotically locally $\AdS_D$, at least at one endpoint of the interval, which will play the role of conformal boundary. This requirement guarantees the existence of a region in which the defect can be described using the standard holographic dictionary of the ambient SCFT$_d$, with $d=D-1$. In this regime, the defect is holographically encoded as a position-dependent deformation of the initial theory, breaking part of the superconformal symmetry while preserving the isometries along the defect worldvolume. More concretely, the symmetry group of the (Minkowskian) ambient CFT is broken to the direct product of the symmetry group preserved by a $(p-1)$-dimensional defect CFT and the symmetry group of the transverse space,
\begin{equation}
	\SO{d,2} \; \rightarrow \; \SO{p-1,2} \times \SO{q+1} \,,
\end{equation}
where the two factors on the right-hand side encode the isometries of $\AdS_p$ and $S^q$ in~\eqref{generalJanus}, respectively.
From the perspective of the supergravity solution \eqref{generalFG-intro}, this regime corresponds to extracting SCFT observables at the boundary of the higher-dimensional $\AdS_D$ spacetime.

The crucial question at this stage is the following: what is the holographic boundary of the AdS$_p \times S^q\times I$ warped geometry \eqref{generalJanus}? To answer this question, recall that an asymptotically $\AdS_D$ geometry can always be described near the boundary through the Fefferman--Graham (FG) expansion \cite{fefferman1985conformal,Henningson:1998gx},
\begin{equation}\label{generalFG-intro}
	d s^2_D = \frac{d z^2}{z^2} + \frac{1}{z^2} \, g_{ij}(z,x)\, d x^i d x^j \, ,
\end{equation}
where $z$ denotes the holographic coordinate and the conformal boundary is located at $z=0$, which is described by the metric $g$.

Casting an asymptotically AdS solution in the FG form is generically straightforward, especially when the asymptotic geometry is AdS written in global coordinates. In the cases that we discuss, however, the situation is more subtle: in the coordinates naturally adapted to the warped product structure \eqref{generalJanus}, the metric does not have a manifest preferred FG form.

In fact, there are two AdS radial coordinates: one associated with the ambient theory dual to AdS$_D$, and one associated with the defect geometry AdS$_p$. In these coordinates, it is not immediately clear which direction plays the holographic role and where the boundary is located. To resolve this issue and properly identify the holographic direction, a non-trivial coordinate transformation is required in order to recast the geometry into the canonical form \eqref{generalFG-intro}.

\subsection{The defect coordinates and holographic reconstruction}
\label{sec:defectcoordinates}

Let us now examine these aspects in more detail. Given a warped geometry AdS$_p\times S^q$ of the type \eqref{generalJanus}, it is convenient to write the AdS$_p$ metric as
\begin{equation}
d s^2_{\AdS_p} = \frac{ds^2_{\RR^{k,1}} + d\rho^2}{\rho^2} \,,
\end{equation}
with $k=p-2$ the number of spatial dimensions of the defect worldvolume and $\rho>0$. The geometry \eqref{generalJanus} can then be rewritten as
\begin{equation}
	\begin{split}\label{generalMinkSliced}
		ds^2_D &= e^{2U} \rho^{-2} \left( ds^2_{\mathbb{R}^{k,1}} + d\rho^2 + \rho^2 \, ds^2_\mathcal{D} \right) \, , \\
		ds^2_\mathcal{D} &= e^{2(W-U)} ds^2_{S^q} + e^{2(V-U)} d\theta^2 \, .
	\end{split}
\end{equation}
Writing the metric in this form makes the role of the defect more transparent. In fact, the metric manifestly describes the backreaction of a $(k+1)$-dimensional object in AdS$_D$, located at $\rho=0$.

We now introduce the holographic coordinate $z$ for the defect geometry, which we will refer to as the {\itshape defect coordinate} throughout this work. In order to reconstruct the desired flat boundary in the FG expansion, this coordinate needs to be a non-trivial combination of the radial coordinates $\theta$ and $\rho$ associated with the two AdS spaces appearing in the background, the $\AdS_D$ asymptotics and the $\AdS_p$ factor.
Consider the coordinate transformation $(\theta,\rho) \mapsto (z,y)$ defined by
\begin{equation}
	\theta = \theta(x) \,,  \qquad  \rho = z \, t(x) \,,  \qquad\quad  \text{with}  \qquad\quad  x \equiv \frac{y}{z} \,.
\end{equation}
Provided that the functions $\theta(x)$ and $t(x)$ satisfy the differential constraints
\begin{equation} \label{eqChangeGen}
	\left(\frac{d\theta}{dx}\right)^2 = \frac{e^{-2V} \bigl( 1 - e^{-2U} \bigr)}{x^2} \,,  \qquad\qquad
	\frac{d\log t}{dx} = \frac{1 - e^{-2U}}{x} \,,
\end{equation}
metric~\eqref{DefectMetricGen} takes the form
\begin{equation}\label{DefectMetricGen}
	d s^2_D = \frac{dz^2}{z^2} + \frac{1}{z^2} \left[ \frac{1}{t (t - x\,t')} \, ds^2_{\mathbb{R}^{k,1}} + \frac{t'}{x (t - x\,t')} \, d y^2 + z^2e^{2W(\theta(x))} \, d s^2_{S^q} \right] \,,
\end{equation}
with primed quantities denoting derivatives with respect to $x$.
The general form of~\eqref{DefectMetricGen} was originally constructed in \cite{Papadimitriou:2004rz} within the context of the Hamilton--Jacobi approach to holographic renormalization of Janus solutions.%
\footnote{We refer to Appendix B in \cite{Papadimitriou:2004rz} for a detailed analysis, where these metrics were constructed in generality.} In the present manuscript we will apply this analysis to defect geometries, whose general structure is given in~\eqref{generalJanus}.
As shown in \cite{Papadimitriou:2004rz}, this geometry admits the symmetry group $\SO{p-1,2} \times \SO{q+1}$, the isometry group of its original $\AdS_p \times S^q$ form, although in the coordinates $(z,y)$ only the subgroup $\ISO{k,1} \times \SO{q+1}$ is manifest.

Before applying this coordinate transformation to explicit defect solutions, as will be done in the next sections, let us first warm up with the simplest case, namely the global $\AdS_D$ vacuum. By choosing appropriate warp factors in \eqref{generalJanus}, the AdS$_D$ metric can be written in the form
\begin{equation}
	d s_D^2 = \frac{1}{\rho^2 \sin^2(2\theta)} \bigl( ds^2_{\mathbb{R}^{k,1}} + d\rho^2 + 4\rho^2 \, d\theta^2 + \rho^2 \cos^2(2\theta) \, d s_{S^q}^2 \bigr) \,,
\end{equation}
with $\rho>0$ and $\theta \in (0,\frac{\pi}{4})$. The vacuum geometry can then be cast in Poincaré coordinates in the standard way by introducing the coordinates $(z,y)$ as
\begin{equation}
	\left\{
	\begin{aligned}
		\theta &= \frac12 \arccot \Bigl(\frac{y}{z}\Bigr) \, , \\
		\rho &= \sqrt{z^2 + y^2} \, ,
	\end{aligned}
	\right.
	\qquad  \iff  \qquad
	\left\{
	\begin{aligned}
		z &= \rho \sin(2\theta) \, , \\
		y &= \rho \cos(2\theta) \, .
	\end{aligned}
	\right.
\end{equation}
The resulting line element is
\begin{equation} \label{AdSD-Poincare}
	d s_D^2 = \frac{dz^2 + ds^2_{\mathbb{R}^{k,1}} + dy^2  + y^2 \, d s_{S^q}^2}{z^2} \,,
\end{equation}
with $z>0$ and $y>0$. In this parametrization, the boundary is located at $z=0$ and manifestly flat. We notice that, as required by consistency, this change of coordinates satisfies equations~\eqref{eqChangeGen}.
As we will see in the next sections, solving \eqref{eqChangeGen} exactly in the bulk is highly non-trivial, and for this reason we will mainly focus on the boundary analysis. In particular, we will solve equations \eqref{eqChangeGen} perturbatively near the AdS$_D$ asymptotic region, a strategy that will allow us to explicitly compute the deformations of the AdS$_D$ vacuum geometry induced by the defect.

\subsection{The defect interpretation}
\label{subsec:FGexpansion}

The form of the metric \eqref{DefectMetricGen} is very suggestive, since it manifestly describes the backreaction of a $(k+1)$-dimensional object in the vacuum AdS$_D$. For example, in the case $p=2$, $q=2$, the bulk geometry is dual to a line or circular defect in the four-dimensional dual field theory. In this case, the metric \eqref{DefectMetricGen} describes the backreaction in AdS$_5$ of a point-like object located at the center of the transverse space, $y=0$. For $p=3$, $q=2$, one instead has a surface defect in the 5D dual theory. In this case, the metric \eqref{DefectMetricGen} describes the backreaction in AdS$_6$ of a one-dimensional object.
More generally, a $(k+1)$-dimensional defect in a $(D-1)$-dimensional field theory can be described by a $D$-dimensional gravitational solution encoding the backreaction of a $(k+1)$-dimensional object located at the center of the transverse space.

Holographically, the presence of these $(+1)$-dimensional objects can be analyzed by computing the observables of the defect SCFT defined in the asymptotic region of the solutions discussed above. To this end, all bulk fields must be expanded consistently near $z=0$. In this limit, the effect of the defect appears as fluctuations of the supergravity fields around the AdS$_D$ vacuum.
A characteristic feature of warped $\AdS_p \times S^q$ solutions is that, after transforming to the defect coordinates \eqref{DefectMetricGen}, the boundary metric is generically not globally flat. Instead, it takes the schematic form
\begin{equation} \label{boundaryDefectMetric}
	d s^2_\mathrm{bdy} = d s^2_{\RR^{k,1}} + dy^2 + w^2 y^2 \, ds^2_{S^q} \, ,
\end{equation}
where the parameter $w$ is determined by the bulk solution. For $w\neq1$, the boundary geometry exhibits a conical singularity localized at $y=0$, corresponding to the position of the defect. A deficit angle arises for $|w|<1$, while an excess angle corresponds to $|w|>1$. This structure reflects the backreaction of the defect degrees of freedom on the ambient CFT.

In the ideal situation, the boundary expansion of the supergravity solution yields a globally flat boundary metric. In that case, the holographic picture coincides with that of a standard conformal defect embedded in a flat SCFT background. However, such cases are rather rare within the class of solutions studied in this paper. In general, the holographic interpretation is more subtle, since one must account for deformations of the ambient SCFT induced by the conical singularity of the boundary geometry. Nevertheless, as we will show explicitly, the presence of conical singularities does not obstruct the holographic dictionary nor the computation of correlation functions.

After performing the transformation to the defect coordinates, the boundary metric admits the standard FG expansion, whose precise structure, however, depends on the boundary dimension and the matter content of the theory. In this manuscript, we focus on geometries with four- and five-dimensional boundaries, for which the expanded metrics take the following respective forms~\cite{fefferman1985conformal,Henningson:1998gx}%
\footnote{Strictly speaking, the FG coefficients are functions not only of $y$, but may also depend on the remaining boundary coordinates. However, since these terms must preserve the isometries of both $\RR^{k,1}$ and $S^q$, we retained only the explicit dependence on~$y$, with a slight abuse of notation.}
\begin{align} \label{gFGexpa}
	g_{ij}(z,y) &= g_{0\,ij}(y) + z^2 g_{2\,ij}(y) + z^4 \bigl( g_{4\,ij}(y) + \log z \, \tilde g_{4\,ij}(y) + \log^2\!z \, \hat g_{4\,ij}(y) \bigr) + \ldots \, , \nonumber \\
	g_{ij}(z,y) &= g_{0\,ij}(y) + z^2 g_{2\,ij}(y) + z^4 g_{4\,ij}(y) + z^5 g_{5\,ij}(y) + \ldots \, .
\end{align}
We observe the presence of logarithmic terms in the first formula, which arise in even boundary dimensions and are related to the conformal anomaly.

Scalar fields admit analogous asymptotic expansions. For a canonically normalized real scalar field $\Phi$ of mass squared $m_\Phi^2$ in $\AdS_D$, one finds \cite{Bianchi:2001kw}
\begin{equation}\label{XFGexpa}
	\Phi(z, y) = z^{d-\Delta_\Phi} \Phi_{(d-\Delta_\Phi)}(y) + z^{\Delta_\Phi} \bigl( \Phi_{(\Delta_\Phi)}(y) + \log z \, \tilde{\Phi}_{(\Delta_\Phi)}(y) \bigr) + \ldots \, ,
\end{equation}
where the scaling dimension $\Delta_\Phi$ satisfies the standard relation $\Delta_\Phi(\Delta_\Phi-d)=m_X^2 L^2$, with $L$ the AdS$_D$ radius. Also in this case, logarithmic terms are associated with conformal anomalies, and are present only when the boundary has even dimension.
Generically, $\Phi_{(d-\Delta_\Phi)}(y)$ represents the non-normalizable mode, while $\Phi_{(\Delta_\Phi)}(y)$ encodes the normalizable one. An exception to the rule occurs when $\Delta_\Phi=d/2$, with $d$ even, in which case the role of $\Phi_{(d-\Delta_\Phi)}(y)$ is played by $\tilde{\Phi}_{(\Delta_\Phi)}(y)$ \cite{Bianchi:2001kw}. This is exactly the situation we will encounter in Section~\ref{sec:Defects5D}, where a scalar field of conformal weight $\Delta_\Phi=2$ sources an operator in a four-dimensional SCFT.

The FG expansion of higher-form gauge fields is similar.
For a bulk $n$-form gauge potential $B$ in $\AdS_D$, the asymptotic behavior of the field is determined by its mass squared $m_B^2$ through $(\Delta_B - n)(\Delta_B + n - d) = m_B^2 L^2$.
As a concrete example, consider a bulk two-form $B$ in the case $p\geq2$ and/or $q\geq2$. In the defect coordinates $(z,y)$, it decomposes into components tangent to the boundary and components extended along the $z$ direction,
\begin{equation}\label{genBFG}
	B = B^{\mathrm{tan}}(z,y) + B^{\mathrm{mix}}(z,y) \wedge dz \, .
\end{equation}
Near $z=0$, both components admit an asymptotic expansion in powers of $z$, possibly including negative powers and, in addition, logarithmic terms when the boundary dimension is even.
The precise powers appearing in these expansions are determined by the bulk equations of motion and depend on the spacetime dimension and on the specific form of the higher-form dynamics.

\subsection{One-point correlators}
\label{sec:displacement}

Once the FG coordinates have been identified, defect observables can be obtained through the standard AdS/CFT prescription.
In the standard quantization scheme, the leading terms in the FG expansion of the bulk fields define the sources of the associated operators in the dual theory, whereas the coefficients of the normalizable modes determine the corresponding expectation values. According to the AdS/CFT dictionary, these are extracted by functional differentiation of the on-shell action. However, the latter suffers from near-boundary divergences, which can be systematically cured by applying the holographic renormalization prescription \cite{Henningson:1998gx,deHaro:2000vlm,Bianchi:2001kw,Papadimitriou:2004rz}.

Let us briefly review this procedure. The first step consists in imposing a cutoff in the spacetime close to the boundary, in our conventions at $z=\epsilon$, with $\epsilon\ll1$, and integrating the bulk action up to the cutoff. A well-posed variational problem further requires the addition of the Gibbons--Hawking--York (GHY) boundary term, which must also be evaluated at the cutoff. The resulting regularized action comprises a finite part and a set of terms diverging when the cutoff is removed.
Applying a standard QFT prescription, these singularities can be removed by introducing appropriate local covariant counterterms. Remarkably, these counterterms can be entirely expressed in terms of the induced boundary metric and of the bulk fields, all evaluated at the cutoff. The sum of the regularized action and the counterterms admits a smooth $\epsilon\to0$ limit, thus yielding a finite renormalized on-shell action.

Given a scalar operator $\mathcal{O}_\Phi$, its dual bulk field $\Phi(z,y)$ in the defect geometry has the near-boundary behavior~\eqref{XFGexpa}. The leading coefficient $\Phi_{(d-\Delta_\Phi)}$ acts as a source for $\mathcal{O}_\Phi$, while its one-point function is obtained by functional differentiation of the renormalized on-shell action with respect to the corresponding source,
\begin{equation}
	\langle\mathcal{O}_\Phi\rangle = \frac{1}{\sqrt{-g_0}} \frac{\delta \mathcal{S}_\mathrm{ren}}{\delta \Phi_{(d-\Delta_X)}} \,.
\end{equation}
The renormalized on-shell action is defined as the sum of the following three contributions~\cite{Henningson:1998gx,deHaro:2000vlm,Bianchi:2001kw},
\begin{equation}
	\mathcal{S}_{\mathrm{ren}} = \lim_{\epsilon\to0} \left( \mathcal{S}_{\mathrm{bulk}} + \mathcal{S}_{\mathrm{GHY}} + \mathcal{S}_{\mathrm{c.t.}} \right) \, ,
\end{equation}
where $\mathcal{S}_{\mathrm{bulk}}$ is the on-shell action computed with a cutoff at $z=\epsilon$, while $\mathcal{S}_{\mathrm{GHY}}$ and $\mathcal{S}_{\mathrm{c.t.}}$ denote, respectively, the GHY term and the holographic counterterms, both evaluated at the cutoff $\epsilon$.

The leading term $g_{0}$ in the expansion \eqref{gFGexpa} determines the background geometry of the defect theory, and the one-point function of its dual operator, the holographic stress--energy tensor, is canonically defined as
\begin{equation}
	\langle T_{ij} \rangle = -\frac{2}{\sqrt{-g_0}} \frac{\delta \mathcal{S}_{\mathrm{ren}}}{\delta g_0^{ij}} \, .
\end{equation}
Similarly, higher-form gauge potentials provide sources for tensor currents in the dual theory.
For a bulk two-form $B$, the source is identified with the coefficient of the non-normalizable mode of the tangential component $B^{\mathrm{tan}}$ in \eqref{genBFG}, while the corresponding one-point function is obtained by functional differentiation of the renormalized action with respect to this source,
\begin{equation}
	\langle J^{ij} \rangle = \frac{1}{\sqrt{-g_0}} \frac{\delta \mathcal{S}_{\mathrm{ren}}}{\delta B^{\mathrm{tan}}_{\text{(source)}\,ij}} \, .
\end{equation}

The Ward identities of the defect theory are encoded in the trace and the divergence of the stress tensor. In this work, these quantities will be computed explicitly from various supergravity setups.
Logarithmic counterterms generate a Weyl anomaly in even boundary dimension, namely
\begin{equation}
	\langle T^i_{\ i} \rangle = \mathcal{A} + \text{matter contributions} \, .
\end{equation}
On the other hand, deriving the divergence of the stress--energy tensor allows one to study violations of momentum conservation, which encode the breaking of translations in the directions transverse to the defect. Starting from defect solutions in supergravity and applying holographic renormalization, we will explicitly compute $\nabla^0_i \langle T^{ij}\rangle$, where $\nabla^0$ denotes the covariant derivative associated with the boundary metric $g_0$.

\section{Line and surface defects in SCFT$_5$}
\label{sec:Defects6D}

The rest of the paper is devoted to the application of the general ideas and results of the previous section to a selected list of concrete examples. Specifically, in this section we will focus on two classes of half-BPS solutions to six-dimensional Romans supergravity, whose holographic interpretation is that of line and surface defects in SCFT$_5$, respectively.
The first family of solutions we consider was constructed in~\cite{Chen:2019qib}, has the structure of a warped product $\AdS_2 \times S^3 \times I$ and features a non-trivial scalar field and a two-form. These geometries interpolate between a regular interior and an $\AdS_6$ asymptotic region. In~\cite{Chen:2019qib} it was proposed that they provide the holographic realization of a line defect inside a five-dimensional SCFT.
The second class of backgrounds has not appeared previously in the literature and will be constructed building upon the results of~\cite{Dibitetto:2018gtk}. Similarly to its companion, this solution has a smooth interior and is asymptotically $\AdS_6$, is characterized by non-trivial scalar and two-form, but its metric is given by the warped product of $\AdS_3 \times S^2$ over an interval. As we will show, such geometries have a natural interpretation in the dual field theory in terms of surface defects in a 5d SCFT.

Because the construction of Section~\ref{ref:GeneralSection} holds in complete generality, the analysis for both line and surface defects will be carried out in parallel.
In particular, after presenting the two classes of solutions, we derive the expression of the renormalized on-shell action, identify the sources of the dual operators, here represented by a scalar operator $\mathcal{O}_X$, a current~$J^{ij}$ and the stress--energy tensor~$T_{ij}$, and compute the associated one-point functions. In order to specialize this analysis to our geometries, we adopt the defect coordinates appropriate to each case, which allow us to reconstruct a background with flat boundary and, as a consequence, to consistently compute the desired holographic observables.

\subsection{The 6D Romans $\mathrm{F}(4)$ supergravity}

Before going into the details of the geometries that will be studied, we review the key features of the bosonic content of six-dimensional $\mathcal{N}=(1,1)$ Romans $\mathrm{F}(4)$ gauged supergravity~\cite{Romans:1985tw}. In addition to the graviton, the supergravity multiplet includes a two-form $B_{\mu\nu}$, a real scalar $\varphi$, three $\mathrm{SU}(2)$ and one $\mathrm{U}(1)$ vectors. In the rest of the paper, we will set all these one-forms to zero, in which case the bosonic part of the action reads~\cite{Romans:1985tw}%
\footnote{In this section we focus on the branch of parameters in which both the gauge coupling $g$ and the mass $m$ are positive, in which case we can set $g = \frac{3m}{\sqrt2}$.}
\begin{equation} \label{action6D}
	\begin{split}
	  \mathcal{S} = \frac{1}{16\pi G^{(6)}_\mathrm{N}} & \int \Bigl[ ( R-\mathcal{V}) \star\!1 - 4X^{-2} \star\dd X \wedge \dd X - \frac12 X^4 \star\!H \wedge H \\
		& - m^2 X^{-2} \star\!B \wedge B - \frac13 \, m^2 B \wedge B \wedge B \Bigr] \,,
	\end{split}
\end{equation}
where we defined the field strength $H = \dd B$ and the auxiliary scalar $X=\ee^{-\frac{\varphi}{2\sqrt2}}$. Additionally, $m>0$ is the mass parameter and the scalar potential is given by
\begin{equation} \label{scalar-pot6D}
	\mathcal{V} = m^2 \left( X^{-6} - 12X^{-2} - 9X^2 \right) \,.
\end{equation}
The theory admits a maximally supersymmetric vacuum in $X=1$ and $B=0$, corresponding to $\AdS_6$ with radius $L=m^{-1}$.

The equations of motion that follow from action~\eqref{action6D} are the Einstein equations
\begin{equation} \label{einstein-eq6D}
	\begin{split}
		R_{\mu\nu} &= 4X^{-2} \partial_\mu X \, \partial_\nu X + \frac14 X^4 \Bigl( H_{\mu\rho\sigma} H^{\,\;\rho\sigma}_\nu - \frac16 g_{\mu\nu} H_{\rho\sigma\tau} H^{\rho\sigma\tau} \Bigr) \\
		& + m^2 X^{-2} \Bigl( B_{\mu\rho} B^{\,\;\rho}_\nu - \frac18 g_{\mu\nu} B_{\rho\sigma} B^{\rho\sigma} \Bigr) + \frac14 g_{\mu\nu} \mathcal{V} \,,
	\end{split}
\end{equation}
the Klein--Gordon equation for the scalar
\begin{equation} \label{scalar-eq6D}
	\dd\bigl( X^{-1} \star\dd X \bigr) = -\frac14 X^4 \star\!H \wedge H + \frac14 \, m^2 X^{-2} \star\!B \wedge B - \frac18 X \, \partial_X \mathcal{V}(X) \star\!1 \,,
\end{equation}
and the Maxwell equation
\begin{equation} \label{maxwell-eq6D}
	\dd\bigl( X^4 \star\!H \bigr) = -2 m^2 X^{-2} \star\!B - m^2 B \wedge B \,.
\end{equation}
These equations must be supplemented with the following condition coming from truncating away the gauge vectors
\begin{equation}
	\dd\bigl( X^{-2} \star\!B \bigr) = -B \wedge H \,.
\end{equation}
This additional equation is implied by Maxwell's equation~\eqref{maxwell-eq6D} and, for this reason, we will not consider it anymore.

\subsection{Warped $\mrm{AdS}_2\times S^3$ and $\mrm{AdS}_3\times S^2$ solutions}

The geometries that we consider in this section are two families of warped ${\AdS_2 \times S^3}$ and $\AdS_3 \times S^2$ backgrounds, whose holographic interpretation is that of line defects and surface defects in a 5d SCFT. We now review the first of these two classes, previously derived in~\cite{Chen:2019qib}, and present the construction of the second.

\subsubsection{The line defect}

The first solution that we discuss is the $\AdS_2 \times S^3 \times I$ geometry presented in~\cite{Chen:2019qib}, whose holographic interpretation is that of a line defect within a five-dimensional CFT. This construction builds on the results of~\cite{Dibitetto:2018gtk}, where the BPS equations of the theory were derived and whose analysis shows that the background under exam preserves half of the total number of initial supersymmetries.
The metric, scalar field and two-form are given by~\cite{Chen:2019qib}%
\footnote{With respect to~\cite{Chen:2019qib}, we redefined the parameter $p$ as $p=1-\lambda$, fixed $q=p-1=-\lambda$, so to have a smooth space in the bulk, and set $r=1/2$, as required by the algebraic constraint~(2.21) of~\cite{Dibitetto:2018gtk}.}
\begin{equation} \label{6Dline}
 \begin{split}
	 \dd s^2_6 &= \frac{1}{m^2 X^2 \sin^2(2\theta)} \left[ \frac{1}{(1-\lambda)^2} \, \dd s^2_{\AdS_2} + 4X^8 \dd\theta^2 + \frac{4\cos^2(2\theta)}{(2+\lambda)^2} \, \dd s^2_{S^3} \right] \,, \\
	 X &= \left[ 1 + \lambda \tan^2(2\theta) \, \bigl(1 - \sin(2\theta)\bigr) \right]^{-1/4} \,, \\
	 B &= \frac{\lambda \left( 1 - \sin^3(2\theta) \right)}{m^2 (1-\lambda)^2 \sin(2\theta) \cos^2(2\theta)} \, \vol_{\AdS_2} \,,
 \end{split}
\end{equation}
where $\lambda$ is a real parameter, the coordinate $\theta$ ranges in the interval $\theta \in (0,\frac{\pi}{4})$, and $\dd s^2_{\AdS_2}$ and $\dd s^2_{S^3}$ denote the unit radius metrics on $\AdS_2$ and $S^3$, respectively.

First, we notice that the $\AdS_6$ vacuum is reproduced taking $\lambda=0$, with the six-dimensional AdS space written as
\begin{equation}
	 \dd s^2_6 = \frac{1}{m^2 \sin^2(2\theta)} \left( \dd s^2_{\AdS_2} + 4\dd\theta^2 + \cos^2(2\theta) \, \dd s^2_{S^3} \right) \,.
\end{equation}
Moreover, as pointed out in~\cite{Chen:2019qib}, the background~\eqref{6Dline} is smooth in the interval $(0,\frac{\pi}{4})$, as it produces a regular $\AdS_2 \times \RR^4$ space at $\theta=\frac{\pi}{4}$ and is asymptotically AdS for $\theta\to0$. Indeed, in a neighborhood of the former point we have
\begin{equation}
	\begin{split}
		X &\sim \frac{2^{1/4}}{(2+\lambda)^{1/4}} \equiv X_0 \,, \\
		\dd s^2_6 &\sim \frac{1}{m^2 (1-\lambda)^2 X_0^2} \, \dd s^2_{\AdS_2} + \frac{4X_0^6}{m^2} \left( \dd\theta^2 + \Bigl(\theta-\frac{\pi}{4}\Bigr)^2 \, \dd s^2_{S^3} \right) \,,
	\end{split}
\end{equation}
with the term in parentheses parametrizing $\RR^4$ in spherical coordinates.
Approaching the boundary at $\theta=0$, the metric matches that of $\AdS_6$ asymptotically
\begin{equation}
	\dd s^2_6 \sim \frac{1}{4m^2 \theta^2} \left( \frac{1}{(1-\lambda)^2} \, \dd s^2_{\AdS_2} + 4 \dd\theta^2 + \frac{4}{(2+\lambda)^2} \, \dd s^2_{S^3} \right) \,,
\end{equation}
with a possible conical singularity at the location of the defect, as will become clear in Section~\ref{subsec:defect-coord6D}. We point out that this conical singularity is not present when the warp factors of $\AdS_2$ and $S^3$ are equal, \ie\ for $\lambda=0$, global $\AdS_6$, and $\lambda=4$. The existence of a particular solution without a conical defect renders this class of solutions rather unique among the defect solutions considered in this paper, and more generally among solutions preserving four supercharges, namely $\mathcal N=(4,0)$ AdS$_3$ and $\mathcal N=4$ AdS$_2$.

We close this section with a brief analysis of the values that $\lambda$ is allowed to take. In order for the scalar $X$ to be regular within the whole interval $(0,\frac{\pi}{4})$ we need $\lambda>-2$. Additionally, the $\AdS_2$ warp factor in~\eqref{6Dline} dictates $\lambda\neq1$, thus leaving two possible intervals, $\lambda\in(-2,1)$ and $\lambda\in(1,\infty)$, the former being connected to the $\AdS_6$ vacuum.

\subsubsection{The surface defect: a new solution}

We now move to the construction of a new family of solutions within Romans $\mathrm{F}(4)$ supergravity, whose holographic interpretation is that of half-BPS superconformal surface defects. To this end, we consider an ansatz for the metric comprising an $\AdS_3$ factor and a two-sphere, both of them fibered over an interval,
\begin{equation}
	\dd s^2_6 = \ee^{2U(\alpha)} \dd s^2_{\AdS_3} + \ee^{2V(\alpha)} \dd\alpha^2 + \ee^{2W(\alpha)} \dd s^2_{S^2} \,,
\end{equation}
where $\dd s^2_{\AdS_3}$ and $\dd s^2_{S^2}$ denote the unit radius metrics on $\AdS_3$ and $S^2$, respectively. We will assume a non-trivial profile for the scalar field along the interval and an expression for the two-form which does not break the isometries of the system,
\begin{equation}
	X = X(\alpha) \,,  \qquad\quad  B = b(\alpha) \, \vol_{S^2} \,.
\end{equation}
The BPS equations for this ansatz were derived and studied in~\cite{Dibitetto:2018iar}, where it was shown that for a solution preserving half of the total initial supercharges, namely eight out of sixteen, the following system of first-order equations must be satisfied%
\footnote{We set $L=R=1$ in~\cite{Dibitetto:2018iar} because we consider metrics for $\AdS_3$ and $S^2$ with unit radius.}
\begin{equation} \label{floweq-AdS3S2}
	\begin{split}
	  U' &= \frac{\ee^V}{2\cos(2\theta)} \bigl( -(3+\cos(4\theta)) \, \mathcal{W} - 2\sin^2(2\theta) \, X\,\partial_X \mathcal{W} - e^{-U} \sin(2\theta) \bigr) \,, \\
	  W' &= \frac{\ee^V}{2\cos(2\theta)} \bigl((-5+\cos(4\theta)) \, \mathcal{W} + 2\sin^2(2\theta) \, X\,\partial_X \mathcal{W} - e^{-U} \sin(2\theta) \bigr) \,, \\
	  X' &= \frac{X \ee^V}{2\cos(2\theta)} \bigl( (3+\cos(4\theta)) \, X\,\partial_X \mathcal{W} + 2\sin^2(2\theta) \, \mathcal{W} + e^{-U} \sin(2\theta) \bigr) \,, \\[1ex]
		\theta' &= \ee^V \sin(2\theta) \, (\mathcal{W} - X\,\partial_X \mathcal{W}) \,,
  \end{split}
\end{equation}
where $\theta$ is a dynamical variable associated to the Killing spinor.
Equations \eqref{floweq-AdS3S2} have to be supplemented with two algebraic conditions, the first of which provides an expression for~$b$ in terms of the other fields
\begin{equation} \label{b-AdS3S2}
  b = -\frac{2\,\ee^{2W} X}{m \cos(2\theta)} \bigl( e^{-U} + 2\sin(2\theta) \, (\mathcal{W} + X\,\partial_X \mathcal{W}) \bigr) \,,
\end{equation}
while the second is a constraint among the warp factors $U$ and $W$ and the scalar field~$X$,
\begin{equation} \label{constraint-AdS3S2}
  2\,\ee^{-U+W} \cos^{-1}(2\theta) + 2\,\ee^W \tan(2\theta) \, (3\mathcal{W} + X\,\partial_X \mathcal{W}) = 1 \,.
\end{equation}
Here, $\mathcal{W}$ is the superpotential, defined for the 6D Romans $\mathrm{F}(4)$ supergravity as
\begin{equation}
	\mathcal{W}(X) = \frac{m}{8} \bigl( X^{-3} + 3X \bigr) \,,
\end{equation}
which allows us to rewrite the scalar potential~\eqref{scalar-pot6D} as $\mathcal{V} = 16 \bigl[ X^2 (\partial_X \mathcal{W})^2 - 5\mathcal{W}^2 \bigr]$.

To solve the system of equations~\eqref{floweq-AdS3S2}, we follow the strategy outlined in~\cite{Chen:2019qib}.
First, we exploit the freedom to redefine the coordinate~$\alpha$ to pick a ``gauge'' in which
\begin{equation} \label{gaugechoiceAdS3S2}
	\ee^{-V} = \sin(2\theta) \, (\mathcal{W} - X\,\partial_X \mathcal{W}) \,.
\end{equation}
The last equation in~\eqref{floweq-AdS3S2} simplifies to $\theta'=1$, which is solved, without loss of generality, by $\theta(\alpha)=\alpha$. In what follows, we will trade the coordinate~$\alpha$ for~$\theta$, and express everything in terms of the latter.
Combining the first and third equations in~\eqref{floweq-AdS3S2} we get
\begin{equation}
  U' + \frac{X'}{X} + \frac{2\cos(2\theta)}{\sin(2\theta)} = 0 \,,
\end{equation}
whose most general solution is
\begin{equation}
	\ee^{-U} = m (\lambda-1) \, X \sin(2\theta) \,,
\end{equation}
with $\lambda$ real integration constant.
Plugging back into~\eqref{floweq-AdS3S2} we obtain a first-order equation for $X$
\begin{equation}
	X' = \frac{X}{2\cos(2\theta) \sin(2\theta)} \, \bigl[ -3+2\sin^2(2\theta) + \bigl(3 + 2(\lambda-1) \sin^2(2\theta)\bigr) X^4 \bigr] \,,
\end{equation}
which is solved by
\begin{equation}
	X = \frac{\cos^{1/4}(2\theta)}{\bigl[ \cos(2\theta) \, \bigl( 1 + 2\lambda \sin^2(2\theta) \bigr) + q \sin^3(2\theta) \bigr]^{1/4}} \,,
\end{equation}
where $q$ is a real integration constant.
Finally, the warp factor $W$ and the function~$b$ can be obtained from the algebraic relations~\eqref{constraint-AdS3S2} and~\eqref{b-AdS3S2}, respectively,
\begin{equation}
  \ee^{-W} = m (1 + 2\lambda) \, X \tan(2\theta) \,,  \qquad\qquad
  b = \frac{1 - (1+2\lambda) X^4}{m^2 (1 + 2\lambda)^2 X^4 \tan(2\theta)} \,.
\end{equation}

Before writing down the full solution, we analyze its behavior in the range of definition of the coordinate $\theta$, which we take to be $(0,\frac{\pi}{4})$. In order for the physical scalar~$\varphi$ to be well-defined, $X$ needs to be strictly positive and finite.
We begin studying the scalar $X$ in a neighborhood of $\theta=\frac{\pi}{4}$, where
\begin{equation}
	X \sim \frac{2^{1/4}}{q^{1/4}} \Bigl(\frac{\pi}{4} - \theta\Bigr)^{1/4} 
	+ \mathcal{O}\Bigl(\frac{\pi}{4} - \theta\Bigr)^{9/4} \,.
\end{equation}
We observe that setting $q=0$ the scalar field becomes regular at this point, namely
\begin{equation}
	X \sim \frac{1}{(1+2\lambda)^{1/4}} \equiv X_0 \,.
\end{equation}
With this choice made, the expansion of the line element around $\theta=\frac{\pi}{4}$ gives
\begin{equation}
	\dd s^2_6 \sim \frac{1}{m^2 (1-\lambda)^2 X_0^2} \, \dd s^2_{\AdS_3} + \frac{4X_0^6}{m^2} \left( \dd\theta^2 + \Bigl(\theta-\frac{\pi}{4}\Bigr)^2 \, \dd s^2_{S^2} \right) \,,
\end{equation}
where the term in parentheses represents the three-dimensional Euclidean space in spherical coordinates.

The final expression of our $\AdS_3 \times S^2 \times I$ surface defect solution is
\begin{equation} \label{6Dsurface}
  \begin{split}
    \dd s^2_6 &= \frac{1}{m^2 X^2 \sin^2(2\theta)} \left[ \frac{1}{(1-\lambda)^2} \, \dd s^2_{\AdS_3} + 4X^8 \dd\theta^2 + \frac{\cos^2(2\theta)}{(1+2\lambda)^2} \, \dd s^2_{S^2} \right] \,, \\
    X &= \left[ 1 + 2\lambda \sin^2(2\theta) \right]^{-1/4} \,, \\
    B &= -\frac{2\lambda \cos^3(2\theta)}{m^2 (1+2\lambda)^2 \sin(2\theta)} \, \vol_{S^2} \,.
  \end{split}
\end{equation}
This background describes a one-parameter family of deformations of the global $\AdS_6$ vacuum, which is recovered for~$\lambda=0$, with the AdS metric written in a similar parametrization as for the line defect.
Analogously to the previous case, regularity of metric and scalar field put constraints on the range of validity of $\lambda$, namely $\lambda>-1/2$, with $\lambda\neq1$.

As already mentioned, in $\theta=\frac{\pi}{4}$ the metric reduces to a regular $\AdS_3 \times \RR^3$ geometry. At the other endpoint of the interval, $\theta=0$, which plays the role of conformal boundary, the solutions behaves as
\begin{equation}
  \begin{split}
    \dd s^2_6 &\sim \frac{1}{4m^2 \theta^2} \left( \frac{1}{(1-\lambda)^2} \, \dd s^2_{\AdS_3} + 4 \dd\theta^2 + \frac{1}{(1+2\lambda)^2} \, \dd s^2_{S^2} \right) \,, \\
    X &\sim 1 \,, \\
    B &\sim -\frac{2\lambda}{2m^2 (1+2\lambda)^2 \, \theta} \, \vol_{S^2} \,.
  \end{split}
\end{equation}
Our geometry is therefore asymptotically $\AdS_6$, but only locally, because the presence of a non-vanishing two-form $B$ breaks the AdS isometries. Additionally, we observe the presence of a conical singularity on the boundary unless $\lambda=0$, corresponding to the $\AdS_6$ vacuum, or $\lambda=-2$, which, however, lies outside the allowed range of~$\lambda$.

The surface defect geometry here presented can be immediately compared with the original solution constructed in~\cite{Dibitetto:2018iar}, which can be retrieved by setting ${\lambda=3}$ in~\eqref{6Dsurface}, once the relation $g=\frac{3m}{\sqrt2}$ is imposed.
As a closing remark, background~\eqref{6Dsurface} can be equivalently obtained applying the method of~\cite{Conti:2024rwd} or performing a suitable double analytic continuation of the $\AdS_2 \times S^3 \times I$ solution of~\cite{Chen:2019qib}, \cf\ equation~\eqref{6Dline}.

\subsection{The on-shell action}
\label{sec:boundaryexp_6D}

As discussed in Section \ref{ref:GeneralSection}, to start the holographic analysis we first need to expand the fields in a region close to the boundary. Since our solutions are asymptotically locally $\mrm{AdS}_6$, we can use the Fefferman--Graham expansion to study the 6D metric near the $z=0$ boundary:
\begin{equation}
  \label{FGds6D}
  ds_6^2 = \frac{1}{z^2} \left( dz^2 + g_{ij}(z,x) \, dx^i dx^j \right) \, ,
\end{equation}
where $z$ is the FG coordinate, while the metric tensor $g_{ij}$ can be expanded as
\begin{equation}
  \label{g:6D}
  g = g_0 + z^2 g_2 + z^4 g_4 + z^5 g_5 + \ma O(z^6) \, .
\end{equation}
As in the pure gravitational case, the terms containing odd powers of $z$ are expected to vanish up to order $z^5$ \cite{Henningson:1998gx}. For the scalar field $X$ and the two-form $B$ we assume
\begin{equation}
  \label{boundaryfields6D}
  \begin{split}
    X & = 1 + z^2 X_2 + z^3 X_3 + \ma O(z^4) \, ,\\
    B & = z^{-1} B_{-1} - A_0 \wedge dx + z \, B_1 - z^2 A_2 \wedge dz + z^2 B_2 + \ma O (z^3) \, ,
  \end{split}
\end{equation}
where $B_{-1}$, $B_1$ and $B_2$ are two-forms living on the boundary and $A_0$ and $A_1$ are one-forms defined on the boundary. Possible terms $B_0$ and $z \, A_1 \wedge dz$ in $B$ can be shown to vanish.

This analysis was already presented in \cite{Alday:2014bta, Chen:2019qib}. Here we extend the computation of~\cite{Alday:2014bta} by adding the derivation of the one-point function of the stress--energy tensor, while we generalize the results obtained in \cite{Chen:2019qib} by considering a more general FG expansion for the two-form $B$. Following the same logic, we first use the equations of motion to rewrite the bulk action \eqref{action6D} as%
\footnote{To avoid cumbersome expressions we set $m=1$, as is done in Appendix~\ref{app:counterterms-6d}.}
\begin{equation}
	\begin{split}
		\mathcal{S}_\mathrm{bulk} = \frac{1}{16\pi G_\mathrm{N}^{(6)}} & \int_M \dd^6x \, \sqrt{-G} \, \biggl( \frac12 \, \mathcal{V} - \frac{1}{12} X^4 H_{\mu\nu\rho} H^{\mu\nu\rho} - \frac14 X^{-2} B_{\mu\nu} B^{\mu\nu} \\
		& + \frac{1}{24} \frac{\varepsilon^{\mu\nu\rho\sigma\tau\lambda}}{\sqrt{-G}} B_{\mu\nu} B_{\rho\sigma} B_{\tau\lambda} \biggr) \,,
        \end{split}
\end{equation}
where $M$ is the bulk manifold and we denote by $G$ the six-dimensional metric, to distinguish it from the near-boundary metric $g$ in \eqref{g:6D}. The GHY term is given by
\begin{equation}
	\mathcal{S}_\mathrm{GHY} = \frac{1}{8\pi G_\mathrm{N}^{(6)}} \int_{\partial M} \dd^5x \, \sqrt{-h} \, h^{ij} K_{ij} \,,
\end{equation}
where $h_{ij}$ is the induced metric on the boundary and $K_{ij}$ its extrinsic curvature.
As dictated by the holographic renormalization prescription, we need to add suitable counterterms to render the on-shell action finite. The details of the construction are presented in Appendix \ref{app:counterterms-6d}, here we only report their expression:
\begin{align}
  \ma S_{\mathrm{c.t.}} & = \frac{1}{16\pi G_\mathrm{N}^{(6)}}  \int_{\partial M} \dd^5x \, \sqrt{-h} \, \biggl\{ -8 - \frac13 R[h] + \frac12 |B|^2_h - \frac19 R_{ij}[h] \, R^{ij}[h] + \frac{5}{144} R[h]^2 \nonumber \\
	& - \frac{7}{48} R[h] \, |B|^2_h - \frac13 \tr(h^{-1} \mathrm{Ric}[h] \, h^{-1} B h^{-1} B) - 8(1-X)^2 + \frac{45}{64} |B|^4_h \\
	& - \frac12 \tr\bigl[(h^{-1} B)^4\bigr] + \frac12 |\dd B|^2_h - \frac14 \, h_{ij} \nabla_k B^{ik} \nabla_l B^{jl} + \frac{\sigma}{8} \frac{\varepsilon^{ijklm}}{\sqrt{-h}} \, h_{mn} B_{ij} B_{kl} \nabla_p B^{np} \biggr\} \nonumber \,,
\end{align}
where all the quantities are to be evaluated at the cutoff $z=\epsilon$.

\subsection{One-point correlators and Ward identities}
\label{subsec:1pt_6D}

Before moving to the discussion of the one-point functions, we need to the identify which are the sources of the operators dual to the gravity fields in the bulk. We mainly follow \cite{Chen:2019qib}.

The conformal dimensions of operators in the dual CFT corresponding to the scalar field and the two-form field in the bulk can be derived from the linearized equations of motion near the $AdS_6$ boundary. By inserting $X-1 \sim z^{\Delta_X}$ into \eqref{scalar-eq6D}, we obtain
\begin{equation}
  \Delta_X ( \Delta_X - 5) = -6 \, ,
\end{equation}
where $-6$ is the mass squared of $X$ after setting the AdS radius $L=m^{-1}=1$. As pointed out in \cite{Chen:2019qib}, the mass lies in the window where both standard and alternative quantization are allowed \cite{Klebanov:1999tb}, meaning that the scaling dimension of the dual scalar operator $\ma O_X$ could be either $\Delta_X =2$ or $\Delta_X = 3$. Nevertheless, looking at the dual fields in the gravity supermultiplet \cite{Ferrara:1998gv}, we observe that the energy level of the bottom scalar is $E_0=3$. This forces us to choose $\Delta_X=3$ in order for the dual scalar operator to be in the same superconformal multiplet as the stress--energy tensor, the dual partner of the graviton. The source of the scalar operator in the CFT will be then given by the coefficient of the term $z^{5-\Delta_X}$ in the FG expansion, corresponding to $X_2$ in \eqref{boundaryfields6D}, while its vacuum expectation value (VEV) will be related to the $z^{\Delta_X}$ term, given by $X_3$.

In the same spirit, the insertion of $B = z^{\Delta_B-2} dx^1 \wedge dx^2$ in \eqref{maxwell-eq6D} gives
\begin{equation}
  (\Delta_B-2)(\Delta_B-3)=2
\end{equation}
and thus we have $\Delta_B=4$, in agreement with \cite{Ferrara:1998gv}. The source of the current dual to~$B$ can be then identified with $B_{-1}$.

Once we have determined the sources of the CFT operators, we can compute the associated one-point functions, which are defined in the standard way~\cite{deHaro:2000vlm,Bianchi:2001kw} as
\begin{equation} \label{1-pt_6D}
	\begin{split}
		\langle\mathcal{O}_X\rangle &= \frac{1}{\sqrt{-g_0}} \frac{\delta \mathcal{S}_\mathrm{ren}}{\delta X_2} =
		\lim_{\epsilon\to0} \biggl( \frac{1}{\epsilon^3} \frac{1}{\sqrt{-h}} \frac{\delta \mathcal{S}_\mathrm{sub}}{\delta X} \biggr) \,, \\
		\langle J^{ij} \rangle &= \frac{1}{\sqrt{-g_0}} \frac{\delta \mathcal{S}_\mathrm{ren}}{\delta B_{-1\,ij}} =
		\lim_{\epsilon\to0} \biggl( \frac{1}{\epsilon^6} \frac{1}{\sqrt{-h}} \frac{\delta \mathcal{S}_\mathrm{sub}}{\delta B_{ij}} \biggr) \,, \\
		\langle T_{ij} \rangle &= \frac{-2}{\sqrt{-g_0}} \frac{\delta \mathcal{S}_\mathrm{ren}}{\delta g_0^{ij}} = \lim_{\epsilon\to0} \biggl( \frac{1}{\epsilon^3} \frac{-2}{\sqrt{-h}} \frac{\delta \mathcal{S}_\mathrm{sub}}{\delta h^{ij}} \biggr) \,,
	\end{split}
\end{equation}
where $\mathcal{S}_\mathrm{sub} = \mathcal{S}_\mathrm{reg} + \mathcal{S}_\mathrm{c.t.}$ and $\mathcal{S}_\mathrm{reg}$ is the sum of the bulk action and GHY term computed with the cutoff $\epsilon$. Their explicit form can be derived using the FG expansions~\eqref{g:6D} and \eqref{boundaryfields6D}.

\paragraph{Scalar operator}
To compute the one-point function for the scalar, we make use of the following relation (see Section 6.3 of \cite{Faedo:2025kjf}):
\begin{equation}
  \label{Sreg:X_6D}
  \frac{1}{\sqrt{-h}} \frac{\delta \mathcal{S}_\mathrm{reg}}{\delta X} = \frac{1}{16\pi G_\mathrm{N}^{(6)}} \bigl( 8X^{-2} \epsilon \, \partial_\epsilon X \bigr) \,.
\end{equation}
On the other hand, the contribution from $\mathcal{S}_\mathrm{c.t.}$ is given by
\begin{equation}
	\frac{1}{\sqrt{-h}} \frac{\delta \mathcal{S}_\mathrm{c.t.}}{\delta X} = \frac{1}{16\pi G_\mathrm{N}^{(6)}} \, 16(1-X)  \,.
\end{equation}
Expanding in $z$ and computing the limit~\eqref{1-pt_6D}, we find
\begin{equation} \label{1pt_scalar_6D}
	\langle\mathcal{O}_X\rangle = \frac{1}{16\pi G_\mathrm{N}^{(6)}} \, 8X_3 \,,
\end{equation}
in agreement with what we expected from the discussion on the conformal dimension of the scalar operator made at the beginning of this section.

\paragraph{Two-form current}

In order to compute the one-point function of the current associated with the two-form, we can use the relation
\begin{equation}
  \frac{1}{\sqrt{-h}} \frac{\delta \mathcal{S}_\mathrm{reg}}{\delta B_{ij}} = \frac{1}{16\pi G_\mathrm{N}^{(6)}} \biggl( \frac12 \, X^4 g^{ik} g^{jl} \epsilon^5 H_{\epsilon kl} \biggr) \,,
\end{equation}
that can be derived along the same lines as \eqref{Sreg:X_6D}. The contribution of the counterterms follows straightforwardly and expanding in $z$ and computing the limit~\eqref{1-pt_6D}, we obtain
\begin{equation} \label{1pt_current_6D}
	\langle J^{ij} \rangle = \frac{1}{16\pi G_\mathrm{N}^{(6)}} \, g_0^{ik} g_0^{jl} \biggl( \frac32 B_{2\,kl} - 2X_3 B_{-1\,kl} \biggr) \,.
\end{equation}

\paragraph{Holographic stress--energy tensor}

Finally, we can compute the one-point function of the boundary stress tensor. All the divergent contributions can be shown to vanish in the limit~\eqref{1-pt_6D}, leaving only the finite terms:
\begin{equation} \label{stress_tensor_6D}
	\begin{split}
		\langle T_{ij} \rangle &= \frac{1}{16\pi G_\mathrm{N}^{(6)}} \biggl( 5g_{5\,ij} - 5g_{0\,ij} \tr(g_0^{-1} g_5) - 16g_{0\,ij} X_2 X_3 \\
		& + \bigl( B_{-1} g_0^{-1} B_2 + B_2 g_0^{-1} B_{-1} \bigr)_{ij} - \frac12 \, g_{0\,ij} \tr(g_0^{-1} B_{-1} g_0^{-1} B_2) \biggr) \,.
	\end{split}
\end{equation}
Taking the trace with the metric $g_0$ we obtain
\begin{equation} \label{stress_trace_6D}
	\langle T^k_{\ k} \rangle = \frac{1}{16\pi G_\mathrm{N}^{(6)}} \biggl( 16X_2 X_3 + \frac32 \tr(g_0^{-1} B_{-1} g_0^{-1} B_2) + 4X_3 \, |B_{-1}|^2_{g_0} \biggr) \,,
\end{equation}
while its covariant derivative, computed with the metric $g_0$ and the associated connection $\nabla^0$, is given by
\begin{align} \label{stress_divergence_6D}
	\nabla^0_i \langle T^{ij} \rangle &= \frac{1}{16\pi G_\mathrm{N}^{(6)}} \bigg[ 8X_3 \nabla^{0\,j} X_2 - 4B_{-1}^{jk} \nabla^{0\,l} (X_3 B_{-1\,kl}) + 3B_{-1}^{jk} \nabla^{0\,l} B_{2\,kl} \nonumber \\
	& - 2X_3 B_{-1\,kl} (\dd B_{-1})^{jkl} + \frac32 B_{2\,kl} (\dd B_{-1})^{jkl} \biggr] \,.
\end{align}

\paragraph{Ward identities}

The arguments in this subsection borrow the ideas from Section~4.5 of~\cite{Bianchi:2001kw}.
The renormalized action is a functional of the sources $g_0$, $X_2$ and~$B_{-1}$. Any variation of the sources produces a variation of the renormalized action given by (\cf\ definitions~\eqref{1-pt_6D} and see~\cite{Bianchi:2001kw})
\begin{equation} \label{variation}
	\delta \mathcal{S}_\mathrm{ren}[g_0,X_2,B_{-1}] = \int_{\partial M} \dd^5x \, \sqrt{-g_0} \, \biggl( -\frac12 \langle T_{ij} \rangle \delta g_0^{ij} + \langle\mathcal{O}_X\rangle \delta X_2 + \langle J^{ij} \rangle \delta B_{-1\,ij} \biggr) \,.
\end{equation}
Performing a Weyl transformation of the boundary metric, a rescaling of the FG coordinate and a suitable rescaling of the boundary values of the fields%
\footnote{Also the fields playing the role of VEVs, in our case $(g_5,X_3,B_2$), must change appropriately.}
\begin{equation}
	\delta g_0 = 2\omega g_0 \,,  \qquad  \delta z = \omega z \,,  \qquad  \delta X_2 = -2\omega X_2 \,,  \qquad  \delta B_{-1} = \omega B_{-1} \,,
\end{equation}
the metric and fields in the bulk are left invariant. The bulk action is therefore invariant ($\delta \mathcal{S}_\mathrm{reg} = 0$), and the same happens to the counterterms ($\delta \mathcal{S}_\mathrm{c.t.} = 0$). From~\eqref{variation} we derive the following identity for the trace of the stress--energy tensor
\begin{equation}
	\langle T^k_{\ k} \rangle = 2 \langle\mathcal{O}_X\rangle X_2 - \langle J^{ij} \rangle B_{-1\,ij} \,,
\end{equation}
which is identically satisfied by means of~\eqref{1pt_scalar_6D}, \eqref{1pt_current_6D} and~\eqref{stress_trace_6D}.

\vskip 1em

The bulk action is invariant also under diffeomorphisms generated, through Lie derivative, by a generic change of coordinates $\delta x^i = \xi^i(x)$, \ie
\begin{equation}
	\begin{gathered}
		\delta g_0^{ij} = -( \nabla^i \xi^j + \nabla^j \xi^i) \,,  \qquad  \delta X_2 = \xi^k \nabla_k X_2 \,, \\
		\delta B_{-1\,ij} = \xi^k \nabla_k B_{-1\,ij} + \nabla_i \xi^k \, B_{-1\,kj} + \nabla_j \xi^k \, B_{-1\,ik} \,.
	\end{gathered}
\end{equation}
Because the counterterms are written in a covariant way, they are invariant as well, therefore $\delta \mathcal{S}_\mathrm{ren} = 0$. Inserting the explicit form of the diffeomorphisms into~\eqref{variation}, we obtain the following identity
\begin{equation} \label{ward2_6D}
	g_0^{ij} \nabla^0_i \langle T_{jk} \rangle = \langle\mathcal{O}_X\rangle \nabla^0_k X_2 - 2B_{-1\,kj} \nabla^0_i \langle J^{ij} \rangle + \langle J^{ij} \rangle (\dd B_{-1})_{ijk} \,.
\end{equation}
Inserting~\eqref{1pt_scalar_6D}, \eqref{1pt_current_6D} and~\eqref{stress_divergence_6D}, it is possible to show that this identity is satisfied, representing a strong cross-check of our construction.

\subsection{The defect coordinates and holographic reconstruction}
\label{subsec:defect-coord6D}

We now want to apply the above analysis to the $\AdS_2$ solution \eqref{6Dline} and to the $\AdS_3$ solution \eqref{6Dsurface}. To do so, we first need to introduce the defect coordinates, as discussed in Section \ref{sec:defectcoordinates}.

\subsubsection{The line defect}

Let us start from the $\AdS_2\times S^3 \times I$ background describing a line defect, given in~\eqref{6Dline}, of which we recall the line element in the case $m=1$,
\begin{equation}
  \dd s_6^2  = \frac{1}{X^2 \sin^2(2\theta)} \left[ \frac{1}{(1-\lambda)^2} \, \dd s^2_{\AdS_2} + 4X^8 \dd\theta^2 + \frac{4\cos^2(2\theta)}{(2+\lambda)^2} \, \dd s^2_{S^3} \right] \,.
\end{equation}
To move to the defect coordinates, we begin writing the metric on $\AdS_2$ in Poincaré coordinates as
\begin{equation}
	\dd s^2_{\AdS_2} = \frac{-\dd\tau^2 + \dd\rho^2}{\rho^2} \,.
\end{equation}
We then perform the change of coordinates $(\theta,\rho) \mapsto (z,y)$ obtained through
\begin{equation}
	\theta = \theta(x) \,,  \qquad  \rho = z \, t(x)  \qquad\quad  \text{with}  \qquad\quad  x = \frac{y}{z} \,,
\end{equation}
where the two functions $\theta(x)$ and $t(x)$ satisfy~\eqref{eqChangeGen}
\begin{align}
	\label{eqFG1_6D}
	(\theta')^2 &= \frac{\sin^2(2\theta) \bigl[ 1 - (1-\lambda)^2 \sin^2(2\theta) \, X^2 \bigr]}{4x^2 X^6} \,, \\
	\label{eqFG2_6D}
	\frac{t'}{t} &= \frac{1 - (1-\lambda)^2 \sin^2(2\theta) \, X^2}{x} \,,
\end{align}
with $X(\theta)$ given in~\eqref{6Dline}. In the new coordinates, our metric takes the form of \eqref{DefectMetricGen}:
\begin{equation}
	\dd s_6^2 = \frac{\dd z^2}{z^2} + \frac{1}{z^2} \left[ -\frac{1}{t (t - x\,t')} \, \dd\tau^2 + \frac{t'}{x (t - x\,t')} \, \dd y^2 + \frac{4z^2 \cot^2(2\theta)}{(2+\lambda)^2 X^2} \, \dd s^2_{S^3} \right] \,.
\end{equation}
As shown in Appendix B of~\cite{Papadimitriou:2004rz}, this metric is invariant under the $\mathrm{O}(1,2)$ group, the same symmetry group of $\AdS_2$, thus providing a consistent set of FG coordinates for the initial metric.

Equation~\eqref{eqFG1_6D} can be integrated expanding in series for small values of $\theta$. The result, up to a rescaling of $x$, is%
\footnote{Equation~\eqref{eqFG1_6D} also admits a solution of the type $\theta = x + \ldots$, but we discard it because we want $x\to\infty$ close to the boundary at $\theta=0$.}
\begin{equation}
	\begin{split}
		\theta(x) &= \frac{1}{x} - \frac{8-21\lambda+6\lambda^2}{6x^3} - \frac{2\lambda}{x^4} + \frac{16-65\lambda+105\lambda^2-45\lambda^3+5\lambda^4}{5x^5} \\
		& + \frac{4\lambda (11-32\lambda+10\lambda^2)}{5x^6} + \mathcal{O}\Bigl(\frac{1}{x^7}\Bigr) \,.
	\end{split}
\end{equation}
With the expression of $\theta(x)$, we can integrate equation~\eqref{eqFG2_6D} to get
\begin{equation}
	\begin{split}
		t(x) &= x + \frac{2(1-\lambda)^2}{x} - \frac{(1-\lambda)^2 (2-\lambda)}{x^3} \\
		& + \frac{(1-\lambda)^2 (24-60\lambda+17\lambda^2+8\lambda^3-4\lambda^4)}{6x^5} + \mathcal{O}\Bigl(\frac{1}{x^6}\Bigr)
	\end{split}
\end{equation}
and obtain the full metric. Writing, schematically, the line element as
\begin{equation}
	\dd s_6^2 = \frac{\dd z^2}{z^2} + \frac{1}{z^2} \bigl( -f_1 \, \dd\tau^2 + f_2 \, \dd y^2 + f_3 \, \dd s_{S^3}^2 \bigr) \,,
\end{equation}
we have
\begin{equation}
	\begin{split}
		f_1 &= \frac{1}{4(1-\lambda)^2} + \frac{\lambda (3-2\lambda) z^2}{4(1-\lambda)^2 y^2} + \frac{\lambda^2 (67-68\lambda+20\lambda^2) z^4}{16(1-\lambda)^2 y^4} - \frac{4\lambda (3-\lambda) z^5}{5(1-\lambda)^2 y^5} + \mathcal{O}(z^6) \,, \\
		f_2 &= \frac{1}{4(1-\lambda)^2} + \frac{\lambda (3-2\lambda) z^2}{4(1-\lambda)^2 y^2} + \frac{\lambda (24+3\lambda-12\lambda^2+4\lambda^3) z^4}{16(1-\lambda)^2 y^4} - \frac{4\lambda (3-\lambda) z^5}{5(1-\lambda)^2 y^5} + \mathcal{O}(z^6) \,, \\
		f_3 &= \frac{y^2}{(2+\lambda)^2} - \frac{\lambda (5-2\lambda) z^2}{(2+\lambda)^2} - \frac{\lambda (8-3\lambda+12\lambda^2-4\lambda^3) z^4}{4(2+\lambda)^2 y^2} + \frac{16\lambda z^5}{5(2+\lambda) y^3} + \mathcal{O}(z^6) \,.
	\end{split}
\end{equation}

From the expressions above we can derive the leading behavior of the metric at the boundary,
\begin{equation}
  d s^2_6 \sim \frac{d\tilde{z}^2}{\tilde{z}^2} + \frac{1}{\tilde{z}^2} \left( -d\tau^2 + dy^2 + w^2 y^2 \, ds^2_{S^3} \right) \qquad \text{with} \qquad w = \frac{2(1-\lambda)}{2+\lambda} \, ,
\end{equation}
where we introduced the rescaled FG coordinate $\tilde{z}^2 = 4(1-\lambda)^2 z^2$. This metric makes manifest that the boundary geometry has a conical deficit at its center $y=0$ for $|w|<1$, while it has an excess for $|w|>1$.

\subsubsection{The surface defect}

For what concerns the surface defect, we start from the metric of the $\AdS_3 \times S^2 \times I$ background given in~\eqref{6Dsurface}, which we recall for the reader's convenience with $m=1$,
\begin{equation}
    \dd s_6^2  = \frac{1}{X^2 \sin^2(2\theta)} \left[ \frac{1}{(1-\lambda)^2} \left( \frac{-\dd\tau^2 + \dd u^2 + \dd\rho^2}{\rho^2} \right) + 4X^8 \dd\theta^2 + \frac{\cos^2(2\theta)}{(1+2\lambda)^2} \, \dd s^2_{S^2} \right] \,.
\end{equation}
We then perform the change of coordinates $(\theta,\rho) \mapsto (z,y)$ obtained through
\begin{equation}
	\theta = \theta(x) \,,  \qquad  \rho = z \, t(x)  \qquad\quad  \text{with}  \qquad\quad  x = \frac{y}{z} \,,
\end{equation}
with the two functions $\theta$ and $t$ satisfying again equations~\eqref{eqFG1_6D} and~\eqref{eqFG2_6D}, where now $X(\theta)$ is given in~\eqref{6Dsurface}. As a result, our metric takes the form \eqref{DefectMetricGen}
\begin{equation}
	\dd s_6^2 = \frac{\dd z^2}{z^2} + \frac{1}{z^2} \left[ \frac{-\dd\tau^2 + \dd u^2}{t (t - x\,t')} + \frac{t'}{x (t - x\,t')} \, \dd y^2 + \frac{z^2 \cot^2(2\theta)}{(1+2\lambda)^2 X^2} \, \dd s^2_{S^2} \right] \,.
\end{equation}

Equation~\eqref{eqFG1_6D} can be integrated expanding in series for small values of $\theta$. The result, up to a rescaling of $x$, is%
\footnote{As in the previous case, we do not consider a possible solution of the type $\theta = x + \mathcal{O}(x^2)$.}
\begin{equation}
	\begin{split}
		\theta(x) &= \frac{1}{x} - \frac{4-15\lambda+3\lambda^2}{3x^3} + \frac{16-120\lambda+200\lambda^2-70\lambda^3+5\lambda^4}{5x^5} + \mathcal{O}\Bigl(\frac{1}{x^7}\Bigr) \,.
	\end{split}
\end{equation}
With the expression of $\theta(x)$, we can integrate equation~\eqref{eqFG2_6D} to get
\begin{equation}
	t(x) = x + \frac{2(1-\lambda)^2}{x} - \frac{2(1-\lambda)^3}{x^3} + \frac{2(1-\lambda)^2 (6-22\lambda+15\lambda^2-\lambda^4)}{3 x^5} + \mathcal{O}\Bigl(\frac{1}{x^7}\Bigr)
\end{equation}
and obtain the full metric. Writing, schematically, the line element as
\begin{equation}
	\dd s_6^2 = \frac{\dd z^2}{z^2} + \frac{1}{z^2} \bigl( f_1 (-\dd\tau^2 + \dd u^2) + f_2 \, \dd y^2 + f_3 \, \dd s_{S^2}^2 \bigr) \,,
\end{equation}
we have
\begin{equation}
	\begin{split}
		f_1 &= \frac{1}{4(1-\lambda)^2} - \frac{\lambda z^2}{2(1-\lambda) y^2} + \frac{\lambda (4+7\lambda-14\lambda^2+5\lambda^3) z^4}{4(1-\lambda)^2 y^4} + \mathcal{O}(z^6) \,, \\
		f_2 &= \frac{1}{4(1-\lambda)^2} - \frac{\lambda z^2}{2(1-\lambda) y^2} + \frac{\lambda (8-5\lambda-2\lambda^2+\lambda^3) z^4}{4(1-\lambda)^2 y^4} + \mathcal{O}(z^6) \,, \\
		f_3 &= \frac{y^2}{4(1+2\lambda)^2} - \frac{\lambda (3-\lambda) z^2}{2(1+2\lambda)^2} - \frac{\lambda (8+5\lambda+2\lambda^2-\lambda^3) z^4}{4(1+2\lambda)^2 y^2} + \mathcal{O}(z^6) \,.
	\end{split}
\end{equation}
Similarly to the line defect, the leading behavior of the metric at the boundary is thus given by
\begin{equation}
  ds^2_6 \sim \frac{d\tilde{z}^2}{\tilde{z}^2} + \frac{1}{\tilde{z}^2} \left( -d\tau^2 + du^2 + dy^2 + w^2 y^2 \, ds^2_{S^2}  \right) \qquad \text{with} \qquad w = \frac{1-\lambda}{1+2\lambda} \, ,
\end{equation}
using the rescaled FG coordinate $\tilde{z}^2 = 4(1-\lambda)^2 z^2$. As in the previous case, the boundary geometry has a conical deficit in $y=0$ for $|w|<1$ and an excess for $|w|>1$.

\subsection{The defect observables}

We can now compute the one-point correlators derived in Section \ref{subsec:1pt_6D} in the FG coordinates presented above, for the line and surface defects.

\subsubsection{The line defect}

Let us start considering the line defect. The scalar $X$ is given by the following boundary expansion
\begin{equation}
	X = 1 - \frac{\lambda z^2}{y^2} + \frac{2\lambda z^3}{y^3} - \frac{\lambda^2 (9-4\lambda) z^4}{2y^4} + \mathcal{O}(z^5) \,, \\
\end{equation}
while the components of the two-form
\begin{equation}
	B = b \, \frac{\dd\tau \wedge \dd\rho}{\rho^2} = \hat{b} \, \dd\tau \wedge \dd y + \hat{a} \, \dd\tau \wedge \dd z
\end{equation}
read
\begin{equation}
	\begin{split}
		\hat{a} &= \frac{2\lambda}{y^2} + \frac{\lambda^2 (11-6\lambda) z^2}{y^4} - \frac{4\lambda (4-\lambda) z^3}{y^5} + \mathcal{O}(z^4) \,, \\
		\hat{b} &= \frac{\lambda}{2(1-\lambda)^2 y z} + \frac{\lambda^2 (17-10\lambda) z}{4(1-\lambda)^2 y^3} - \frac{\lambda (4-\lambda) z^2}{(1-\lambda)^2 y^4} + \mathcal{O}(z^3) \,.
	\end{split}
\end{equation}
From these expansions we can derive the one-point correlators of the defect CFT dual to our supergravity solution. The scalar one-point function~\eqref{1pt_scalar_6D} is given by
\begin{equation}
	\langle\mathcal{O}_X\rangle = \frac{1}{16\pi G_\mathrm{N}^{(6)}} \, \frac{16\lambda}{y^3} \,.
\end{equation}
When $\lambda=0$ this correlator vanishes, as expected from the fact that, on the gravity side, for $\lambda=0$ we recover the $\AdS_5$ vacuum, dual to an unbroken CFT$_4$, with no defect present.
For $\lambda\neq0$ the structure of $\langle\mathcal{O}_X\rangle$ agrees with what predicted by the literature. Indeed, in our setup the defect extends along the time direction~$\tau$, and its transverse space is parametrized in spherical coordinates with $y$ as radial coordinate. The defect is located at $y=0$, therefore $y$ also measures the distance from the defect itself.
Given a scalar operator $\mathcal{O}_X$ of conformal dimension $\Delta_X$ in such a background, its expectation value has a precise form fixed by the symmetries of the system, namely $\langle\mathcal{O}_X\rangle = c_X y^{-\Delta_X}$, where $c_X$ is a fixed coefficient~\cite{Billo:2016cpy}.
As explained in Section~\ref{subsec:1pt_6D}, in the case under exam $\mathcal{O}_X$ has scaling dimension $\Delta_X=3$, so the behavior of its one-point function $\langle\mathcal{O}_X\rangle \sim 1/y^3$ agrees with what expected.

The current one-point function~\eqref{1pt_current_6D} associated with the two-form is
\begin{equation}
	\langle J\rangle = -\frac{1}{16\pi G_\mathrm{N}^{(6)}} \, \frac{\lambda (12+\lambda)}{2(1-\lambda)^2 y^4} \, \dd\tau \wedge \dd y \,.
\end{equation}
Finally, from equation~\eqref{stress_tensor_6D}, we can compute the correlator of the holographic stress--energy tensor:
\begin{equation}
	\langle T_{ij} \rangle = \frac{1}{16\pi G_\mathrm{N}^{(6)}}
	\begin{pmatrix}
		T_{\tau\tau} & 0 & 0 \\
		0 & -T_{\tau\tau} & 0 \\
		0 & 0 & T_{S^3}
	\end{pmatrix}_{ij} \,,
\end{equation}
where we defined
\begin{equation}
	T_{\tau\tau} = \frac{2\lambda (6+8\lambda-\lambda^2)}{(1-\lambda)^2 y^5} \,,  \qquad\quad
	T_{S^3} = \frac{8\lambda (2-\lambda)}{(2+\lambda) y^3} \, g_{S^3} \,.
\end{equation}
The trace of~$\langle T_{ij} \rangle$ can be derived from~\eqref{stress_trace_6D}, which gives
\begin{equation}
	\langle T^k_{\ k} \rangle = \frac{1}{16\pi G_\mathrm{N}^{(6)}} \biggl( -\frac{8\lambda^2 (16+\lambda)}{y^5} \biggr) \,.
\end{equation}
Lastly, using~\eqref{stress_divergence_6D}, we can obtain the divergence of the stress--energy tensor
\begin{equation}
	\nabla^0_i \langle T^{ik} \rangle = \frac{1}{16\pi G_\mathrm{N}^{(6)}} \, \frac{16\lambda^2 (1-\lambda) (16+\lambda)}{y^6} \, v^k \,,
\end{equation}
where $v = 2(1-\lambda) \, \partial_y$ is a unit vector with respect to the metric $g_0$, orthogonal to the defect located at $y=0$ and extended along~$\tau$.

\subsubsection{The surface defect}

In the solution associated with the surface defect, the scalar $X$ is given by
\begin{equation}
	X = 1 - \frac{2\lambda z^2}{y^2} + \frac{2\lambda (4-5\lambda+2\lambda^2) z^4}{y^4} + \mathcal{O}(z^6) \,, \\
\end{equation}
while the component of the two-form
\begin{equation}
	B = b \, \vol_{S^2}
\end{equation}
reads
\begin{equation}
	b = -\frac{\lambda y}{(1+2\lambda)^2 z} + \frac{\lambda (4+5\lambda-\lambda^2) z}{(1+2\lambda)^2 y} - \frac{\lambda (16-16\lambda-7\lambda^2+4\lambda^3) z^3}{(1+2\lambda)^2 y^3} + \mathcal{O}(z^3) \,.
\end{equation}
 All the one-point correlators can be shown to vanish. This is an immediate consequence of the fact that all the terms playing the role of VEVs in~\eqref{1pt_scalar_6D}, \eqref{1pt_current_6D} and~\eqref{stress_tensor_6D}, namely $g_5$, $X_3$ and $B_2$, as indicated by the discussion at the beginning of Section~\ref{subsec:1pt_6D}, are identically zero.
Despite the vanishing of the one-point functions, the conformal isometries of the ambient field theory are explicitly broken by non-zero sources for the operators dual to the fields in the bulk. We then observe that the dual interpretation of this solution appears to be different from that of the line defect discussed above. This deserves further investigation for future work.

\section{Line defects in SCFT$_4$}\label{sec:Defects5D}

In this section we apply the general formulation given in Section \ref{ref:GeneralSection} to a class of supersymmetric line defects in $\mathrm{SCFT}_4$ using five-dimensional Romans supergravity as a holographic dual description. These solutions were originally obtained in \cite{Faedo:2025kjf}. The bulk geometries we consider are warped $\AdS_2\times S^2\times I$ backgrounds supported by a running scalar $X$ and a complex two-form $B$. A key feature of this family is that it is completely regular in the interior, while in the asymptotic region the metric becomes locally AdS$_5$. In \cite{Faedo:2025kjf} these backgrounds were accordingly proposed to describe line defects in the $\mathrm{SCFT}_4$ holographically dual to the AdS$_5$ vacuum.

As a first step, we construct the Fefferman--Graham expansion and the renormalized on-shell action, identifying the sources for the dual operators and the corresponding one-point functions $\langle\mathcal O_X\rangle$, $\langle J^{ij}_\alpha\rangle$ and $\langle T_{ij}\rangle$. We then rewrite the resulting boundary data in the defect coordinates introduced in Section \ref{sec:defectcoordinates}, where the boundary geometry approaches $\RR^{4,1}$, with a conical defect at $y=0$. Finally, we extract the Ward identities governing the trace anomaly and the breaking of transverse translation invariance.

\subsection{The 5D Romans supergravity}

The starting point of this section is the review of the bosonic field content of five-dimensional Romans supergravity~\cite{Romans:1985ps}. In this theory the gravitational field is coupled to a doublet of two-forms $B_{\mu\nu}^\alpha$, with $\alpha=1, 2$, and a real scalar $X$. In general, the $\ma N=4$ supergravity multiplet involves also three $\text{SU(2)}$ vectors and a $\text{U(1)}$ vector, which will be set to zero in this paper.
The bosonic part of the action is given by~\cite{Lu:1999bw, Gauntlett:2007sm}%
\footnote{Following the original notation of~\cite{Romans:1985ps}, this theory is also called $\ma N=4^+$ supergravity and can be obtained setting $g_1=g$ and $g_2=\sqrt2\,g$ in the general setup.}
\begin{equation} \label{action5D}
  \mathcal{S} = \frac{1}{16\pi G^{(5)}_\mathrm{N}} \int \Bigl[ (R-\mathcal{V}) \star 1 - 3X^{-2} \star\,d X \wedge d X - \frac12 X^{-2} \star B^\alpha \wedge B^\alpha + \frac{1}{2g} \, \varepsilon_{\alpha\beta} B^\alpha \wedge d B^\beta \Bigr] ,
\end{equation}
where $g>0$ is the gauge coupling, $\varepsilon_{\alpha\beta}$ is the totally antisymmetric symbol and the scalar potential reads
\begin{equation} \label{scalar-pot5D}
	\mathcal{V} = -4 g^2 \bigl(  X^2 + 2  X^{-1} \bigr) \,.
\end{equation}
This potential has a $\ma N=4$ AdS$_5$ extremum in $X=1$ and $B^\alpha=0$, which corresponds to the maximally supersymmetric vacuum.
The above theory has a natural description in Type IIB supergravity, as it was shown in~\cite{Lu:1999bw}.
A convenient representation of this theory can be obtained introducing a complex notation for the two-forms $B=B^1+iB^2$ and the associated field strength $H=dB$, in which case action \eqref{action5D} can be written as
\begin{equation}
  \mathcal{S} = \frac{1}{16\pi G^{(5)}_\mathrm{N}} \int \Bigl[( R-\mathcal{V}) \star\!1 - 3X^{-2} \star d X \wedge d X - \frac12 X^{-2} \star\!B \wedge \bar{B} + \frac{i}{2g} B \wedge \bar{H}  \Bigr] \,.
\end{equation}
The associated equations of motion comprise the Einstein and Klein--Gordon equations
\begin{equation} \label{einstein-eq5D}
	\begin{gathered}
		R_{\mu\nu} = 3X^{-2} \partial_\mu X \, \partial_\nu X  + \frac12 X^{-2} \Bigl( B_{(\mu|\lambda|} \bar{B}_{\nu)}^{\ \lambda} - \frac16 g_{\mu\nu} B_{\rho\sigma} \bar{B}^{\rho\sigma} \Bigr) + \frac13 g_{\mu\nu} \mathcal{V} \,, \\
		d\bigl(X^{-1} \star d X) =  - \frac16 X^{-2} \star\!B \wedge \bar{B} + \frac16 X \, \partial_X \mathcal{V}(X) \star\!1 \,,
	\end{gathered}
\end{equation}
together with the Maxwell equations for the two-form fields
\begin{equation} \label{maxwell-eq5D}
	X^2 \star H = -i g B \,.
\end{equation}
These equations must be supplemented with the condition
\begin{equation} \label{constraint-B}
	\sum_\alpha B^\alpha \wedge B^\alpha = 0 \,,
\end{equation}
which is required after truncating the gauge vectors in a consistent way.

\subsection{Warped $\mrm{AdS}_2\times S^2$ solutions}
\label{subsec:janus-sol}

The main focus of this section is the family of $\AdS_2 \times S^2$ solutions derived in~\cite{Faedo:2025kjf}. These are AdS$_2\times S^2$ geometries warped over an interval and defined by a self-dual two-form, filling the AdS$_2$ and $S^2$ directions. In addition, the solution features a non-trivial flow for the real scalar $X$. The explicit form of metric and matter fields is~\cite{Faedo:2025kjf}
\begin{equation}\label{5DJanus}
  \begin{split}
    d s^2_5 &= \frac{1}{g^2 X^2 \sin^2(2\theta)} \left[ \frac{1}{(1-\lambda)^2} d s^2_{\AdS_2} + 4X^6 d\theta^2 + \frac{\cos^2(2\theta)}{(1+\lambda)^2} d s^2_{S^2} \right] \, ,\\
    X &= \left[ 1 + \lambda \, \frac{\sin^2(2\theta)}{\cos(2\theta)} \log(\cot\theta) \right]^{-1/3}\, ,\\
    B &= -\frac{\sqrt2 \lambda}{g(1-\lambda)^2} \sin^{-1}(2\theta) \left[ 1+  \frac{\sin^2(2\theta)}{\cos(2\theta)} \log(\cot\theta) \right] \vol_{\AdS_2} \\
		& - i\,\frac{\sqrt2 \lambda}{g(1+\lambda)^2} \cot(2\theta) \left[ 1 - \frac{\sin^2(2\theta)}{\cos(2\theta)} \log(\cot\theta) \right] \vol_{S^2} \, .
  \end{split}
\end{equation}
These backgrounds are a one-parameter family of solutions defined for $\theta \in (0, \frac{\pi}{4})$ and for $\lambda>-1$, with $\lambda\neq1$. As shown in~\cite{Faedo:2025kjf}, these geometries preserve four supercharges, so they are $\ma N=4$ AdS$_2$ solutions in $D=5$. An interpretation of these solutions as line defect within $\ma N=4$ SYM (or its $\ma N=2$ orbifolds) was proposed in~\cite{Faedo:2025kjf}.

These properties can be directly inferred from solutions \eqref{5DJanus}.
First of all, we point out that for $\lambda=0$ the above geometries reproduce the $\AdS_5$ vacuum,
\begin{equation}
  d s^2_5 = \frac{1}{g^2 \sin^2(2\theta)} \left( d s^2_{\AdS_2}  + 4 d\theta^2 +  \cos^2(2\theta) \, d s^2_{S^2} \right) \, ,
\end{equation}
with radius $L = g^{-1}$. When $\lambda \neq 0$, the solution is locally AdS$_5$ for $\theta\rightarrow 0$,
\begin{equation}
\begin{split}
  d s^2_5 &\sim \frac{1}{4g^2 \theta^2} \left( \frac{1}{(1-\lambda)^2}\,d s^2_{\AdS_2}  + 4 d\theta^2 +  \frac{1}{(1+\lambda)^2}\, d s^2_{S^2} \right) \, ,\\
  X &\sim 1 \,,\\
  B &\sim -\frac{\lambda}{\sqrt2 g\,(1-\lambda)^2\,\theta}\,\vol_{\AdS_2}-\frac{i\lambda}{\sqrt2g\,(1+\lambda)^2\,\theta}\,\vol_{S^2}\,.
  \end{split}
\end{equation}
In this limit, one can show that the Ricci curvature is $R\sim -20 g^2$, which is that of AdS$_5$ vacua. We point out that the two-form is subleading, but not zero, which implies that the vacua isometries are broken in the asymptotics and the background is only locally AdS$_5$ for $\theta\rightarrow 0$. At the other endpoint $\theta=\frac{\pi}{4}$, the geometry closes off smoothly; indeed, in this region the metric describes an AdS$_2\times \mathbb{R}^3$ space parametrized as
\begin{equation}
	ds^2_5 \sim \frac{(1+\lambda)^{2/3}}{g^2 (1-\lambda)^2} \left[ ds_{\mathrm{AdS}_2}^2 + \frac{4(1-\lambda)^2}{(1+\lambda)^2} \left( d\theta^2 + \left(\theta - \frac{\pi}{4}\right)^2 ds_{S^2}^2 \right) \right]\,.
\end{equation}
It is important to point out that this class of solutions is featured by a conical singularity in the asymptotics $\theta \rightarrow 0$. Holographically, this behavior is associated to the backreaction of the defect within the ambient space.

\subsection{The on-shell action}
\label{sec:boundaryexp_5D}

As we did for the defect solutions in the previous section, the first step in our analysis is to compute the renormalized on-shell action. To do so, we begin with the standard prescription for the FG expansion of a five-dimensional metric near the $z=0$ boundary
\begin{equation}
  \label{FGds5D}
	d s_5^2 = \frac{1}{z^2} \bigl( d z^2 + g_{ij}(z,x) \, d x^i d x^j \bigr) \,,
\end{equation}
where $z$ is the FG radial coordinate. The metric tensor $g_{ij}$ reads
\begin{equation} \label{boundarymetric5D}
	g = g_0 + z^2 g_2 + z^4 \big( g_4 + \log z \, \tilde{g}_4 + \log^2\!z \, \hat{g}_4 \bigr) + \mathcal{O}(z^6) \,,
\end{equation}
while for the other fields we assume
\begin{align} \label{boundaryfields5D}
	X &= 1 + z^2 \bigl( X_2 + \log z \, \tilde{X}_2 \bigr) + \mathcal{O}(z^4) \,, \\
	B^\alpha &= z^{-1} B^\alpha_{-1} - A^\alpha_0 \wedge d z + z \bigl( B^\alpha_1 + \log z \, \tilde{B}^\alpha_1 \bigr) - z^2 \bigl( A^\alpha_2 + \log z \, \tilde{A}^\alpha_2 \bigr) \wedge d z + \mathcal{O}(z^3) \nonumber \,,
\end{align}
where $B^\alpha_{-1}$ and $B^\alpha_1$ are two-forms defined on the boundary. This prescription extends the results obtained in \cite{Faedo:2025kjf}, including the contributions along the $z$ direction given by the one-forms $A^\alpha_0$, $A^\alpha_2$ and $\tilde{A}^\alpha_2$, defined at the boundary. As we will observe later, these terms are necessary to move from the AdS$_2\times S^2$ boundary to the flat one.

As in the six-dimensional case, we start rewriting the gravity action employing the equations of motion, following the same logic of \cite{Faedo:2025kjf}. Using the Maxwell equation and substituting the expression for the Ricci scalar $R$ coming from the Einstein equations, the bulk action \eqref{action5D} can be written as%
\footnote{To keep the notation simpler, we set $g=1$ for the rest of the section.}
\begin{equation}
	\mathcal{S}_\mathrm{bulk} = \frac{1}{16\pi G^{(5)}_\mathrm{N}} \int_M d^5x \, \sqrt{-G} \, \biggl( \frac{1}{12} X^{-2} B^\alpha_{\mu\nu} B^{\alpha\mu\nu}+ \frac23 \, \mathcal{V} \biggr) \,,
\end{equation}
where $M$ is the bulk manifold and we called $G$ the five-dimensional metric to avoid confusion with the near-boundary metric \eqref{boundarymetric5D}. The GHY term has the form
\begin{equation} \label{GHYterm5D}
	\mathcal{S}_\mathrm{GHY} = \frac{1}{8\pi G^{(5)}_\mathrm{N}} \int_{\partial M} d^4x \, \sqrt{-h} \, h^{ij} K_{ij}\,,
\end{equation}
where $h_{ij}$ is the induced metric at the boundary and $K_{ij}$ its extrinsic curvature.
Lastly, the counterterms required to renormalize the action read (see Appendix~\ref{app:counterterms-5d} for the detailed construction)
\begin{align}
	\mathcal{S}_{\mathrm{c.t.}} & = \frac{1}{16\pi G^{(5)}_\mathrm{N}}  \int_{\partial M} d^4x \, \sqrt{-h} \, \biggl\{ -6 - \frac12 R[h] - 6(1-X)^2 + \frac14 \, |B^\alpha|^2_h \\
	& + \log\epsilon \biggl[ \frac14 R_{ij}[h] R^{ij}[h] - \frac{1}{12} R[h]^2 + \frac14 \tr(h^{-1} \mathrm{Ric}[h] \, h^{-1} B^\alpha h^{-1} B^\alpha) + \frac18 R[h] \, |B^\alpha|^2_h \nonumber \\
	& + \frac{1}{16} \tr\bigl[(h^{-1} B^\alpha)^2 (h^{-1} B^\beta)^2\bigr] - \frac18 \, \varepsilon_{\alpha\beta} \, \frac{\varepsilon^{ijkl}}{\sqrt{-h}} \, B^\alpha_{ij} \nabla_m (d B^\beta)^m_{\;\ kl} \biggr] - \frac{3}{\log\epsilon}(1-X)^2 \biggr\} \,, \nonumber
\end{align}
where all the quantities are to be evaluated at the cutoff $z=\epsilon$.
Notice the presence of logarithmic terms, caused by the Weyl anomalies of the theory.

\subsection{One-point correlators and Ward identities}
\label{subsec:1pt_5D}

Similarly to what we did in $D=6$, to derive the one-point functions for the dual CFT operators we first need to identify their sources. Here, the conformal dimension of the operator dual to the scalar field can be obtained inserting $X-1 \sim z^{\Delta_X}$ in~\eqref{einstein-eq5D}, yielding
\begin{equation}
  \Delta_X(\Delta_X-4)=-4 \, ,
\end{equation}
which implies that the dual scalar operator can only have conformal dimension ${\Delta_X = 2}$. As it was explained in Section~\ref{subsec:FGexpansion}, because $\Delta_X = d/2$, in this specific case the source of the dual operator is given by the term $\tilde{X}_2$ in the FG expansion~\eqref{boundaryfields5D}, and its one-point function is expected to be related to the coefficient of the $z^{\Delta_X}$ factor, namely $X_2$.

For what concerns the two-forms, plugging $B_{ij} \sim z^{\Delta_B-2}$ in \eqref{maxwell-eq5D} we obtain
\begin{equation}
  ( \Delta_B-2)^2=1
\end{equation}
and thus $\Delta_B=3$ for the current operator dual to $B$, whose source can be identified with $B^\alpha_{-1}$. The conformal dimensions here derived agree with the energy levels of the fields in the dual supergravity multiplet~\cite{Ferrara:1998ur}, following the same reasoning of Section~\ref{subsec:1pt_6D}.

The one-point functions associated with these operators are then given by~\cite{deHaro:2000vlm,Bianchi:2001kw}
\begin{equation} \label{1-pt_5D}
	\begin{split}
		\langle\mathcal{O}_X\rangle &= \frac{1}{\sqrt{-g_0}} \frac{\delta \mathcal{S}_\mathrm{ren}}{\delta \tilde{X}_2} =
		\lim_{\epsilon\to0} \biggl( \frac{\log\epsilon}{\epsilon^2} \frac{1}{\sqrt{-h}} \frac{\delta \mathcal{S}_\mathrm{sub}}{\delta X} \biggr) \,, \\
		\langle J^{ij}_\alpha \rangle &= \frac{1}{\sqrt{-g_0}} \frac{\delta \mathcal{S}_\mathrm{ren}}{\delta B^\alpha_{-1\,ij}} =
		\lim_{\epsilon\to0} \biggl( \frac{1}{\epsilon^5} \frac{1}{\sqrt{-h}} \frac{\delta \mathcal{S}_\mathrm{sub}}{\delta B^\alpha_{ij}} \biggr) \,, \\
		\langle T_{ij} \rangle &= \frac{-2}{\sqrt{-g_0}} \frac{\delta \mathcal{S}_\mathrm{ren}}{\delta g_0^{ij}} = \lim_{\epsilon\to0} \biggl( \frac{1}{\epsilon^2} \frac{-2}{\sqrt{-h}} \frac{\delta \mathcal{S}_\mathrm{sub}}{\delta h^{ij}} \biggr) \,,
	\end{split}
\end{equation}
where $\mathcal{S}_\mathrm{sub} = \mathcal{S}_\mathrm{reg} + \mathcal{S}_\mathrm{c.t.}$, with $\mathcal{S}_\mathrm{reg}$ being the sum of the bulk and GHY actions computed with the cutoff $\epsilon$. We can now derive the explicit form of the above correlators using the boundary expansions \eqref{boundarymetric5D} and \eqref{boundaryfields5D}.

\paragraph{Scalar operator}

In order to compute $\langle\mathcal{O}_X\rangle$ in \eqref{1-pt_5D} we recall the relation\footnote{See Section 6.3 of \cite{Faedo:2025kjf} for the explicit derivation.\label{fn}}
\begin{equation}
	\begin{split}
		\frac{1}{\sqrt{-h}} \frac{\delta \mathcal{S}_\mathrm{reg}}{\delta X} &= \frac{1}{16\pi G^{(5)}_\mathrm{N}} \bigl( 6X^{-2} \epsilon \, \partial_\epsilon X \bigr) \,,
	\end{split}
\end{equation}
while the contribution from the counterterms is given by
\begin{equation}
	\frac{1}{\sqrt{-h}} \frac{\delta \mathcal{S}_\mathrm{c.t.}}{\delta X} = \frac{1}{16\pi G^{(5)}_\mathrm{N}} \biggl[ 12(1-X) + \frac{6}{\log\epsilon} (1-X) \biggr] \,.
\end{equation}
Using the expansion \eqref{boundaryfields5D} and extracting the limit~\eqref{1-pt_5D}, we obtain
\begin{equation} \label{1pt_scalar5D}
	\langle\mathcal{O}_X\rangle = -\frac{6X_2}{16\pi G^{(5)}_\mathrm{N}} \,,
\end{equation}
as expected from its conformal dimension.

\paragraph{Two-form current}

We can now compute the one-point function of the current associated to the two-forms. For $\mathcal{S}_\mathrm{reg}$ we may use the relation$^{\ref{fn}}$
\begin{equation}
	\frac{1}{\sqrt{-h}} \frac{\delta \mathcal{S}_\mathrm{reg}}{\delta B^\alpha_{ij}} = \frac{1}{16\pi G^{(5)}_\mathrm{N}} \biggl( \frac{3}{4g_1^2} \, X^2 g^{ik} g^{jl} \epsilon^5 \, \partial_{[\epsilon} B^\alpha_{kl]} \biggr) \,,
\end{equation}
whereas the contribution of $\mathcal{S}_\mathrm{c.t.}$ to the one-point function of the two-forms $B^\alpha$ is
\begin{align}
  \frac{1}{\sqrt{-h}} \frac{\delta \mathcal{S}_\mathrm{c.t.}}{\delta B^\alpha_{ij}} &= \frac{1}{16\pi G^{(5)}_\mathrm{N}} \, h^{ik} h^{jl} \biggl[ \frac14 B^\alpha_{kl} - \log\epsilon \biggl( \frac14 \bigl( \mathrm{Ric}[h] \, h^{-1} B^\alpha + B^\alpha h^{-1} \mathrm{Ric}[h] \bigr)_{kl} \nonumber \\
  & - \frac18 R[h] B^\alpha_{kl} + \frac18 \bigl( B^\alpha h^{-1} B^\beta h^{-1} B^\beta + B^\beta h^{-1} B^\beta h^{-1} B^\alpha \bigr)_{kl} \biggr) \biggr] \, .
\end{align}
Expanding in $z$ and computing the limit~\eqref{1-pt_5D}, we have
\begin{equation} \label{1pt_current5D}
	\langle J^{ij}_\alpha \rangle = \frac{1}{16\pi G^{(5)}_\mathrm{N}} \, g_0^{ik} g_0^{jl} \biggl( \frac12 B^\alpha_{1\,kl} + \frac14 \tilde{B}^\alpha_{1\,kl} - \frac12 X_2 B^\alpha_{-1\,kl} - \frac14 (d A^\alpha_0)_{kl} \biggr) \,,
\end{equation}
where $\tilde{B}^\alpha_1$ and $\dd A^\alpha_0$ are given in~\eqref{5d:B1t} and~\eqref{5d:dA0}, respectively.

\paragraph{Holographic stress--energy tensor}

The one-point function of the stress--energy tensor is defined as
\begin{equation}
	\langle T_{ij} \rangle = \lim_{\epsilon\to0} \biggl( \frac{1}{\epsilon^2} \, T_{ij}[h] \biggr) \,,
\end{equation}
where $T_{ij}[h]$ is the boundary stress tensor. This quantity can be decomposed as
\begin{equation}
	T_{ij}[h] = T^\mathrm{reg}_{ij}[h] + T_{ij}^\mathrm{c.t.}[h] \,,  \qquad  \text{with}  \qquad
	T^\mathrm{reg}_{ij}[h] = -\frac{1}{8\pi G^{(5)}_\mathrm{N}} \bigl( K_{ij} - K h_{ij} \bigr) \,,
\end{equation}
where $K=h^{kl} K_{kl}$ is the trace of the extrinsic curvature tensor.

It can be shown that all divergent contributions vanish in the limit~\eqref{1-pt_5D}, leaving only the finite terms. More details about the derivation of $\langle T_{ij} \rangle$ are included in Appendix~\ref{app:counterterms-5d}. The account to the total stress--energy tensor is
\begin{align} \label{stress_tensor5D}
	\langle T_{ij} \rangle &= \frac{1}{16\pi G^{(5)}_\mathrm{N}} \biggl( 4g_{4\,ij} + \tilde{g}_{4\,ij} + g_{2\,ij} \tr(g_0^{-1} g_2) - 4g_{0\,ij} \tr(g_0^{-1} g_4) - g_{0\,ij} \tr(g_0^{-1} \tilde{g}_4) \nonumber \\
	& + g_{0\,ij} \tr(g_0^{-1} g_2 g_0^{-1} g_2) - \frac12 \, g_{0\,ij} \tr^2(g_0^{-1} g_2) - 6g_{0\,ij} \bigl( X_2^2 + X_2 \tilde{X}_2 \bigr) + \Delta R_{ij} \nonumber \\
	& + X_2 \bigl( B^\alpha_{-1} g_0^{-1} B^\alpha_{-1} \bigr)_{ij} - \frac12 \tilde{X}_2 \bigl( B^\alpha_{-1} g_0^{-1} B^\alpha_{-1} \bigr)_{ij} - \frac12 \bigl( B^\alpha_{-1} g_0^{-1} g_2 g_0^{-1} B^\alpha_{-1} \bigr)_{ij} \nonumber \\
	& + \frac14 \bigl( B^\alpha_{-1} g_0^{-1} d A^\alpha_0 + d A^\alpha_0 g_0^{-1} B^\alpha_{-1} \bigr)_{ij} - \frac14 \, g_{0\,ij} |A^\alpha_0|^2_{g_0} \biggr) \,.
\end{align}
The trace of the stress--energy tensor, computed with the metric $g_0$, reads
\begin{align} \label{stress_trace5D}
	\langle T^k_{\ k} \rangle &= \frac{1}{16\pi G^{(5)}_\mathrm{N}} \biggl( \frac14 R_{ij}[g_0] R^{ij}[g_0] - \frac{1}{12} R[g_0]^2 + 3\tilde X^2_2 - 12X_2 \tilde{X}_2  + \frac12 \tr(g_0^{-1} B^\alpha_{-1} g_0^{-1} B^\alpha_1) \nonumber \\
  & + \frac12 \tr(g_0^{-1} \mathrm{Ric}[g_0] \, g_0^{-1} B^\alpha_{-1} g_0^{-1} B^\alpha_{-1}) + \frac{3}{16} \tr\bigl[(g_0^{-1} B^\alpha_{-1})^2 (g_0^{-1} B^\beta_{-1})^2\bigr] - \frac12 \, |d B^\alpha_{-1}|^2_{g_0} \biggr) \,,
\end{align}
while its covariant derivative, again with respect to the metric~$g_0$, is given by
\begin{equation} \label{stress_divergence5D}
	\nabla^0_i \langle T^{ij} \rangle = \frac{1}{16\pi G^{(5)}_\mathrm{N}} \big( -6X_2 \, g_0^{jk} \nabla^0_k \tilde{X}_2 + \mathbb{B}^{\alpha\,ml} g_{0\,kl} \nabla^0_m B^{\alpha\,jk}_{-1} - B^{\alpha\,jk}_{-1} g_{0\,kl} \nabla^0_m \mathbb{B}^{\alpha\,ml} \bigr) \,,
\end{equation}
where, for convenience, we defined the two-form
\begin{equation}
	\mathbb{B}^\alpha \equiv B^\alpha_1 + \frac12 \tilde{B}^\alpha_1 - X_2 B^\alpha_{-1} - \frac12 \, \dd A^\alpha_0 \,.
\end{equation}

\paragraph{Ward identities}

The argument that leads to the Ward identity relating the one-point functions that we just derived is the same as the one presented in Section~\ref{subsec:1pt_6D}. The main difference worth noticing is that for even-dimensional boundaries, under the Weyl transformation, the counterterms might not be invariant, $\delta \mathcal{S}_\mathrm{c.t.} \neq 0$. Indeed this happens when logarithmic contributions (\ie\ proportional to $\log\epsilon$) are present. In this case a Weyl anomaly appears, and from~\eqref{variation} we derive the following identity for the trace of the stress--energy tensor
\begin{equation}
	\langle T^k_{\ k} \rangle = 2 \langle\mathcal{O}_X\rangle \tilde{X}_2 - \langle J^{ij}_\alpha \rangle B^\alpha_{-1\,ij} + \mathcal{A} \,,
\end{equation}
where $\mathcal{A}$ is the conformal anomaly.
In the case at hand, putting together~\eqref{1pt_scalar5D}, \eqref{1pt_current5D} and~\eqref{stress_trace5D} we obtain
\begin{align} \label{anomaly}
	\mathcal{A} &= \frac{1}{16\pi G^{(5)}_\mathrm{N}} \biggl( \frac14 R_{ij}[g_0] R^{ij}[g_0] - \frac{1}{12} R[g_0]^2 + 3\tilde X^2_2 + \frac14 \tr(g_0^{-1} \mathrm{Ric}[g_0] \, g_0^{-1} B^\alpha_{-1} g_0^{-1} B^\alpha_{-1}) \nonumber \\
	& + \frac{1} {16}\tr\bigl[(g_0^{-1} B^\alpha_{-1})^2 (g_0^{-1} B^\beta_{-1})^2\bigr] - \frac12 \, |d B^\alpha_{-1}|^2_{g_0} \biggr) \,.
\end{align}
This result can be compared with what already present in the literature. The gravitational anomaly, represented by the first two terms, reproduces the standard expression~\cite{Henningson:1998gx}. The other contributions come from the coupling of the theory to matter fields, and, in particular, the scalar anomaly agrees with \cite{Bianchi:2001kw}.

The Ward identity for the conservation of the stress--energy tensor closely resembles~\eqref{ward2_6D}
\begin{equation}
	g_0^{ij} \nabla^0_i \langle T_{jk} \rangle = \langle\mathcal{O}_X\rangle \nabla^0_k \tilde{X}_2 - 2B^\alpha_{-1\,kj} \nabla^0_i \langle J^{ij}_\alpha \rangle + \langle J^{ij}_\alpha \rangle (d B^\alpha_{-1})_{ijk} \,.
\end{equation}
Inserting~\eqref{1pt_scalar5D}, \eqref{1pt_current5D} and~\eqref{stress_divergence5D}, it is possible to show that this identity is satisfied, representing a strong cross-check of our construction.%
\footnote{In order to prove this we used the fact that $2g_{0\,ij} \langle J_\alpha^{jk} \rangle A^\alpha_{0\,k} = \langle J_\alpha^{jk} \rangle (d B^\alpha_{-1})_{ijk}$.}

\subsection{The defect coordinates and holographic reconstruction}

We now apply the above analysis to the AdS$_2$ solution \eqref{5DJanus}, whose metric reads
\begin{equation}
	d s_5^2 = \frac{1}{(1-\lambda)^2 X^2 \sin^2(2\theta)} \, \left(\frac{-d\tau^2 + d\rho^2}{\rho^2}\right) + \frac{4X^4 \, d\theta^2}{\sin^2(2\theta)} + \frac{\cot^2(2\theta)}{(1+\lambda)^2 X^2} \, d s^2_{S^2} \,,
\end{equation}
where we set the gauge coupling $g=1$ and we explicitly wrote the $\AdS_2$ line element in Poincaré coordinates $(\tau,\rho)$, with $\rho>0$.
As we explained in Section~\ref{sec:defectcoordinates}, we perform the change of coordinates $(\theta,\rho) \mapsto (z,y)$ defined as follows
\begin{equation}\label{changecoord5D}
	\theta = \theta(x) \,,  \qquad  \rho = z \, t(x)  \qquad\quad  \text{with}  \qquad\quad  x = \frac{y}{z} \,.
\end{equation}
The functions $\theta(x)$ and $t(x)$ satisfy the following differential constraints~\eqref{eqChangeGen}
\begin{align}
	\label{eqFG1_5D}
	(\theta')^2 &= \frac{\sin^2(2\theta) \bigl[ 1 - (1-\lambda)^2 \sin^2(2\theta) \, X^2 \bigr]}{4x^2 X^4} \,, \\
	\label{eqFG2_5D}
	\frac{t'}{t} &= \frac{1 - (1-\lambda)^2 \sin^2(2\theta) \, X^2}{x} \,,
\end{align}
where the expression of the scalar field $X(\theta)$ is given in~\eqref{5DJanus}.
In this new coordinates the metric takes the form of \eqref{DefectMetricGen},
\begin{equation}
	ds_5^2 = \frac{dz^2}{z^2} + \frac{1}{z^2} \left[ -\frac{1}{t (t - x\,t')} \, d\tau^2 + \frac{t'}{x (t - x\,t')} \, d y^2 + \frac{z^2 \cot^2(2\theta)}{(1+\lambda)^2 X^2} \, d s^2_{S^2} \right] \,.
\end{equation}
This metric is $\mathrm{O}(1,2)$ invariant, thus providing a consistent set of FG coordinates for the initial metric.

Integrating equations~\eqref{eqFG1_5D} and \eqref{eqFG2_5D} in a closed form is impossible, except in the case $\lambda=0$, corresponding to the global $\AdS_5$ vacua. However, equation~\eqref{eqFG1_5D} can be integrated perturbatively for small values of $\theta$ and large values of $x$. The result is given by
\begin{align}
	\theta(x) &= \frac{1}{x} - \frac{1}{x^3} \biggl[ \frac{4 - 8\lambda + 3\lambda^2}{3} - \frac{4\lambda}{3} \log x \biggr] + \frac{1}{x^5} \biggl[ \frac{288 - 900\lambda + 955\lambda^2 - 480\lambda^3 + 90\lambda^4}{90}	\nonumber \\
	& - \frac{2\lambda (24 - 43\lambda + 18\lambda^2)}{9} \log x + \frac{20\lambda^2}{9} \log^2\!x \biggr] + \mathcal{O}\Bigl(\frac{1}{x^6}\Bigr) \,.
\end{align}
With this expression of $\theta(x)$, we can integrate equation~\eqref{eqFG2_5D} to get
\begin{equation}
	t(x) = x + \frac{2 (1-\lambda)^2}{x} -\frac{2 (1-\lambda)^2 (3-2\lambda)}{3x^3}  + \mathcal{O}\Bigl(\frac{1}{x^4}\Bigr)
\end{equation}
and obtain the full metric. Writing schematically the line element as follows
\begin{equation}
	ds_5^2 = \frac{dz^2}{z^2} + \frac{1}{z^2} \bigl( -f_1 \, d\tau^2 + f_2 \, d y^2 + f_3 \, d s_{S^2}^2 \bigr) \,,
\end{equation}
the warp factors $f_1$, $f_2$ and $f_3$ are given by
\begin{align}
	f_1 &= \frac{1}{4 (1-\lambda)^2} + \frac{\lambda (4-3\lambda) z^2}{6 (1-\lambda)^2 y^2} + \frac{\lambda z^4}{36 (1-\lambda)^2 y^4} \bigl[ -3 (12 - 47\lambda + 48\lambda^2 - 15\lambda^3) \\
	& + 12 (4-\lambda) \log y - 8\lambda \log^2\!y - 4 (12 - 3\lambda - 4\lambda \log y) \log z - 8\lambda \log^2\!z \bigr] + \mathcal{O}(z^5) \,, \nonumber \\
	f_2 &= \frac{1}{4 (1-\lambda)^2} + \frac{\lambda (4-3\lambda) z^2}{6 (1-\lambda)^2 y^2} + \frac{\lambda z^4}{36 (1-\lambda)^2 y^4} \bigl[ 3 (4 + 3\lambda - 8\lambda^2 + 3\lambda^3) \nonumber \\
	& + 12 (4-\lambda) \log y - 8\lambda \log^2\!y - 4 (12 - 3\lambda - 4\lambda \log y) \log z - 8\lambda \log^2\!z \bigr] + \mathcal{O}(z^5) \,, \nonumber \\
		f_3 &= \frac{y^2}{4 (1+\lambda)^2} - \frac{\lambda (8-3\lambda) z^2}{6 (1+\lambda)^2} + \frac{\lambda z^4}{36 (1+\lambda)^2 y^2} \bigl[ 3 (4 + 3\lambda - 8\lambda^2 + 3\lambda^3) \nonumber \\
	& - 12 (4+\lambda) \log y - 8\lambda \log^2\!y + 4 (12 + 3\lambda + 4\lambda \log y) \log z - 8\lambda \log^2\!z \bigr] + \mathcal{O}(z^5) \,. \nonumber
\end{align}
From these expression we can immediately extract the leading behavior of the metric at the boundary,
\begin{equation}
 d s_5^2 \sim \frac{d\tilde{z}^2}{\tilde{z}^2} + \frac{1}{\tilde{z}^2} \left( -d\tau^2 + d y^2 + w^2 y^2 \, d s_{S^2}^2 \right) \qquad \text{with} \qquad w = \frac{1-\lambda}{1+\lambda} \,,
\end{equation}
where we introduced the rescaled FG coordinate $\tilde{z}^2=4(1-\lambda)^2 z^2$.
From this metric is manifest that the boundary geometry has a conical deficit at its center $y=0$ for $|w|<1$ ($\lambda>0$) and an excess for $|w|>1$ ($\lambda<0$).

\subsection{The defect observables}

Let us explicitly compute the one-point correlators derived in Section \ref{subsec:1pt_5D} in the defect FG coordinates.
As first ingredient, the scalar $X$ is characterized by the following boundary expansion
\begin{equation}
	X = 1 - \frac{4\lambda (\log y \!-\! \log z) z^2}{3y^2} - \frac{4\lambda [ 3 \!-\! 8\lambda \!+\! 3\lambda^2 \!-\! 6(1\!-\!\lambda)^2 \log y \!+\! 6(1\!-\!\lambda)^2 \log z ] z^4}{9y^4} + \mathcal{O}(z^5) \,.
\end{equation}
The components of the two-forms have the form
\begin{equation}
	B^1 = b \, \frac{d\tau \wedge d\rho}{\rho^2} = \hat{b} \, d\tau \wedge d y + \hat{a} \, d\tau \wedge dz \,,  \qquad\qquad  B^2 = c \, \vol_{S^2} \,,
\end{equation}
where the functions $\hat{a}$, $\hat{b}$ and $c$ have the following expansion
\begin{align}
	\hat{a} &= -\frac{2\sqrt2 \lambda}{y^2} + \frac{2\sqrt2 \lambda [ 12 - 20\lambda + 9\lambda^2 - 4(3-\lambda) \log y + 4(3-\lambda) \log z ] z^2}{3y^4} + \mathcal{O}(z^3) \,, \nonumber \\
	\hat{b} &= -\frac{\lambda}{\sqrt2 (1-\lambda)^2 y z} + \frac{\lambda [ 12 - 28\lambda + 15\lambda^2 - 4(3-\lambda) \log y + 4(3-\lambda) \log z ] z}{3\sqrt2 (1-\lambda)^2 y^3} + \mathcal{O}(z^2) \,, \nonumber \\
	c &= -\frac{\lambda y}{\sqrt2 (1+\lambda)^2 z} + \frac{\lambda [ 8\lambda - 3\lambda^2 + 4(3+\lambda) \log y - 4(3+\lambda) \log z ] z}{3\sqrt2 (1+\lambda)^2 y} + \mathcal{O}(z^2) \,.
\end{align}
We now summarize the holographic observables associated with the defect.
The scalar one-point function~\eqref{1pt_scalar5D} is given by
\begin{equation}
	\langle\mathcal{O}_X\rangle = \frac{1}{16\pi G^{(5)}_\mathrm{N}} \, \frac{8\lambda \log y}{y^2} \,.
\end{equation}
As for the line defect solutions studied previously in $D=6$, this expectation value has the correct power-law scaling for a scalar operator with conformal weight $\Delta_X$, namely $\langle\mathcal{O}_X\rangle \sim 1/y^{\Delta_X}$, with $\Delta_X=2$. However, in this case we observe the presence of a novel logarithmic term, which, although unusual in the defect CFT literature, has already appeared in the holographic study of mass deformations of $\mathcal{N}=4$ SYM~\cite{Arav:2020obl} and monodromy defects within $\mathcal{N}=4$ SYM and in the $\mathcal{N}=1$ Leigh--Strassler SCFT~\cite{Arav:2024exg}. We do not have a clear understanding of how to interpret this logarithmic dependence in the correlation function. Such a behavior is reminiscent of logarithmic CFTs \cite{Gurarie:1993xq,Gaberdiel:2001tr}, although its physical interpretation remains unclear. One possible interpretation is that this behavior is related to a conformal anomaly in the defect SCFT, although we do not currently have evidence supporting this picture.

From equation~\eqref{1pt_current5D} we can extract the one-point functions associated with the two-form
\begin{equation}
	\begin{split}
		\langle J_1 \rangle &= \frac{\lambda (2-2\lambda+\lambda^2 - 4\log y)}{2\sqrt2 \, (1-\lambda)^2 y^3} \, \dd \tau \wedge \dd y \,, \\
		\langle J_2 \rangle &= -\frac{\lambda (2-2\lambda+\lambda^2 - 4\log y)}{2\sqrt2 \, (1+\lambda)^2 y} \, \vol_{S^2} \,.
	\end{split}
\end{equation}
The holographic stress--energy tensor can be computed using equation~\eqref{stress_tensor5D}, and its correlator has the general form
\begin{equation}
	\langle T_{ij} \rangle = \frac{1}{16\pi G^{(5)}_\mathrm{N}}
	\begin{pmatrix}
		T_{\tau\tau} & 0 & 0 \\
		0 & T_{yy} & 0 \\
		0 & 0 & T_{S^2}
	\end{pmatrix}_{ij} \,,
\end{equation}
where the boundary expansion for the three independent components is the following
\begin{equation}
\begin{split}
T_{\tau\tau} &= \frac{\lambda \left( 24 - 24\lambda + 12\lambda^2 + 3\lambda^3 \right)}{3(1-\lambda)^2 y^4} - \frac{8\lambda \left( 2 + 2\lambda - \lambda^2 \right)}{3(1-\lambda)^2 y^4} \log y \, , \\[3pt]
T_{yy} &= -\frac{\lambda \left( 8 + 8\lambda - 4\lambda^2 + 3\lambda^3 \right)}{3(1-\lambda)^2 y^4} + \frac{8\lambda \left( 2 + 2\lambda - \lambda^2 \right)}{3(1-\lambda)^2 y^4} \log y \, , \\[3pt]
T_{S^2} &= \left[ \frac{\lambda \left( 16 - 8\lambda + 4\lambda^2 - 3\lambda^3 \right)}{3(1+\lambda)^2 y^2} - \frac{8\lambda \left( 2 - 2\lambda - \lambda^2 \right)}{3(1+\lambda)^2 y^2} \log y \right] g_{S^2} \, .
\end{split}
\end{equation}
Using these expressions we can compute the trace of~$\langle T_{ij} \rangle$, whose general form was written in~\eqref{stress_trace5D},
\begin{equation}
	\langle T^k_{\ k} \rangle = \frac{1}{16\pi G^{(5)}_\mathrm{N}} \left( -\frac{16\lambda^4}{y^4} + \frac{256\lambda^2}{3y^4} \log y \right).
\end{equation}
The expression for the conformal anomaly~\eqref{anomaly} is given by
\begin{equation}
	\mathcal{A} = \frac{1}{16\pi G^{(5)}_\mathrm{N}} \, \frac{32\lambda^2 (1-\lambda)}{y^4} \,.
\end{equation}
Notice that $\mathcal{A}$ vanishes for the vacuum solution $\lambda=0$, even though it includes also non-matter terms; this is due to the fact that in the defect coordinates $(z,y)$ the $\AdS_5$ vacuum has a flat boundary.
Finally, using \eqref{stress_divergence5D} we can obtain the covariant derivative of the stress tensor,
\begin{equation}
	\nabla^0_i \langle T^{ik} \rangle = \frac{1}{16\pi G^{(5)}_\mathrm{N}} \left( \frac{32(1-\lambda)\lambda^2 (12 - 6\lambda + 3\lambda^2)}{3y^5} - \frac{512(1-\lambda)\lambda^2}{3y^5}\log y \right) n^k \,,
\end{equation}
where $n = 2(1-\lambda) \, \partial_y$ is a unit vector with respect to the metric $g_0$, orthogonal to the defect, which is located at $y=0$, and extended along~$\tau$.

\section{Conclusions}

The main focus of this paper is to formulate a comprehensive analysis of conformal defects from the dual supergravity perspective. We consider warped AdS$_p\times S^q$ solutions coupled to higher-form gauge fields and scalar fields, which are dual to line and surface defect SCFTs, and develop a systematic approach to the derivation of defect observables for this class of backgrounds. In these solutions, the higher-form and scalar fields are directly related to the backreaction of the defect on the ambient theory.

A crucial step in our procedure is to perform a change of coordinates that allows us to move to a frame with a flat conformal boundary. This makes it possible to reconstruct the bulk geometry in a coordinate system where the boundary is manifestly flat. In this frame, the standard holographic dictionary can be applied to extract observables of the dual defect field theory.

In this work we considered three explicit examples, corresponding to line and surface defects in five- and six-dimensional Romans supergravity theories. Two of these solutions were already present in the literature and describe an $\AdS_2 \times S^3 \times I$ family of geometries in six-dimensional supergravity~\cite{Chen:2019qib} and an $\AdS_2 \times S^2 \times I$ class in five dimensions~\cite{Faedo:2025kjf}. The remaining one was constructed in the present paper and represents a new family of solutions of $D=6$ Romans $\mathrm{F}(4)$ gauged supergravity, with the structure of an $\AdS_3 \times S^2$ geometry warped over an interval. This new background shares several properties with the previous solutions. In particular, it preserves half of the original supersymmetry and features a non-trivial profile for a scalar field and two-form fields. Moreover, it is regular at one endpoint of the interval, with an $\AdS_3 \times \RR^3$ structure, and asymptotically locally $\AdS_6$ at the other endpoint. The asymptotic $\AdS$ boundary presents a conical singularity at the location of the defect. Nevertheless, this does not prevent the holographic interpretation of the solution as a surface defect within an ambient SCFT$_5$, nor the computation of field-theory one-point functions.

To set the stage for the analysis of the dual field theory, we described the holographic renormalization procedure in both five and six dimensions, extending the results of~\cite{Faedo:2025kjf} in 5D and~\cite{Alday:2014bta,Chen:2019qib} in 6D. This prescription begins with the construction of the counterterms needed to cancel the divergences of the on-shell action and obtain a finite result. The knowledge of the renormalized on-shell action then enables us to derive the one-point functions of the dual operators, namely the scalar operator, the currents, the holographic stress tensor, and the corresponding Ward identities.

Finally, we applied the above procedure to the three geometries under consideration, after performing the coordinate transformation required to obtain an asymptotically flat boundary. The identification of the correct defect coordinates is crucial for obtaining the appropriate Fefferman--Graham expansion, from which the holographic observables characterizing the dual field theory can be extracted.
In general, we found that the scalar one-point function exhibits the expected power-law scaling with the distance from the defect predicted by defect CFT. However, in $D=5$ an additional logarithmic dependence appears, which had already been observed in \cite{Arav:2020obl,Arav:2024exg}. The origin of this logarithmic behavior remains unclear and suggests that the holographic interpretation of these solutions is not yet complete.
More generally, a major challenge for the future is to connect the holographic study of conformal defects arising in string theory with the broader literature on defects in quantum field theory and effective field theories. In this context, a central question is how the ambient theory constrains the possible RG flows on the defect \cite{Cuomo:2021rkm}. Our current understanding of defect theories engineered by D-brane intersections is still far from this level of detail. Nevertheless, it would be a very ambitious and interesting goal to study defect RG flows holographically and relate them to the underlying brane dynamics.

\section*{Acknowledgements}

We would like to thank Lorenzo Bianchi, Marco Billò, Giuseppe Dibitetto, Francesco Galvagno, Marco Meineri, and Matthew Roberts for interesting discussions and comments. During the first part of this work N. P. was supported by INFN Turin with a GSS postdoctoral fellowship. N.P. thanks the members of the hep-th group at the University of Turin for their kind hospitality while parts of this work were carried out.


\appendix

\section{Conventions}
\label{app:conventions}

 Given a generic $p$-form $\omega$ in $D$ spacetime dimensions, the components of its Hodge dual are defined as
\begin{equation}
	(\star\,\omega)_{\mu_1\ldots\mu_q} = \frac{1}{p!} \, g_{\mu_1\nu_1} \ldots g_{\mu_q\nu_q} \frac{\varepsilon^{\nu_1\ldots\nu_q\rho_1\ldots\rho_p}}{\sqrt{-g}} \, \omega_{\rho_1\ldots\rho_p} \,,
\end{equation}
where $q=D-p$ and the Levi--Civita symbol has $\varepsilon^{012\ldots D-1} = -1$.

The norm of a $p$-form with respect to a given metric $g$ is denoted as
\begin{equation}
	|\omega|^2_g \equiv \frac{1}{p!} \, g^{i_1 j_1} \ldots g^{i_p j_p} \, \omega_{i_1 \ldots i_p} \, \omega_{j_1 \ldots j_p} \,.
\end{equation}

\section{Counterterms in $D=6$}
\label{app:counterterms-6d}

In this appendix we summarize the computation of the counterterms necessary for the holographic renormalization procedure. Assuming a Fefferman--Graham expansion of the fields as in \eqref{FGds6D}--\eqref{boundaryfields6D} (or \eqref{FGds5D}--\eqref{boundaryfields5D} in $D=5$), the $D$-dimensional Ricci tensor decomposes as
\begin{align}
  R_{ij} &= R_{ij}[g] + \frac12 \, g'_{ik} g^{kl} g'_{lj} - \frac14 \, g'_{ij} \tr(g^{-1} g') - \frac12 \, g_{ij}'' + \frac12 \, z^{-1} \biggl( (d-1) g'_{ij} +  g_{ij} \tr(g^{-1} g') \biggr) \nonumber \\
  & - d\,z^{-2} g_{ij} \,, \nonumber \\
	R_{iz} &= \frac12 \, g^{jk} \nabla_k g'_{ij} - \frac12 \, g^{jk} \nabla_i g'_{jk} \,, \\[0.2em]
	R_{zz} &= \frac14 \tr(g^{-1} g' g^{-1} g') - \frac12 \tr(g^{-1} g'') + \frac12 \, z^{-1} \tr(g^{-1} g') - d\,z^{-2} \nonumber \,,
\end{align}
where $d=D-1$, the prime denotes the derivative with respect to $z$ and $R_{ij}[g]$ and $\nabla_i$ are constructed using the lower-dimensional metric $g$. Another key formula in the derivation of the counterterms is the expansion of the determinant, which in $d=5$ reads
\begin{align}
	\sqrt{-g} &= \sqrt{-g_0} \biggl[ 1 + \frac12 z^2 \tr(g_0^{-1} g_2) + \frac12 z^4 \biggl( \! \tr(g_0^{-1} g_4) - \frac12 \tr(g_0^{-1} g_2 g_0^{-1} g_2) + \frac14 \tr^2(g_0^{-1} g_2) \! \biggr) \nonumber \\
	& + \frac12 z^5 \tr(g_0^{-1} g_5) \biggr] + \mathcal{O}(z^6) \,.
\end{align}

\subsection{Order-by-order equations of motion}

In the Fefferman--Graham expansion, most of the higher-order contributions are fixed in terms of the leading-order ones through the equations of motion. These relations are derived solving the equations of motion~\eqref{einstein-eq6D}--\eqref{maxwell-eq6D} order-by-order in $z$, where, for the sake of simplicity, we set $m=1$. From the $(iz)$ and $(ij)$ components of~\eqref{maxwell-eq6D} we get the expressions of $A_0$ and $B_1$:
\begin{align} \label{A0_6D}
	A_{0\,i} &= -\frac12 \, g_0^{jk} \nabla^0_j B_{-1\,ik} + \frac{1}{8} \sqrt{-g_0} \, \varepsilon_{ijklm} B_{-1}^{jk} B_{-1}^{lm} \,, \\
	\begin{split} \label{B1_6D}
		B_{1\,ij} &= -\frac12 \tr(g_0^{-1} g_2) B_{-1\,ij} + 2X_2 B_{-1\,ij} + \bigl( g_2 g_0^{-1} B_{-1} + B_{-1} g_0^{-1} g_2 \bigr)_{ij} \\
		& + \frac12 \, g_0^{kl} \nabla^0_k (\dd B_{-1})_{lij} + \frac{1}{2} \sqrt{-g_0} \, \varepsilon_{ijklm} B_{-1}^{kl} A_0^m \,.
	\end{split}
\end{align}
In addition, we also get
\begin{equation}
	A_{3\,i} = g_0^{jk} \nabla^0_j B_{2\,ik} - 2g_0^{jk} \nabla^0_j (X_3 B_{-1\,ik}) + 2X_3 A_{0\,i} + \frac{1}{4} \sqrt{-g_0} \, \varepsilon_{ijklm} B_{-1}^{jk} B_2^{lm} \,.
\end{equation}
In doing so we took $\varepsilon_{(6)}^{zijklm} = \varepsilon_{(5)}^{ijklm}$. From~\eqref{A0_6D} we obtain
\begin{equation} \label{dA0_6D}
	\begin{split}
		(\dd A_0)_{ij} &= -\frac12 \, g_0^{kl} (\nabla^0_i \nabla^0_k B_{-1\,jl} - \nabla^0_j \nabla^0_k B_{-1\,il}) \\
		& - \frac{1}{8} \sqrt{-g_0} \, \bigl[ \varepsilon_{iklmn} \nabla^0_j (B_{-1}^{kl} B_{-1}^{mn}) - \varepsilon_{jklmn} \nabla^0_j (B_{-1}^{kl} B_{-1}^{mn}) \bigr] \,.
	\end{split}
\end{equation}
From the $(zz)$ component of~\eqref{einstein-eq6D} we derive the following series of traces of higher-order components of metric:
\begin{equation} \label{traces_6D}
	\begin{split}
		\tr(g_0^{-1} g_2) &= -\frac18 R[g_0] + \frac{1}{16} |B_{-1}|^2_{g_0} \,, \\
		\tr(g_0^{-1} g_4) &= \frac14 \tr(g_0^{-1} g_2 g_0^{-1} g_2) - \frac52 X_2^2 - \frac38 X_2 \, |B_{-1}|^2_{g_0} - \frac18 \tr(g_0^{-1} B_{-1} g_0^{-1} B_1) \\
		& + \frac{1}{16} \tr(g_0^{-1} B_{-1} g_0^{-1} \dd A_0) + \frac{1}{16} |\dd B_{-1}|^2_{g_0} - \frac{3}{16} |A_0|^2_{g_0} \,, \\
		\tr(g_0^{-1} g_5) &= -\frac{24}{5} X_2 X_3 - \frac15 X_3 \, |B_{-1}|^2_{g_0} - \frac{1}{10} \tr(g_0^{-1} B_{-1} g_0^{-1} B_2) \,.
	\end{split}
\end{equation}
From the $(iz)$ components of~\eqref{einstein-eq6D} we can obtain the covariant divergences of the higher-order components of the metric, of which we report only the necessary one:
\begin{align} \label{div-g_6D}
	g_0^{jk} \nabla^0_k g_{5\,ij} &= \nabla^0_i \tr(g_0^{-1} g_5) + \frac{24}{5} X_3 \nabla^0_i X_2 + \frac{16}{5} X_2 \nabla^0_i X_3 - \frac{2}{5} X_3 (\dd B_{-1})_{ijk} B_{-1}^{jk} \\
	& - \frac{1}{10} (\dd B_2)_{ijk} B_{-1}^{jk} + \frac15 (\dd B_{-1})_{ijk} B_2^{jk} - \frac45 X_3 B_{-1\,ij} A_0^j + \frac25 B_{2\,ij} A_0^j + \frac25 B_{-1\,ij} A_3^j \nonumber \,.
\end{align}
The higher-order components of the metric can be obtained from the $(ij)$ components of~\eqref{einstein-eq6D}:
\begin{align} \label{g2+g4_6D}
		g_{2\,ij} &= -\frac13 \biggl( R_{ij}[g_0] - \frac18 \, g_{0\,ij} R[g_0] \biggr) - \frac{3}{16} \, g_{0\,ij} |B_{-1}|^2_{g_0} - \frac12 \bigl( B_{-1} g_0^{-1} B_{-1} \bigr)_{ij} \,, \\
		g_{4\,ij} &= -\bigl( g_2 g_0^{-1} g_2 \bigr)_{ij} + \frac14 \, g_{0\,ij} \tr(g_0^{-1} g_2 g_0^{-1} g_2) - \frac12 \, g_{0\,ij} X_2^2 - \frac14 \, g_{2\,ij} |B_{-1}|^2_{g_0} \nonumber \\
		& + \frac18 \, g_{0\,ij} X_2 |B_{-1}|^2_{g_0} + \frac34 \bigl( B_{-1} g_0^{-1} g_2 g_0^{-1} B_{-1} \bigr)_{ij} - \frac14 \, g_{0\,ij} \tr(g_0^{-1} B_{-1} g_0^{-1} B_{-1} g_0^{-1} g_2) \nonumber \\
		& - \frac14 \bigl( B_{-1} g_0^{-1} B_1 + B_1 g_0^{-1} B_{-1} \bigr)_{ij} + \frac18 \, g_{0\,ij} \tr(g_0^{-1} B_{-1} g_0^{-1} B_1) \nonumber\\
		& - \frac14 \bigl( B_{-1} g_0^{-1} \dd A_0 + \dd A_0 g_0^{-1} B_{-1} \bigr)_{ij} + \frac{1}{16} \, g_{0\,ij} \tr(g_0^{-1} B_{-1} g_0^{-1} \dd A_0) + \frac12 A_{0\,i} A_{0\,j} \nonumber\\
		& + \frac{1}{16} \, g_{0\,ij} |A_0|^2_{g_0} + \frac18 \, g_0^{kl} g_0^{mn} (\dd B_{-1})_{ikm} (\dd B_{-1})_{jln} - \frac{3}{16} \, g_{0\,ij} |\dd B_{-1}|^2_{g_0} - \frac12 \Delta R_{ij} \,. \nonumber
\end{align}
Here, $\Delta R_{ij}$ is defined as
\begin{align}
	\Delta R_{ij} &= \frac12 \biggl[ \frac23 R_{kilj} R^{kl} - \frac23 R_{ik} g_0^{kl} R_{lj} - \frac18 \nabla^0_i \nabla^0_j R + \frac13 \nabla^2 R_{ij} - \frac{1}{24} \, g_{0\,ij} \nabla^2 R + R_{kilj} \mathcal{B}^{kl} \nonumber \\
	& - \frac12 \bigl( R_{ik} g_0^{kl} \mathcal{B}_{lj} + \mathcal{B}_{ik} g_0^{kl} R_{lj} \bigr) - \frac12 \bigl( \nabla^0_i \nabla^0_k \mathcal{B}_j^{\ k} + \nabla^0_j \nabla^0_k \mathcal{B}_i^{\ k} \bigr) + \frac12 \nabla^2 \mathcal{B}_{ij} \\
	& - \frac{7} {16} \nabla^0_i \nabla^0_j |B_{-1}|^2_{g_0} + \frac{3}{16} \, g_{0\,ij} \nabla^2 |B_{-1}|^2_{g_0} \biggr] \nonumber \,,
\end{align}
with $\mathcal{B} = B_{-1} g_0^{-1} B_{-1}$ symmetric tensor introduced for brevity, and the reason for the name $\Delta R_{ij}$ will become clear shortly. All the quantities are referred to the boundary metric~$g_0$, \eg\ $R_{ij} \equiv R_{ij}[g_0]$ and $\nabla^2 \equiv g_0^{kl} \nabla^0_k \nabla^0_l$, which is also used to raise the indices.
A convenient expression for the Ricci scalar $R[g]$ can be obtained contracting the $(ij)$~components of the Einstein equations with $g^{ij}$ (\cf~\cite{Chen:2019qib})
\begin{equation} \label{Rg_6d}
	\begin{split}
		R[g] &= R[g_0] + \! z^2 \Bigl[ 3\tr (g_0^{-1} g_2 g_0^{-1} g_2) \! + \! \tr^2 (g_0^{-1} g_2) \! + \! 2X_2 |B_{-1}|^2_{g_0} \! + \! \frac12 \! \tr(g_0^{-1} B_{-1} g_0^{-1} B_{-1} g_0^{-1} g_2) \\
		& + \frac12 \tr(g_0^{-1} B_{-1} g_0^{-1} B_1) - \frac12 \tr(g_0^{-1} B_{-1} g_0^{-1} \dd A_0) + 2|A_0|^2_{g_0} - \frac12 |\dd B_{-1}|^2_{g_0} \Bigr] + \mathcal{O}(z^4) \,,
	\end{split}
\end{equation}
whereas the expansion of the Ricci tensor
\begin{equation}
	R_{ij}[g] = R_{ij}[g_0] + z^2 \Delta R_{ij} + \mathcal{O}(z^3)
\end{equation}
follows from the rather general formula
\begin{equation}
	\Delta R_{ij} = \frac12 \, g_0^{kl} \bigl( \nabla^0_k \nabla^0_j g_{2\,il} + \nabla^0_k \nabla^0_i g_{2\,jl} - \nabla^0_i \nabla^0_j g_{2\,kl} - \nabla^0_k \nabla^0_l g_{2\,ij} \bigr) \,.
\end{equation}

\subsection{Renormalized on-shell action}

We have now all the ingredients to construct the counterterms that will make the on-shell action finite. As explained in Section \ref{sec:boundaryexp_6D} (and more in details in Section~6 of~\cite{Faedo:2025kjf}), the strategy is to set a cutoff at $z = \epsilon$ to compute the regularized action. The latter is given by the sum of bulk action, integrated up to the cutoff, and GHY term evaluated at the cutoff. The regularized action will have power-law divergences, that can be removed order-by-order introducing suitable counterterms \cite{Bianchi:2001kw}. Notice that a counterterm could introduce new, even if less severe, divergences, and thus they need to be taken into account in the remaining part of the analysis.

Using the equations of motion, the bulk on-shell action can be written as
\begin{equation}
	\begin{split}
		\mathcal{S}_\mathrm{bulk} = \frac{1}{16\pi G_\mathrm{N}^{(6)}} & \int_M \dd^6x \, \sqrt{-G} \, \biggl( \frac12 \, \mathcal{V} - \frac{1}{12} X^4 H_{\mu\nu\rho} H^{\mu\nu\rho} - \frac14 X^{-2} B_{\mu\nu} B^{\mu\nu} \\
		& + \frac{1}{24} \frac{\varepsilon^{\mu\nu\rho\sigma\tau\lambda}}{\sqrt{-G}} B_{\mu\nu} B_{\rho\sigma} B_{\tau\lambda} \biggr) \,.
	\end{split}
\end{equation}
The general expression for the Gibbons--Hawking--York term is given by
\begin{equation}
	\mathcal{S}_\mathrm{GHY} = \frac{1}{8\pi G_\mathrm{N}^{(6)}} \int_{\partial M} \dd^5x \, \sqrt{-h} \, h^{ij} K_{ij} \,,
\end{equation}
where $h_{ij}$ is the induced metric on the boundary and $K_{ij}$ its extrinsic curvature. In the Fefferman--Graham coordinates, these two read
\begin{equation}
	h_{ij} = z^{-2} g_{ij} \,,  \qquad  K_{ij} = -\frac{z}{2} \, \partial_z h_{ij} \,,
\end{equation}
and the GHY term takes the form
\begin{equation}
	\mathcal{S}_\mathrm{GHY} = \frac{1}{16\pi G_\mathrm{N}^{(6)}} \int_{\partial M} \dd^5x \, \bigl( -2z \, \partial_z \sqrt{-h} \bigr) \,.
\end{equation}
\paragraph{Order $\boldsymbol{\epsilon^{-5}}$}
The most severe divergences appear at order $\mathcal{O}(\epsilon^{-5})$ and are given by%
\footnote{Here and in the rest of the analysis, in the schematic results we will omit a prefactor of $\frac{1}{16\pi G^{(6)}_\mathrm{N}}$, the multiplication by $\sqrt{-g_0}$ and the integration in $\int d^5x$.}
\begin{align*}
	& \mathcal{S}_\mathrm{pot}: && -2 \\
	& \mathcal{S}_\mathrm{GHY}: && 10
\end{align*}
The total divergence is
\begin{equation}
	\mathcal{S}_\mathrm{div}^{(5)} = \frac{\epsilon^{-5}}{16\pi G^{(6)}_\mathrm{N}}  \int d^5x \, \sqrt{-g_0} \, \bigl[ 8 \bigr] \,,
\end{equation}
which can be canceled by the covariant counterterm
\begin{equation}
	\mathcal{S}_\mathrm{c.t.}^{(5)} = \frac{1}{16\pi G^{(6)}_\mathrm{N}}  \int_{\partial M} d^5x \, \sqrt{-h} \, \bigl[ -8 \bigr] \,.
\end{equation}

\paragraph{Order $\boldsymbol{\epsilon^{-3}}$}
The next divergences that we encounter are of order $\mathcal{O}(\epsilon^{-3})$ and are given by
\begin{align*}
	& \mathcal{S}_\mathrm{pot}: && -\frac53 \tr(g_0^{-1} g_2) \\
	& \mathcal{S}_{HH}: && -\frac16 |B_{-1}|^2_{g_0} \\
	& \mathcal{S}_{BB}: && -\frac16 |B_{-1}|^2_{g_0} \\
	& \mathcal{S}_\mathrm{GHY}: && 3\tr(g_0^{-1} g_2) \\
	& \mathcal{S}_\mathrm{c.t.}^{(5)}: && -4\tr(g_0^{-1} g_2)
\end{align*}
The total divergence is then
\begin{equation}
	\begin{split}
		\mathcal{S}_\mathrm{div}^{(3)} &= \frac{\epsilon^{-3}}{16\pi G_\mathrm{N}^{(6)}} \int \dd^5x \, \sqrt{-g_0} \, \biggl[ -\frac83 \tr(g_0^{-1} g_2) - \frac13 |B_{-1}|^2_{g_0} \biggr] \\
		&= \frac{\epsilon^{-3}}{16\pi G_\mathrm{N}^{(6)}} \int \dd^5x \, \sqrt{-g_0} \, \biggl[ \frac13 R[g_0] - \frac12 |B_{-1}|^2_{g_0} \biggr] \,,
	\end{split}
\end{equation}
that can be canceled by the covariant counterterm
\begin{equation}
	\mathcal{S}_\mathrm{c.t.}^{(3)} = \frac{1}{16\pi G_\mathrm{N}^{(6)}} \int_{\partial M} \dd^5x \, \sqrt{-h} \, \biggl[ -\frac13 R[h] + \frac12 |B|^2_h \biggr] \,.
\end{equation}

\paragraph{Order $\boldsymbol{\epsilon^{-1}}$}
The divergences of order $\mathcal{O}(\epsilon^{-1})$ are given by
\begingroup
\allowdisplaybreaks
\begin{align*}
	& \mathcal{S}_\mathrm{pot}: && -5\tr(g_0^{-1} g_4) + \frac52 \tr(g_0^{-1} g_2 g_0^{-1} g_2) - \frac54 \tr^2(g_0^{-1} g_2) - 12X_2^2 \\
	& \mathcal{S}_{HH}: && -\frac12 \tr(g_0^{-1} B_{-1} g_0^{-1} B_1) - \frac12 \tr(g_0^{-1} B_{-1} g_0^{-1} B_{-1} g_0^{-1} g_2) - 2X_2 |B_{-1}|^2_{g_0} \\
	&&& - \frac14 \tr(g_0^{-1} g_2) |B_{-1}|^2_{g_0} - \frac12 |\dd B_{-1}|^2_{g_0} + \frac12 \tr(g_0^{-1} B_{-1} g_0^{-1} \dd A_0) \\
	& \mathcal{S}_{BB}: && \frac12 \tr(g_0^{-1} B_{-1} g_0^{-1} B_1) - \frac12 \tr(g_0^{-1} B_{-1} g_0^{-1} B_{-1} g_0^{-1} g_2) + X_2 |B_{-1}|^2_{g_0} \\
	&&& - \frac14 \tr(g_0^{-1} g_2) |B_{-1}|^2_{g_0} - \frac12 |A_0|^2_{g_0} \\
	& \mathcal{S}_{BBB}: && \frac{\sigma}{4} \frac{\varepsilon^{ijklm}}{\sqrt{-g_0}} \, A_{0\,i} B_{-1\,jk} B_{-1\,lm} \\
	& \mathcal{S}_\mathrm{GHY}: && \tr(g_0^{-1} g_4) - \frac12 \tr(g_0^{-1} g_2 g_0^{-1} g_2) + \frac14 \tr^2(g_0^{-1} g_2) \\
	& \mathcal{S}_\mathrm{c.t.}^{(5)}: && -4\tr(g_0^{-1} g_4) + 2\tr(g_0^{-1} g_2 g_0^{-1} g_2) - \tr^2(g_0^{-1} g_2) \\
	& \mathcal{S}_\mathrm{c.t.}^{(3)}: && \frac83 \tr(g_0^{-1} g_4) - \frac53 \tr(g_0^{-1} g_2 g_0^{-1} g_2) + \tr^2(g_0^{-1} g_2) + \frac{20}{3} X_2^2 + \frac16 \tr(g_0^{-1} g_2) |B_{-1}|^2_{g_0} \\
	&&& - \frac13 \tr(g_0^{-1} B_{-1} g_0^{-1} B_1) + \frac13 \tr(g_0^{-1} B_{-1} g_0^{-1} B_{-1} g_0^{-1} g_2) + \frac13 X_2 |B_{-1}|^2_{g_0} - \frac16 |A_0|^2_{g_0}
\end{align*}
\endgroup
and thus the total divergence is
\begin{align}
		\mathcal{S}_\mathrm{div}^{(1)} &= \frac{\epsilon^{-1}}{16\pi G_\mathrm{N}^{(6)}} \int \dd^5x \, \sqrt{-g_0} \, \biggl[ \tr(g_0^{-1} g_2 g_0^{-1} g_2) - \tr^2(g_0^{-1} g_2) + 8X_2^2 - \frac13 \tr(g_0^{-1} g_2) |B_{-1}|^2_{g_0} \nonumber \\
		& - \frac23 \tr(g_0^{-1} B_{-1} g_0^{-1} B_{-1} g_0^{-1} g_2) + \frac43 X_2 |B_{-1}|^2_{g_0} + \frac13 \tr(g_0^{-1} B_{-1} g_0^{-1} B_1) \\
		& + \frac16 \tr(g_0^{-1} B_{-1} g_0^{-1} \dd A_0) + \frac13 |A_0|^2_{g_0} - \frac56 |\dd B_{-1}|^2_{g_0} + \frac{\sigma}{4} \frac{\varepsilon^{ijklm}}{\sqrt{-g_0}} \, A_{0\,i} B_{-1\,jk} B_{-1\,lm} \biggr] \,. \nonumber
\end{align}
Using the explicit forms of $g_2$ and $A_0$ and integrating by part, we obtain the final expression
\begin{align}
	\mathcal{S}_\mathrm{div}^{(1)} &= \frac{\epsilon^{-1}}{16\pi G_\mathrm{N}^{(6)}} \int \dd^5x \, \sqrt{-g_0} \, \biggl[ \frac19 R_{ij}[g_0] \, R^{ij}[g_0] - \frac{5}{144} R[g_0]^2 + 8X_2^2 \\
	& + \frac{7}{48} R[g_0] \, |B_{-1}|^2_{g_0} + \frac13 \tr(g_0^{-1} \mathrm{Ric}[g_0] \, g_0^{-1} \mathcal{B}) - \frac{45}{64} |B_{-1}|^4_{g_0} + \frac12 \tr\bigl[(g_0^{-1} B_{-1})^4\bigr] \nonumber \\
	& - \frac12 |\dd B_{-1}|^2_{g_0} + \frac14 \, g_{0\,ij} \nabla^0_k B_{-1}^{ik} \nabla^0_l B_{-1}^{jl} - \frac{\sigma}{8} \frac{\varepsilon^{ijklm}}{\sqrt{-g_0}} \, g_{0\,mn} B_{-1\,ij} B_{-1\,kl} \nabla^0_p B_{-1}^{np} \biggr] \nonumber \,,
\end{align}
that can be reabsorbed introducing the following counterterm:
\begin{equation}
  \begin{split}
    \mathcal{S}_\mathrm{c.t.}^{(1)} & = \frac{1}{16\pi G_\mathrm{N}^{(6)}} \int_{\partial M} \dd^5x \, \sqrt{-h} \, \biggl[ -\frac19 R_{ij}[h] \, R^{ij}[h] + \frac{5}{144} R[h]^2 - 8(1-X)^2 \\
		& - \frac{7}{48} R[h] \, |B|^2_h - \frac13 \tr(h^{-1} \mathrm{Ric}[h] \, h^{-1} B h^{-1} B) + \frac{45}{64} |B|^4_h - \frac12 \tr\bigl[(h^{-1} B)^4\bigr] \\
		& + \frac12 |\dd B|^2_h - \frac14 \, h_{ij} \nabla_k B^{ik} \nabla_l B^{jl} + \frac{\sigma}{8} \frac{\varepsilon^{ijklm}}{\sqrt{-h}} \,h_{mn} B_{ij} B_{kl} \nabla_p B^{np} \biggr] \,.
  \end{split}
\end{equation}

\paragraph{Order $\boldsymbol{\log\epsilon}$}
Also some divergences of order $\mathcal{O}(\log\epsilon)$ appear, given by
\begin{align*}
	& \mathcal{S}_\mathrm{pot}: && 5\tr(g_0^{-1} g_5) + 24X_2 X_3 \\
	& \mathcal{S}_{HH}: && 2X_3 \, |B_{-1}|^2_{g_0} + \tr(g_0^{-1} B_{-1} g_0^{-1} B_2) \\
	& \mathcal{S}_{BB}: && -X_3 \, |B_{-1}|^2_{g_0} - \frac12 \tr(g_0^{-1} B_{-1} g_0^{-1} B_2)
\end{align*}
The total divergence is
\begin{equation}
  \begin{split}
    \mathcal{S}_\mathrm{div}^{(\ell)} & = \frac{\log\epsilon}{16\pi G_\mathrm{N}^{(6)}} \int \dd^5x \, \sqrt{-g_0} \, \biggl[ 5\tr(g_0^{-1} g_5) + 24X_2 X_3 + X_3 \, |B_{-1}|^2_{g_0} \\
    & + \frac12 \tr(g_0^{-1} B_{-1} g_0^{-1} B_2) \biggr] \,,
  \end{split}
\end{equation}
which vanishes due to~\eqref{traces_6D}.

\paragraph{Total counterterm}
Summing all the counterterms we obtain
\begin{align}
  \ma S_{\mathrm{c.t.}} & = \frac{1}{16\pi G_\mathrm{N}^{(6)}}  \int_{\partial M} \dd^5x \, \sqrt{-h} \, \biggl\{ -8 - \frac13 R[h] + \frac12 |B|^2_h - \frac19 R_{ij}[h] \, R^{ij}[h] + \frac{5}{144} R[h]^2 \nonumber \\
	& - \frac{7}{48} R[h] \, |B|^2_h - \frac13 \tr(h^{-1} \mathrm{Ric}[h] \, h^{-1} B h^{-1} B) - 8(1-X)^2 + \frac{45}{64} |B|^4_h \\
	& - \frac12 \tr\bigl[(h^{-1} B)^4\bigr] + \frac12 |\dd B|^2_h - \frac14 \, h_{ij} \nabla_k B^{ik} \nabla_l B^{jl} + \frac{\sigma}{8} \frac{\varepsilon^{ijklm}}{\sqrt{-h}} \, h_{mn} B_{ij} B_{kl} \nabla_p B^{np} \biggr\} \nonumber \,,
\end{align}
where all the quantities are to be evaluated at the cutoff.

\section{Counterterms in $D=5$}
\label{app:counterterms-5d}

In this appendix we outline the calculation of the counterterms needed for the holographic renormalization procedure in the five-dimensional case. The logic is the same as in six dimensions, with some differences due to the fact that here the boundary is even-dimensional and due to the specific form of the solution under exam.

\subsection{Order-by-order equations of motion}

As in the previous case, we begin our analysis by solving the equations of motion order-by-order in~$z$, so to derive the expressions of the higher-order contributions in terms of the leading-order ones. In order to have a cleaner notation, we set $g=1$ from now on.

From the $(zz)$ component of the Einstein equations, expanded up to order $\mathcal{O}(z^2)$, we obtain
\begin{equation} \label{5d:BB=0}
	\sum_\alpha |B^\alpha_{-1}|^2_{g_0} = 0 \,,
\end{equation}
and the following series of trace relations (sum over $\alpha$ is understood)
\begin{equation} \label{5d:trg04}
	\begin{split}
		\tr(g_0^{-1} g_4) &= \frac14 \tr(g_0^{-1} g_2 g_0^{-1} g_2) - 2X_2^2 - \frac14 \tilde{X}_2^2 - \frac{1}{24} \tr(g_0^{-1} B^\alpha_{-1} g_0^{-1} B^\alpha_1) \\
		& + \frac{1}{32} \tr(g_0^{-1} B^\alpha_{-1} g_0^{-1} \tilde{B}^\alpha_1) + \frac{1}{24} \tr(g_0^{-1} B^\alpha_{-1} g_0^{-1} B^\alpha_{-1} g_0^{-1} g_2) - \frac{1}{12} |A^\alpha_0|^2_{g_0} \,, \\
		\tr(g_0^{-1} \tilde{g}_4) &= -4X_2 \tilde{X}_2 - \frac{1}{24} \tr(g_0^{-1} B^\alpha_{-1} g_0^{-1} \tilde{B}^\alpha_1) \,, \\
		\tr(g_0^{-1} \hat{g}_4) &= -2\tilde{X}_2^2 \,.
	\end{split}
\end{equation}
The $(ij)$ components of the Einstein equations, together with~\eqref{5d:BB=0}, gives
\begin{gather}
	\label{5d:g2}
	g_{2\,ij} = -\frac12 \biggl( R_{ij}[g_0] - \frac16 \, g_{0\,ij} R[g_0] \biggr) - \frac14 \bigl( B^\alpha_{-1} g_0^{-1} B^\alpha_{-1} \bigr)_{ij} \,, \\
	\label{5d:trg02}
	\tr(g_0^{-1} g_2) = -\frac16 R[g_0] \,.
\end{gather}
The equation for the scalar does not provide any additional constraint, whereas Maxwell's equations impose a ``self-duality'' condition on $B^\alpha_{-1}$ and fix the term $A^\alpha_0$ as
\begin{equation} \label{5d:sdBm1+A0}
	B^\alpha_{-1} = \varepsilon_{\alpha\beta} \star_{g_0} \! B^\beta_{-1} \,,  \qquad\qquad  A^\alpha_0 = -\varepsilon_{\alpha\beta} \star_{g_0} \! \dd B^\beta_{-1} \,.
\end{equation}
The second relation implies
\begin{equation} \label{5d:dA0}
	\dd A^\alpha_0 = -\varepsilon_{\alpha\beta} \, \dd \star_{g_0} \! \dd B^\beta_{-1}  \qquad  \iff  \qquad  (\dd A^\alpha_0)_{ij} = \frac12 \, \varepsilon_{\alpha\beta} \sqrt{-g_0} \, \varepsilon_{ijkl} \nabla^0_m (\dd B^\beta_{-1})^{mkl} \,,
\end{equation}
which, combined with the first in~\eqref{5d:sdBm1+A0}, gives
\begin{equation} \label{5d:Bm1dA0}
	\tr(g_0^{-1} B^\alpha_{-1} g_0^{-1} \dd A^\alpha_0) = B^\alpha_{-1\,kl} \nabla^0_n (\dd B^\alpha_{-1})^{nkl} \,.
\end{equation}
Moreover, the second equation in~\eqref{5d:sdBm1+A0} also yields
\begin{equation} \label{5d:A0square}
	|A^\alpha_0|^2_{g_0} = -|\dd B^\alpha_{-1}|^2_{g_0} \,.
\end{equation}
One last constraint involving the components of the two-form can be obtained from the ``square'' of the Maxwell equation in~\eqref{maxwell-eq5D}, namely
\begin{equation} \label{os-max}
	\star \, d\bigl( X^2 \star\! dB^\alpha \bigr) = -X^{-2} B^\alpha \,.
\end{equation}
Indeed, plugging the FG expansion of $B^\alpha$, equation~\eqref{os-max} implies
\begin{equation} \label{5d:B1t}
	\begin{split}
		\tilde{B}^\alpha_{1\,ij} &= \frac12 \tr(g_0^{-1} g_2) B^\alpha_{-1\,ij} + \tilde{X}_2 B^\alpha_{-1\,ij} - \bigl( g_2 g_0^{-1} B^\alpha_{-1} + B^\alpha_{-1} g_0^{-1} g_2 \bigr)_{ij} + \frac12 ( \dd A^\alpha_0 )_{ij} \\
		& - \frac12 \, g_0^{kl} \nabla^0_k (\dd B^\alpha_{-1})_{lij} \,.
	\end{split}
\end{equation}
We close this section with a convenient expression for the Ricci scalar $R[g]$, which can be derived along the same lines as~\eqref{Rg_6d},
\begin{equation} \label{5d:Rg}
  R[g] = R[g_0] + z^2 \Bigl[ 2\tr(g_0^{-1} g_2 g_0^{-1} g_2) + \tr^2(g_0^{-1} g_2) - \frac14 \tr(g_0^{-1} B^\alpha_{-1} g_0^{-1} \tilde{B}^\alpha_1) + \frac12 |A^\alpha_0|^2_{g_0} \Bigr] + \mathcal{O}(z^3) \,.
\end{equation}
In writing this relation, we already made use of the constraints presented above.

\subsection{Renormalized on-shell action}

Now that we have at our disposal some essential relations involving the coefficients in the FG expansion of the fields, we are ready for the construction of the counterterms that will make the on-shell action finite. We follow the same logic presented for the 6D case. The main difference is that now the regularized action will present both power-law and logarithmic divergences in $\epsilon$, the latter due to the fact that the boundary is even-dimensional, that will be removed order-by-order introducing appropriate covariant counterterms~\cite{deHaro:2000vlm}.

The starting point is the bulk action~\eqref{action5D}, with $g=1$. By means of the equations of motion, the on-shell action can be cast in the form
\begin{equation}
	\mathcal{S}_\mathrm{bulk} = \frac{1}{16\pi G^{(5)}_\mathrm{N}} \int_M d^5x \, \sqrt{-G} \, \biggl( \frac23 \, \mathcal{V} + \frac{1}{12} X^{-2} B^\alpha_{\mu\nu} B^{\alpha\,\mu\nu} \biggr) \,.
\end{equation}
The GHY term~\eqref{GHYterm5D} comprises the induced boundary metric~$h_{ij}$ and its extrinsic curvature~$K_{ij}$, which, in the FG coordinates, read
\begin{equation}
	h_{ij} = z^{-2} g_{ij} \,,  \qquad  K_{ij} = -\frac{z}{2} \, \partial_z h_{ij} \,,
\end{equation}
thus yielding the simple form
\begin{equation}
	\mathcal{S}_\mathrm{GHY} = \frac{1}{16\pi G^{(5)}_\mathrm{N}} \int_{\partial M} d^4x \, \bigl( -2z \, \partial_z \sqrt{-h} \bigr) \,.
\end{equation}

\paragraph{Order $\boldsymbol{\epsilon^{-4}}$}
The most severe divergences that we encounter appear at order $\mathcal{O}(\epsilon^{-4})$ and are given by
\begin{align*}
	& \mathcal{S}_\mathrm{pot}: && -2 \\
	& \mathcal{S}_\mathrm{GHY}: && 8
\end{align*}
The total divergence is
\begin{equation}
	\mathcal{S}_\mathrm{div}^{(4)} = \frac{\epsilon^{-4}}{16\pi G^{(5)}_\mathrm{N}}  \int d^4x \, \sqrt{-g_0} \, \bigl[ 6 \bigr] \,,
\end{equation}
which can be canceled by the covariant counterterm
\begin{equation}
	\mathcal{S}_\mathrm{c.t.}^{(4)} = \frac{1}{16\pi G^{(5)}_\mathrm{N}}  \int_{\partial M} d^4x \, \sqrt{-h} \, \bigl[ -6 \bigr] \,.
\end{equation}

\paragraph{Order $\boldsymbol{\epsilon^{-2}}$}
Due to our specific expansion, we do not have divergences at order $\mathcal{O}(\epsilon^{-3})$, therefore the following ones are at $\mathcal{O}(\epsilon^{-2})$
\begin{align*}
	& \mathcal{S}_\mathrm{pot}: && -2\tr(g_0^{-1} g_2) \\
	& \mathcal{S}_{BB}: && \frac{1}{12} |B^\alpha_{-1}|_{g_0}^2 \\
	& \mathcal{S}_\mathrm{GHY}: && 2\tr(g_0^{-1} g_2) \\
	& \mathcal{S}_\mathrm{c.t.}^{(4)}: && -3\tr(g_0^{-1} g_2)
\end{align*}
Making use of~\eqref{5d:BB=0} and~\eqref{5d:trg02}, the total divergence can be written as
\begin{equation}
	\mathcal{S}_\mathrm{div}^{(2)} = \frac{\epsilon^{-2}}{16\pi G^{(5)}_\mathrm{N}}  \int d^4x \, \sqrt{-g_0} \, \biggl[ \frac12 R[g_0] \biggr] \,,
\end{equation}
and the corresponding covariant counterterm is
\begin{equation}
	\mathcal{S}_\mathrm{c.t.}^{(2)} = \frac{1}{16\pi G^{(5)}_\mathrm{N}}  \int_{\partial M} d^4x \, \sqrt{-h} \, \biggl[ -\frac12 R[h] \biggr] \,.
\end{equation}
In order to compute the subsequent divergences introduced by this counterterm, relation~\eqref{5d:Rg} will be essential.

\paragraph{Order $\boldsymbol{\log^2\!\epsilon}$}
Since order $\mathcal{O}(\epsilon^{-1})$ gives no divergences, we move to the family of logarithmic ones. The first examples appear at order $\mathcal{O}(\log^3\!\epsilon)$, but they cancel when~\eqref{5d:trg04} is imposed. The next divergences, at order $\mathcal{O}(\log^2\!\epsilon)$, read
\begin{align*}
	& \mathcal{S}_\mathrm{pot}: && 2\tr(g_0^{-1} \tilde{g}_4) + 8X_2 \tilde{X}_2 \\
	& \mathcal{S}_{BB}: && \frac{1}{12} \tr(g_0^{-1} B^\alpha_{-1} g_0^{-1} \tilde{B}^\alpha_1) + \frac16 \tilde{X}_2 |B^\alpha_{-1}|_{g_0}^2 \\
	& \mathcal{S}_\mathrm{c.t.}^{(4)}: && -3\tr(g_0^{-1} \hat{g}_4)
\end{align*}
Using again~\eqref{5d:BB=0} and~\eqref{5d:trg04}, the total contribution becomes
\begin{equation}
	\mathcal{S}_\mathrm{div}^{(\ell.2)} = \frac{\log^2\!\epsilon}{16\pi G^{(5)}_\mathrm{N}}  \int d^4x \, \sqrt{-g_0} \, \bigl[ 6\tilde{X}_2^2 \bigr] \,,
\end{equation}
which can be canceled by the counterterm
\begin{equation}
	\mathcal{S}_\mathrm{c.t.}^{(\ell.2)} = \frac{1}{16\pi G^{(5)}_\mathrm{N}}  \int_{\partial M} d^4x \, \sqrt{-h} \, \bigl[ -6(1 - X)^2 \bigr] \,.
\end{equation}

\paragraph{Order $\boldsymbol{\log\epsilon}$}
The last set of divergences appear at order $\mathcal{O}(\log\epsilon)$
\begin{align*}
	& \mathcal{S}_\mathrm{pot}: && \tr^2(g_0^{-1} g_2) - 2\tr(g_0^{-1} g_2 g_0^{-1} g_2) + 4\tr(g_0^{-1} g_4) + 8X_2^2 \\
	& \mathcal{S}_{BB}: && \frac16 \tr(g_0^{-1} B^\alpha_{-1} g_0^{-1} B^\alpha_1) - \frac16 \tr(g_0^{-1} B^\alpha_{-1} g_0^{-1} B^\alpha_{-1} g_0^{-1} g_2) + \frac13 X_2 |B^\alpha_{-1}|_{g_0}^2 \\
	&&& - \frac{1}{12} \tr(g_0^{-1} g_2) |B^\alpha_{-1}|_{g_0}^2 - \frac16 |A^\alpha_0|^2_{g_0} \\
	& \mathcal{S}_\mathrm{GHY}: && -2\tr(g_0^{-1} \hat{g}_4) \\
	& \mathcal{S}_\mathrm{c.t.}^{(4)}: && -3\tr(g_0^{-1} \tilde{g}_4) \\
	& \mathcal{S}_\mathrm{c.t.}^{(\ell.2)}: && -12X_2 \tilde{X}_2
\end{align*}
By means of~\eqref{5d:trg04}, all these contributions sum up to
\begin{equation}
	\begin{split}
		\mathcal{S}_\mathrm{div}^{(\ell.1)} &= \frac{\log\epsilon}{16\pi G^{(5)}_\mathrm{N}}  \int d^4x \, \sqrt{-g_0} \, \biggl[ \tr^2(g_0^{-1} g_2) - \tr(g_0^{-1} g_2 g_0^{-1} g_2) + 3\tilde{X}_2^2 \\
		& + \frac14 \tr(g_0^{-1} B^\alpha_{-1} g_0^{-1} \tilde{B}^\alpha_1) - \frac12 |A^\alpha_0|^2_{g_0} \biggr] \,.
	\end{split}
\end{equation}
Substituting the explicit form of $g_2$ and $\tilde{B}^\alpha_1$ in~\eqref{5d:g2} and~\eqref{5d:B1t}, and making use of~\eqref{5d:Bm1dA0} and~\eqref{5d:A0square}, it is possible to show that this logarithmic divergence can be reabsorbed by the following counterterm
\begin{equation}
  \begin{split}
    \mathcal{S}_\mathrm{c.t.}^{(\ell.1)} & = \frac{1}{16\pi G^{(5)}_\mathrm{N}} \int_{\partial M} d^4x \, \sqrt{-h} \, \biggl\{ -\log\epsilon \biggl[ \frac{1}{12} R[h]^2 - \frac14 R_{ij}[h] R^{ij}[h] \\
    & + \frac{1}{16} \tr\bigl[(h^{-1} B^\alpha)^2 (h^{-1} B^\beta)^2\bigr] \biggr] - \frac{3}{\log\epsilon}(1-X)^2 \biggr\} \, .
  \end{split}
\end{equation}
Summing all the counterterms corresponding to the different orders of divergence, we obtain
\begin{equation} \label{5d:Sct-old}
  \begin{split}
    \ma S_{\mathrm{c.t.}} & = \frac{1}{16\pi G^{(5)}_\mathrm{N}}  \int_{\partial M} d^4x \, \sqrt{-h} \, \biggl\{ -6 - \frac12 R[h] -6(1-X)^2 - \log\epsilon \biggl[ \frac{1}{12} R[h]^2 \\
    & - \frac14 R_{ij}[h] R^{ij}[h] + \frac{1}{16} \tr\bigl[(h^{-1} B^\alpha)^2 (h^{-1} B^\beta)^2\bigr] \biggr] - \frac{3}{\log\epsilon}(1-X)^2 \biggr\} \, ,
  \end{split}
\end{equation}
where all the quantities are to be evaluated at the cutoff.

Even though the renormalized action is finite, the current associated with the $B$-field and the stress--energy tensor are still divergent. The reason for this behavior is that property \eqref{5d:BB=0} prevents us from fixing uniquely the counterterms.
Indeed, a possible counterterm proportional to $\int_{\partial M} \dd^4x \, \sqrt{-h} \, \sum_\alpha |B^\alpha|^2_h$ would not yield any divergence at leading order, \ie\ $\mathcal{O}(\epsilon^{-2})$, because $\sum_\alpha |B^\alpha_{-1}|^2_{g_0}=0$, while the subleading divergence at order $\mathcal{O}(\log\epsilon)$, of the form $\tr(g_0^{-1} B^\alpha_{-1} g_0^{-1} \tilde{B}^\alpha_1)$, could be cured by the addition of suitable counterterms.
As we will see explicitly, the freedom of adding such a counterterm with any desired coefficient is not present at the level of two-form current $\langle J^{ij}_\alpha\rangle$, since, at leading order, $\frac{\delta}{\delta B^\alpha_{ij}} \sum_\beta |B^\beta|^2_h \sim g_0^{ik} B^\alpha_{-1\,kl} g_0^{lj}$, which is not vanishing anymore. Removing the divergences in $\langle J^{ij}_\alpha\rangle$ will specify the counterterms in a unique way.

\subsection{Two-form current}

The prescription to construct the one-point function of the current associated with the two-form $B^\alpha$ was presented in Section~\ref{subsec:1pt_5D} and it can be summarized as follows:
\begin{equation} \label{5d:current}
	\langle J^{ij}_\alpha \rangle = \lim_{\epsilon\to0} \biggl( \frac{1}{\epsilon^5} \frac{1}{\sqrt{-h}} \frac{\delta \mathcal{S}_\mathrm{reg}}{\delta B^\alpha_{ij}} + \frac{1}{\epsilon^5} \frac{1}{\sqrt{-h}} \frac{\delta \mathcal{S}_\mathrm{c.t.}}{\delta B^\alpha_{ij}} \biggr) \,,
\end{equation}
where the contribution from the regularized action is
\begin{equation}
	\frac{1}{\sqrt{-h}} \frac{\delta \mathcal{S}_\mathrm{reg}}{\delta B^\alpha_{ij}} = \frac{1}{16\pi G^{(5)}_\mathrm{N}} \biggl( \frac{3}{4g_1^2} \, X^2 g^{ik} g^{jl} \epsilon^5 \, \partial_{[\epsilon} B^\alpha_{kl]} \biggr) \,.
\end{equation}
As we mentioned at the end of the previous subsection, the structure of $\mathcal{S}_\mathrm{c.t.}$ is not completely determined, however the form of the counterterms coming from the curvature and the scalar can be fixed as in~\eqref{5d:Sct-old}, since the freedom only affects the $B$-field,
\begin{equation}
	\begin{split}
		\mathcal{S}_{\mathrm{c.t.}}^{RX} & = \frac{1}{16\pi G^{(5)}_\mathrm{N}} \int_{\partial M} d^4x \, \sqrt{-h} \, \biggl\{ -6 - \frac12 R[h] - 6(1-X)^2 \\
		& + \log\epsilon \biggl[ \frac14 R_{ij}[h] R^{ij}[h] - \frac{1}{12} R[h]^2 \biggr] - \frac{3}{\log\epsilon}(1-X)^2 \biggr\} \,.
	\end{split}
\end{equation}
Inspired by the expression of the divergences in the on-shell action, we assume the presence of the following counterterms and fix the coefficients requiring that the divergent terms cancel order by order
\begin{equation}
	\begin{split}
		\mathcal{S}_{\mathrm{c.t.}}^{B2} &= \frac{\zeta_1}{16\pi G^{(5)}_\mathrm{N}} \int_{\partial M} d^4x \, \sqrt{-h} \, |B^\alpha|^2_h \,, \\
		\mathcal{S}_{\mathrm{c.t.}}^{B4} &= \frac{\zeta_2 \log\epsilon}{16\pi G^{(5)}_\mathrm{N}} \int_{\partial M} d^4x \, \sqrt{-h} \, \tr\bigl[(h^{-1} B^\alpha)^2 (h^{-1} B^\beta)^2\bigr] \,, \\
		\mathcal{S}_{\mathrm{c.t.}}^{RB.1} &= \frac{\zeta_3 \log\epsilon}{16\pi G^{(5)}_\mathrm{N}} \int_{\partial M} d^4x \, \sqrt{-h} \, \tr(h^{-1} \mathrm{Ric}[h] \, h^{-1} B^\alpha h^{-1} B^\alpha) \,, \\
		\mathcal{S}_{\mathrm{c.t.}}^{RB.2} &= \frac{\zeta_4 \log\epsilon}{16\pi G^{(5)}_\mathrm{N}} \int_{\partial M} d^4x \, \sqrt{-h} \, R[h] \, |B^\alpha|^2_h \,.
	\end{split}
\end{equation}
In our analysis, we considered also terms like $\int_{\partial M} \dd^4x \, \sqrt{-h} \, X^p \sum_\alpha |B^\alpha|^2_h$, with $p\in\RR$, and $\int_{\partial M} \dd^4x \, \sqrt{-h} \, \sum_\alpha |\dd B^\alpha|^2_h$, but their presence turned out to be inconsistent.

The schematic structure of the divergences follows the idea of the previous computations regarding the on-shell action, and in each term we have suppressed an $\epsilon^5$ prefactor, which cancels the $1/\epsilon^5$ in~\eqref{5d:current}.

\paragraph{Order $\boldsymbol{\epsilon^{-2}}$}
The leading divergences appear at order $\mathcal{O}(\epsilon^{-2})$ and read
\begin{align*}
	& \mathcal{S}_\mathrm{reg}: && \frac14 \bigl( g_0^{-1} B^\alpha_{-1} g_0^{-1} \bigr)^{ji} \\
	& \mathcal{S}_{\mathrm{c.t.}}^{B2}: && -\zeta_1 \bigl( g_0^{-1} B^\alpha_{-1} g_0^{-1} \bigr)^{ji}
\end{align*}
The total divergence is canceled choosing $\zeta_1=1/4$.

\paragraph{Order $\boldsymbol{\log\epsilon}$}
Upon using the fact that, at leading order, $R[g]_{ij} = R[g_0]_{ij}$, the next divergent terms, at order $\mathcal{O}(\log\epsilon)$, can be written as
\begin{align*}
	& \mathcal{S}_\mathrm{reg}: && \frac12 \tilde{X}_2 \bigl( g_0^{-1} B^\alpha_{-1} g_0^{-1} \bigr)^{ji} - \frac14 \bigl( g_0^{-1} \tilde{B}^\alpha_1 g_0^{-1} \bigr)^{ji} \\
	& \mathcal{S}_{\mathrm{c.t.}}^{B2}: &&  -\frac14 \bigl( g_0^{-1} \tilde{B}^\alpha_1 g_0^{-1} \bigr)^{ji} \\
	& \mathcal{S}_{\mathrm{c.t.}}^{B4}: &&  2\zeta_2 \bigl( g_0^{-1} B^\alpha_{-1} g_0^{-1} B^\beta_{-1} g_0^{-1} B^\beta_{-1} g_0^{-1} + g_0^{-1} B^\beta_{-1} g_0^{-1} B^\beta_{-1} g_0^{-1} B^\alpha_{-1} g_0^{-1} \bigr)^{ji} \\
	& \mathcal{S}_{\mathrm{c.t.}}^{RB.1}: &&  \zeta_3 \bigl( g_0^{-1} \mathrm{Ric}[g_0] \, g_0^{-1} B^\alpha_{-1} g_0^{-1} + g_0^{-1} B^\alpha_{-1} g_0^{-1} \mathrm{Ric}[g_0] \, g_0^{-1} \bigr)^{ji} \\
	& \mathcal{S}_{\mathrm{c.t.}}^{RB.2}: &&  -\zeta_4 R[g_0] \, \bigl( g_0^{-1} B^\alpha_{-1} g_0^{-1} \bigr)^{ji}
\end{align*}
By means of~\eqref{5d:B1t}, setting $\zeta_2=1/16$, $\zeta_3=1/4$ and $\zeta_4=1/8$ removes all the divergences, up to the following remaining factors
\begin{equation*}
	-\frac14 (\dd A^\alpha_0)^{ji} + \frac14 \nabla^0_k (\dd B^\alpha)^{kji} \,.
\end{equation*}
Using~\eqref{5d:sdBm1+A0} and~\eqref{5d:dA0}, it is possible to show that these pieces are canceled by the addition of
\begin{equation}
	\mathcal{S}_{\mathrm{c.t.}}^{\nabla B} = \frac{\log\epsilon}{16\pi G^{(5)}_\mathrm{N}} \int_{\partial M} d^4x \, \sqrt{-h} \biggl[ -\frac18 \, \varepsilon_{\alpha\beta} \, \frac{\varepsilon^{ijkl}}{\sqrt{-h}} \, B^\alpha_{ij} \nabla_m (\dd B^\beta)^m_{\;\ kl} \biggr] \,.
\end{equation}
As a result of this analysis, the full set of counterterms is
\begin{align}
	\mathcal{S}_{\mathrm{c.t.}} & = \frac{1}{16\pi G^{(5)}_\mathrm{N}}  \int_{\partial M} d^4x \, \sqrt{-h} \, \biggl\{ -6 - \frac12 R[h] - 6(1-X)^2 + \frac14 \, |B^\alpha|^2_h \\
	& + \log\epsilon \biggl[ \frac14 R_{ij}[h] R^{ij}[h] - \frac{1}{12} R[h]^2 + \frac14 \tr(h^{-1} \mathrm{Ric}[h] \, h^{-1} B^\alpha h^{-1} B^\alpha) + \frac18 R[h] \, |B^\alpha|^2_h \nonumber \\
	& + \frac{1}{16} \tr\bigl[(h^{-1} B^\alpha)^2 (h^{-1} B^\beta)^2\bigr] - \frac18 \, \varepsilon_{\alpha\beta} \, \frac{\varepsilon^{ijkl}}{\sqrt{-h}} \, B^\alpha_{ij} \nabla_m (d B^\beta)^m_{\;\ kl} \biggr] - \frac{3}{\log\epsilon}(1-X)^2 \biggr\} \,, \nonumber
\end{align}
and, as a consistency check, the renormalized on-shell action can be proven to be finite.

The finite part of~\eqref{5d:current} can be computed straightforwardly, giving the one-point function of the current in~\eqref{1pt_current5D}.

\subsection{Stress--energy tensor}

The construction of the holographic stress tensor is lengthy and not of much interest, however we summarize here the main ingredients required, as they could be helpful for an interested reader.

From Maxwell's equations~\eqref{maxwell-eq5D} we derive the ``self-duality'' conditions
\begin{equation} \label{5d:sdB1t+sdB1}
	\begin{split}
		\tilde{B}^\alpha_1 &= -\varepsilon_{\alpha\beta} \star_{g_0} \! \tilde{B}^\beta_1 + 2\tilde{X}_2 B^\alpha_{-1} \,, \\
		B^\alpha_1 &= -\varepsilon_{\alpha\beta} \star_{g_0}\!\Bigl( B^\beta_1 - \frac12 \, \dd A^\beta_0 \Bigr) + 2X_2 B^\alpha_{-1} - \tilde{X}_2 B^\alpha_{-1} + \frac12 \, \dd A^\alpha_0 \,.
	\end{split}
\end{equation}
These equations, together with~\eqref{5d:dA0}, can be combined with~\eqref{5d:sdBm1+A0} to give a set of identities in the following way
\begin{align}
	\bigl( B^\alpha_{-1} g_0^{-1} \tilde{B}^\alpha_1 \bigr)_{ij} &= B^\alpha_{-1\,ik} g_0^{kl} \biggl( -\frac12 \, \varepsilon_{\alpha\beta} \, \frac{\varepsilon^{mnpq}}{\sqrt{-g_0}} \, g_{0\,lm} g_{0\,jn} \tilde{B}^\beta_{1\,pq} + 2\tilde{X}_2 B^\beta_{-1\,lj} \biggr) \\
	&= -\frac12 \frac{\varepsilon^{knpq}}{\sqrt{-g_0}} \biggl( \frac12 \sqrt{-g_0} \, \varepsilon_{iklm} g_0^{lr} g_0^{ms} B^\beta_{-1\,rs} \! \biggr) g_{0\,jn} \tilde{B}^\beta_{1\,pq} + 2\tilde{X}_2 \bigl( B^\alpha_{-1} g_0^{-1} B^\alpha_{-1} \bigr)_{ij} \nonumber \\
	&= -\bigl( \tilde{B}^\alpha_1 g_0^{-1} B^\alpha_{-1} \bigr)_{ij} + \frac12 \, g_{0\,ij} \tr(g_0^{-1} B^\alpha_{-1} g_0^{-1} \tilde{B}^\alpha_1) + 2\tilde{X}_2 \bigl( B^\alpha_{-1} g_0^{-1} B^\alpha_{-1} \bigr)_{ij} \nonumber \,,
\end{align}
where, in the second step, we used~\eqref{5d:sdBm1+A0} to express $B^\alpha_{-1\,ik}$ in terms of its Hodge dual.
The resulting identities, involving $\tilde{B}^\alpha_1$, $B^\alpha_1$ and $\dd A^\alpha_0$, are
\begingroup
\allowdisplaybreaks
\begin{align} \label{5d:rel-BBdA}
	& \bigl( B^\alpha_{-1} g_0^{-1} \tilde{B}^\alpha_1 + \tilde{B}^\alpha_1 g_0^{-1} B^\alpha_{-1} \bigr)_{ij} - 2\tilde{X}_2 \bigl( B^\alpha_{-1} g_0^{-1} B^\alpha_{-1} \bigr)_{ij} - \frac12 \, g_{0\,ij} \tr(g_0^{-1} B^\alpha_{-1} g_0^{-1} \tilde{B}^\alpha_1) = 0 \,, \nonumber \\
	\begin{split}
		& \bigl( B^\alpha_{-1} g_0^{-1} B^\alpha_1 + B^\alpha_1 g_0^{-1} B^\alpha_{-1} \bigr)_{ij} - \frac12 \bigl( B^\alpha_{-1} g_0^{-1} \dd A^\alpha_0 + \dd A^\alpha_0 g_0^{-1} B^\alpha_{-1} \bigr)_{ij} \\
		& \qquad - 2X_2 \bigl( B^\alpha_{-1} g_0^{-1} B^\alpha_{-1} \bigr)_{ij} + \tilde{X}_2 \bigl( B^\alpha_{-1} g_0^{-1} B^\alpha_{-1} \bigr)_{ij} - \frac12 \, g_{0\,ij} \tr(g_0^{-1} B^\alpha_{-1} g_0^{-1} B^\alpha_1) \\
		& \qquad + \frac14 \, g_{0\,ij} \tr(g_0^{-1} B^\alpha_{-1} g_0^{-1} \dd A^\alpha_0) = 0 \,,
	\end{split} \\
	& \begin{aligned}
		& \bigl( B^\alpha_{-1} g_0^{-1} \dd A^\alpha_0 + \dd A^\alpha_0 g_0^{-1} B^\alpha_{-1} \bigr)_{ij} - (g_{0\,ik} B^\alpha_{-1\,lj} + B^\alpha_{-1\,ik} g_{0\,lj}) \nabla^0_n (\dd B^\alpha_{-1})^{nkl} \\
		& \qquad - g_{0\,ij} B^\alpha_{-1\,kl} \nabla^0_n (\dd B^\alpha_{-1})^{nkl} = 0 \nonumber \,.
	\end{aligned}
\end{align}
\endgroup

An additional piece of information that enters the computation is the Ricci tensor at second order in~$z$
\begin{equation} \label{5d:Ricg}
	R_{ij}[g] = R_{ij}[g_0] + z^2 \Delta R_{ij} + \mathcal{O}(z^3) \,,
\end{equation}
with $\Delta R_{ij}$ defined as
\begin{align}
	\Delta R_{ij} &= \frac12 \biggl[ R_{kilj} R^{kl} - R_{ik} g_0^{kl} R_{lj} - \frac16 \nabla^0_i \nabla^0_j R + \frac12 \nabla^2 R_{ij} - \frac{1}{12} \, g_{0\,ij} \nabla^2 R + \frac12 R_{kilj} \mathcal{B}^{kl} \nonumber \\
	& - \frac14 \bigl( R_{ik} g_0^{kl} \mathcal{B}_{lj} + \mathcal{B}_{ik} g_0^{kl} R_{lj} \bigr) - \frac14 \bigl( \nabla^0_i \nabla^0_k \mathcal{B}_j^{\ k} + \nabla^0_j \nabla^0_k \mathcal{B}_i^{\ k} \bigr) + \frac14 \nabla^2 \mathcal{B}_{ij} \biggr] \,,
\end{align}
where we introduced the symmetric tensor $\mathcal{B} = B^\alpha_{-1} g_0^{-1} B^\alpha_{-1}$ for brevity, and all the quantities are referred to the metric $g_0$ (in particular, $\nabla^2 \equiv g_0^{kl} \nabla^0_k \nabla^0_l$), which is also used to raise the indices.

From the $(ij)$ components of the Einstein equations we can derive
\begin{align} \label{5d:g4}
	\tilde{g}_{4\,ij} &= \bigl( g_2 g_0^{-1} g_2 \bigr)_{ij} - \frac14 \, g_{0\,ij} \tr(g_0^{-1} g_2 g_0^{-1} g_2) - g_{0\,ij} X_2 \tilde{X}_2 - \frac14 \tilde{X}_2 \bigl( B^\alpha_{-1} g_0^{-1} B^\alpha_{-1} \bigr)_{ij} \nonumber \\
	& - \frac14 \bigl( B^\alpha_{-1} g_0^{-1} g_2 g_0^{-1} B^\alpha_{-1} \bigr)_{ij} + \frac18 \, g_{0\,ij} \tr(g_0^{-1} B^\alpha_{-1} g_0^{-1} B^\alpha_{-1} g_0^{-1} g_2) \nonumber \\
	& + \frac{1}{48} \, g_{0\,ij} \tr(g_0^{-1} B^\alpha_{-1} g_0^{-1} \tilde{B}^\alpha_1) + \frac12 \Delta R_{ij} - \frac14 A^\alpha_{0\,i} A^\alpha_{0\,j} \nonumber \\
	& + \frac18 \bigl( B^\alpha_{-1} g_0^{-1} \dd A^\alpha_0 + \dd A^\alpha_0 g_0^{-1} B^\alpha_{-1} \bigr)_{ij} - \frac{1}{16} \, g_{0\,ij} \tr(g_0^{-1} B^\alpha_{-1} g_0^{-1} \dd A^\alpha_0) \,, \nonumber \\
	\hat{g}_{4\,ij} &= -\frac12 \, g_{0\,ij} \tilde{X}_2^2 \,,
\end{align}
where the previous relations have been employed to simplify the expressions.
Consistency of equations \eqref{5d:Ricg} and \eqref{5d:Rg} or, alternatively, \eqref{5d:g4} and \eqref{5d:trg04}, requires that the following relation holds identically
\begin{equation}
	g_0^{ij} \, \Delta R_{ij} = -\frac14 \nabla^0_i \nabla^0_j \mathcal{B}^{ij} = \frac12 |A^\alpha_0|^2_{g_0} - \frac14 \tr(g_0^{-1} B^\alpha_{-1} g_0^{-1} \dd A^\alpha_0) \,.
\end{equation}
In order to prove this relation, we first begin with
\begin{equation} \label{5d:nablaB}
	\nabla^0_m \mathcal{B}_i^{\ m} = (\nabla^0_m B^\alpha_{-1\,ij}) g_0^{jk} B^\alpha_{-1\,kl} g_0^{lm} + B^\alpha_{-1\,ij} g_0^{jk} (\nabla^0_m B^\alpha_{-1\,kl}) g_0^{lm} \,.
\end{equation}
The second term can be easily rewritten combining the two equations in~\eqref{5d:sdBm1+A0}
\begin{equation}
	A^\alpha_0 = \star_{g_0} \dd (\star_{g_0} B^\alpha_{-1})  \qquad  \iff  \qquad  A^\alpha_{0\,j} = g_0^{ik} \nabla^0_k B^\alpha_{-1\,ij} \,.
\end{equation}
In order to deal with the first term, we notice that, since $B^{\alpha\,jm}_{-1} B^\alpha_{-1\,jm}$ is constant,
\begin{equation}
	\nabla^0_i \bigl( B^{\alpha\,jm}_{-1} B^\alpha_{-1\,jm} \bigr) = 0  \qquad  \implies  \qquad  B^\alpha_{-1\,jm} \nabla^0_i B^{\alpha\,jm}_{-1} = B^{\alpha\,jm}_{-1} \, \nabla^0_i B^\alpha_{-1\,jm} = 0 \,.
\end{equation}
This implies
\begin{equation}
	B^{\alpha\,jm}_{-1} \, \nabla^0_m B^\alpha_{-1\,ij} = \frac32 B^{\alpha\,jm}_{-1} \, \nabla^0_{[m} B^\alpha_{-1\,ij]} = \frac12 B^{\alpha\,jm}_{-1} (\dd B^\alpha_{-1})_{mij} = -g_0^{kl} B^\alpha_{-1\,ik} A^\alpha_{0\,l} \,,
\end{equation}
where, in the last step, we used~\eqref{5d:sdBm1+A0} twice. The final result is
\begin{equation}
	\nabla^0_m \mathcal{B}_i^{\ m} = -2g_0^{jk} B^\alpha_{-1\,ij} A^\alpha_{0\,k} \,.
\end{equation}
Applying a second covariant derivative to this relation and using~\eqref{5d:nablaB} we get
\begin{equation} \label{5d:DDBB}
	\nabla^0_i \nabla^0_j \mathcal{B}^{ij} = -2 |A^\alpha_0|^2_{g_0} + \tr(g_0^{-1} B^\alpha_{-1} g_0^{-1} \dd A^\alpha_0) \,.
\end{equation}

We are now ready to move to the derivation of the stress--energy tensor. An explicit computation shows that the most severe divergences, at order $\mathcal{O}(\epsilon^{-4})$, cancel automatically. The following ones, at order $\mathcal{O}(\epsilon^{-2})$, disappear by means of~\eqref{5d:g2} and~\eqref{5d:trg02}. In order to treat the $\mathcal{O}(\log^2\!\epsilon)$ terms we must use~\eqref{5d:trg04} and~\eqref{5d:g4}, which, together with~\eqref{5d:rel-BBdA} and~\eqref{5d:DDBB}, allow us to cancel also the $\mathcal{O}(\log\epsilon)$ divergences.
The one-point function of the stress--energy tensor $\langle T_{ij} \rangle$ is therefore finite and its expression was already presented in~\eqref{stress_tensor5D}.

The computation of the trace of the stress--energy tensor, namely $\langle T^k_{\ k} \rangle$, does not require additional information and can be carried out directly to get~\eqref{stress_trace5D}.
On the other hand, its covariant derivative requires the knowledge of $\tilde{A}^\alpha_2$ and $A^\alpha_2$, whose expressions follow, again, from Maxwell's equations
\begin{equation} \label{5d:A2t+A2}
	\begin{split}
		\tilde{A}^\alpha_2 &= -\varepsilon_{\alpha\beta} \star_{g_0} \! \dd \tilde{B}^\beta_{-1} + 2\tilde{X}_2 A^\alpha_0 \,, \\
		A^\alpha_2 &= -\varepsilon_{\alpha\beta} \star_{g_0} \! \Bigl( \dd B^\beta_1 - \frac12 \tr(g_0^{-1} g_2) \, \dd B^\beta_{-1} \Bigr) + 2X_2 A^\alpha_0 + \bigl( g_2 g_0^{-1} A^\alpha_0 \bigr)_i \, \dd x^i \,.
	\end{split}
\end{equation}
Additionally, we need the covariant derivative of the coefficients in the FG expansion of the metric $g_{ij}$, which can be extracted from the $(iz)$ components of Einstein's equations
\begingroup
\allowdisplaybreaks
\begin{align} \label{5d:div-g}
	g_0^{jk} \nabla^0_k g_{2\,ij} &= \nabla^0_i \tr(g_0^{-1} g_2) + \frac12 \, g_0^{jk} B^\alpha_{-1\,ij} A^\alpha_{0\,k} \,, \nonumber \\
	\begin{split}
		g_0^{jk} \nabla^0_k g_{4\,ij} &= \nabla^0_i \tr(g_0^{-1} g_4) - \frac38 \nabla^0_i \tr(g_0^{-1} g_2 g_0^{-1} g_2) - \frac14 \, g_0^{jk} g_{2\,ij} \nabla^0_k \tr(g_0^{-1} g_2) \\
		& + \frac12 \, g_0^{jk} \nabla^0_k (g_2 g_0^{-1} g_2)_{ij} + 3X_2 \nabla^0_i X_2 + \frac34 \tilde{X}_2 \nabla^0_i X_2 - \frac34 X_2 \nabla^0_i \tilde{X}_2 \\
		& - \frac12 X_2 \, g_0^{jk} B^\alpha_{-1\,ij} A^\alpha_{0\,k} + \frac18 \tilde{X}_2 \, g_0^{jk} B^\alpha_{-1\,ij} A^\alpha_{0\,k} - \frac14 (g_0^{-1} g_2 g_0^{-1})^{jk} B^\alpha_{-1\,ij} A^\alpha_{0\,k} \\
		& + \frac14 \, g_0^{jk} B^\alpha_{1\,ij} A^\alpha_{0\,k} + \frac14 \, g_0^{jk} B^\alpha_{-1\,ij} A^\alpha_{2\,k} - \frac{1}{16} \, g_0^{jk} \tilde{B}^\alpha_{1\,ij} A^\alpha_{0\,k} - \frac{1}{16} \, g_0^{jk} B^\alpha_{-1\,ij} \tilde{A}^\alpha_{2\,k} \,,
	\end{split} \nonumber \\
	\begin{split}
		g_0^{jk} \nabla^0_k \tilde{g}_{4\,ij} &= \nabla^0_i \tr(g_0^{-1} \tilde{g}_4) + 3\nabla^0_i (X_2 \tilde{X}_2) - \frac12 \tilde{X}_2 \, g_0^{jk} B^\alpha_{-1\,ij} A^\alpha_{0\,k} + \frac14 \, g_0^{jk} \tilde{B}^\alpha_{1\,ij} A^\alpha_{0\,k} \\
		& + \frac14 \, g_0^{jk} B^\alpha_{-1\,ij} \tilde{A}^\alpha_{2\,k} \,,
	\end{split} \nonumber \\
	g_0^{jk} \nabla^0_k \hat{g}_{4\,ij} &= \nabla^0_i \tr(g_0^{-1} \hat{g}_4) + 3\tilde{X}_2 \nabla^0_i \tilde{X}_2 \,.
\end{align}
\endgroup
These expressions allow us to compute $\nabla^0_i \langle T^{ij} \rangle$, which, upon some manipulations, takes the form
\begin{align}
	\nabla^0_i \langle T^{ij} \rangle &= \frac{1}{16\pi G^{(5)}_\mathrm{N}} \bigg[ -6X_2 \nabla^{0\,j} \tilde{X}_2 + \nabla^0_i (X_2 \mathcal{B}^{ij}) + \frac12 \tilde{B}^{\alpha\,jk}_1 A^\alpha_{0\,k} + \frac12 B^{\alpha\,jk}_{-1} \tilde{A}^\alpha_{2\,k} \nonumber \\
	& + B^{\alpha\,jk}_1 A^\alpha_{0\,k} + B^{\alpha\,jk}_{-1} A^\alpha_{2\,k} - B^{\alpha\,jk}_{-1} A^\alpha_{0\,k} \Bigl( 2X_2 + \tilde{X}_2 + \frac{1}{12} R[g_0] \Bigr) - B^{\alpha\,jk}_{-1} g_{2\,kl} A^{\alpha\,l}_0 \nonumber \\
	& + \frac12 \nabla^0_i (A^{\alpha\,i} A^{\alpha\,j}) - \frac14 \nabla^{0\,j} |A^\alpha_0|^2_{g_0} \biggr] \,.
\end{align}
Combining equations~\eqref{5d:sdBm1+A0}, \eqref{5d:sdB1t+sdB1} and~\eqref{5d:A2t+A2} we derive the following relations
\begin{equation}
	\begin{split}
		& g_0^{ik} \nabla^0_k (\tilde{B}^\alpha_1 - 2\tilde{X}_2 B^\alpha_{-1})_{ij} = -\tilde{A}^\alpha_{2\,j} + 2\tilde{X}_2 A^\alpha_{0\,j} \,, \\
		& g_0^{ik} \nabla^0_k \Bigl( B^\alpha_1 - 2X_2 B^\alpha_{-1} + \tilde{X}_2 B^\alpha_{-1} - \frac12 \, \dd A^\alpha_0 \Bigr)_{ij} = -A^\alpha_{2\,j} + 2X_2 A^\alpha_{0\,j} + \bigl( g_2 g_0^{-1} A^\alpha_0 \bigr)_j \\
		& \qquad\qquad - \frac12 \tr(g_0^{-1} g_2) A^\alpha_{0\,j} \,,
	\end{split}
\end{equation}
which can be used to obtain the simplified expression of the divergence of $\langle T^{ij} \rangle$ presented in~\eqref{stress_divergence5D}.


 \bibliographystyle{utphys}
  \bibliography{references}

@article{Karch:2000gx,
    author = "Karch, Andreas and Randall, Lisa",
    title = "{Open and closed string interpretation of SUSY CFT's on branes with boundaries}",
    eprint = "hep-th/0105132",
    archivePrefix = "arXiv",
    reportNumber = "MIT-CTP-3146",
    doi = "10.1088/1126-6708/2001/06/063",
    journal = "JHEP",
    volume = "06",
    pages = "063",
    year = "2001"
}

@article{DeWolfe:2001pq,
    author = "DeWolfe, Oliver and Freedman, Daniel Z. and Ooguri, Hirosi",
    title = "{Holography and defect conformal field theories}",
    eprint = "hep-th/0111135",
    archivePrefix = "arXiv",
    reportNumber = "CALT-68-2359, CITUSC-01-041, NSF-ITP-01-172, MIT-CTP-3212",
    doi = "10.1103/PhysRevD.66.025009",
    journal = "Phys. Rev. D",
    volume = "66",
    pages = "025009",
    year = "2002"
}

@inproceedings{fefferman1985conformal,
  title={Conformal invariants},
  author={Fefferman, Charles and Graham, C Robin},
  booktitle={The Mathematical Heritage of {E}lie {C}artan (Lyon, 1984)},
  journal={Ast{\'e}risque},
  number={Hors S{\'e}rie},
  pages={95--116},
  year={1985}
}

@article{Aharony:2003qf,
    author = "Aharony, Ofer and DeWolfe, Oliver and Freedman, Daniel Z. and Karch, Andreas",
    title = "{Defect conformal field theory and locally localized gravity}",
    eprint = "hep-th/0303249",
    archivePrefix = "arXiv",
    reportNumber = "WIS-03-03-DPP, NSF-KITP-03-21, MIT-CTP-3352, UW-PT-03-06",
    doi = "10.1088/1126-6708/2003/07/030",
    journal = "JHEP",
    volume = "07",
    pages = "030",
    year = "2003"
}

@article{Bachas:2001vj,
    author = "Bachas, C. and de Boer, J. and Dijkgraaf, R. and Ooguri, H.",
    title = "{Permeable conformal walls and holography}",
    eprint = "hep-th/0111210",
    archivePrefix = "arXiv",
    reportNumber = "CALT-68-2361, CITUSC-01-045, ITFA-2001-33, LPTENS-01-42",
    doi = "10.1088/1126-6708/2002/06/027",
    journal = "JHEP",
    volume = "06",
    pages = "027",
    year = "2002"
}

@article{Bak:2003jk,
    author = "Bak, Dongsu and Gutperle, Michael and Hirano, Shinji",
    title = "{A Dilatonic deformation of AdS(5) and its field theory dual}",
    eprint = "hep-th/0304129",
    archivePrefix = "arXiv",
    reportNumber = "UCLA-03-TEP-12, UOSTP-03102",
    doi = "10.1088/1126-6708/2003/05/072",
    journal = "JHEP",
    volume = "05",
    pages = "072",
    year = "2003"
}

@article{Clark:2004sb,
    author = "Clark, A. B. and Freedman, D. Z. and Karch, A. and Schnabl, M.",
    title = "{Dual of the Janus solution: An interface conformal field theory}",
    eprint = "hep-th/0407073",
    archivePrefix = "arXiv",
    reportNumber = "CTP-MIT-3514, UW-PT-04-05",
    doi = "10.1103/PhysRevD.71.066003",
    journal = "Phys. Rev. D",
    volume = "71",
    pages = "066003",
    year = "2005"
}

@article{Clark:2005te,
    author = "Clark, A. and Karch, A.",
    title = "{Super Janus}",
    eprint = "hep-th/0506265",
    archivePrefix = "arXiv",
    reportNumber = "UW-PT-05-15",
    doi = "10.1088/1126-6708/2005/10/094",
    journal = "JHEP",
    volume = "10",
    pages = "094",
    year = "2005"
}

@article{DHoker:2006qeo,
    author = "D'Hoker, Eric and Estes, John and Gutperle, Michael",
    title = "{Interface Yang-Mills, supersymmetry, and Janus}",
    eprint = "hep-th/0603013",
    archivePrefix = "arXiv",
    reportNumber = "UCLA-06-TEP-03",
    doi = "10.1016/j.nuclphysb.2006.07.001",
    journal = "Nucl. Phys. B",
    volume = "753",
    pages = "16--41",
    year = "2006"
}

@article{DHoker:2006vfr,
    author = "D'Hoker, Eric and Estes, John and Gutperle, Michael",
    title = "{Ten-dimensional supersymmetric Janus solutions}",
    eprint = "hep-th/0603012",
    archivePrefix = "arXiv",
    reportNumber = "UCLA-06-TEP-02",
    doi = "10.1016/j.nuclphysb.2006.08.017",
    journal = "Nucl. Phys. B",
    volume = "757",
    pages = "79--116",
    year = "2006"
}

@article{DHoker:2007zhm,
    author = "D'Hoker, Eric and Estes, John and Gutperle, Michael",
    title = "{Exact half-BPS Type IIB interface solutions. I. Local solution and supersymmetric Janus}",
    eprint = "0705.0022",
    archivePrefix = "arXiv",
    primaryClass = "hep-th",
    reportNumber = "UCLA-07-TEP-09",
    doi = "10.1088/1126-6708/2007/06/021",
    journal = "JHEP",
    volume = "06",
    pages = "021",
    year = "2007"
}

@article{DHoker:2009lky,
    author = "D'Hoker, Eric and Estes, John and Gutperle, Michael and Krym, Darya",
    title = "{Janus solutions in M-theory}",
    eprint = "0904.3313",
    archivePrefix = "arXiv",
    primaryClass = "hep-th",
    reportNumber = "UCLA-09-TEP-43, CPHT-RR025-0409",
    doi = "10.1088/1126-6708/2009/06/018",
    journal = "JHEP",
    volume = "06",
    pages = "018",
    year = "2009"
}

@article{Aharony:2011yc,
    author = "Aharony, Ofer and Berdichevsky, Leon and Berkooz, Micha and Shamir, Itamar",
    title = "{Near-horizon solutions for D3-branes ending on 5-branes}",
    eprint = "1106.1870",
    archivePrefix = "arXiv",
    primaryClass = "hep-th",
    reportNumber = "WIS-5-11-MAY-DPPA",
    doi = "10.1103/PhysRevD.84.126003",
    journal = "Phys. Rev. D",
    volume = "84",
    pages = "126003",
    year = "2011"
}

@article{Bobev:2020fon,
    author = "Bobev, Nikolay and Gautason, Fridrik Freyr and Pilch, Krzysztof and Suh, Minwoo and van Muiden, Jesse",
    title = "{Holographic interfaces in $ \mathcal{N} $ = 4 SYM: Janus and J-folds}",
    eprint = "2003.09154",
    archivePrefix = "arXiv",
    primaryClass = "hep-th",
    doi = "10.1007/JHEP05(2020)134",
    journal = "JHEP",
    volume = "05",
    pages = "134",
    year = "2020"
}

@article{Ghodsi:2023pej,
    author = "Ghodsi, Ahmad and Kiritsis, Elias and Nitti, Francesco",
    title = "{Holographic CFTs on AdS$_{d}${\texttimes} S$^{n}$ and conformal defects}",
    eprint = "2309.04880",
    archivePrefix = "arXiv",
    primaryClass = "hep-th",
    reportNumber = "CCTP-2023-6, ITCP-2023/6",
    doi = "10.1007/JHEP10(2023)188",
    journal = "JHEP",
    volume = "10",
    pages = "188",
    year = "2023"
}

@article{Erdmenger:2002ex,
    author = "Erdmenger, Johanna and Guralnik, Zachary and Kirsch, Ingo",
    title = "{Four-dimensional superconformal theories with interacting boundaries or defects}",
    eprint = "hep-th/0203020",
    archivePrefix = "arXiv",
    reportNumber = "HU-EP-02-07",
    doi = "10.1103/PhysRevD.66.025020",
    journal = "Phys. Rev. D",
    volume = "66",
    pages = "025020",
    year = "2002"
}

@article{Constable:2002xt,
    author = "Constable, Neil R. and Erdmenger, Johanna and Guralnik, Zachary and Kirsch, Ingo",
    title = "{Intersecting D-3 branes and holography}",
    eprint = "hep-th/0211222",
    archivePrefix = "arXiv",
    reportNumber = "HU-EP-02-32, MIT-CTP-3316",
    doi = "10.1103/PhysRevD.68.106007",
    journal = "Phys. Rev. D",
    volume = "68",
    pages = "106007",
    year = "2003"
}

@article{Kapustin:2005py,
    author = "Kapustin, Anton",
    title = "{Wilson-'t Hooft operators in four-dimensional gauge theories and S-duality}",
    eprint = "hep-th/0501015",
    archivePrefix = "arXiv",
    reportNumber = "CALT-68-2536",
    doi = "10.1103/PhysRevD.74.025005",
    journal = "Phys. Rev. D",
    volume = "74",
    pages = "025005",
    year = "2006"
}

@article{Gaiotto:2008sa,
    author = "Gaiotto, Davide and Witten, Edward",
    title = "{Supersymmetric Boundary Conditions in N=4 Super Yang-Mills Theory}",
    eprint = "0804.2902",
    archivePrefix = "arXiv",
    primaryClass = "hep-th",
    doi = "10.1007/s10955-009-9687-3",
    journal = "J. Statist. Phys.",
    volume = "135",
    pages = "789--855",
    year = "2009"
}

@article{Gaiotto:2008sd,
    author = "Gaiotto, Davide and Witten, Edward",
    title = "{Janus Configurations, Chern-Simons Couplings, And The theta-Angle in N=4 Super Yang-Mills Theory}",
    eprint = "0804.2907",
    archivePrefix = "arXiv",
    primaryClass = "hep-th",
    doi = "10.1007/JHEP06(2010)097",
    journal = "JHEP",
    volume = "06",
    pages = "097",
    year = "2010"
}

@article{Jensen:2015swa,
    author = "Jensen, Kristan and O'Bannon, Andy",
    title = "{Constraint on Defect and Boundary Renormalization Group Flows}",
    eprint = "1509.02160",
    archivePrefix = "arXiv",
    primaryClass = "hep-th",
    reportNumber = "OUTP-15-19P, YITP-SB-15-33",
    doi = "10.1103/PhysRevLett.116.091601",
    journal = "Phys. Rev. Lett.",
    volume = "116",
    number = "9",
    pages = "091601",
    year = "2016"
}

@article{deLeeuw:2015hxa,
    author = "de Leeuw, Marius and Kristjansen, Charlotte and Zarembo, Konstantin",
    title = "{One-point Functions in Defect CFT and Integrability}",
    eprint = "1506.06958",
    archivePrefix = "arXiv",
    primaryClass = "hep-th",
    reportNumber = "NORDITA-2015-72, UUITP-12-15",
    doi = "10.1007/JHEP08(2015)098",
    journal = "JHEP",
    volume = "08",
    pages = "098",
    year = "2015"
}

@article{Billo:2016cpy,
    author = "Bill{\`o}, Marco and Gon{\c{c}}alves, Vasco and Lauria, Edoardo and Meineri, Marco",
    title = "{Defects in conformal field theory}",
    eprint = "1601.02883",
    archivePrefix = "arXiv",
    primaryClass = "hep-th",
    doi = "10.1007/JHEP04(2016)091",
    journal = "JHEP",
    volume = "04",
    pages = "091",
    year = "2016"
}

@article{Estes:2014hka,
    author = "Estes, John and Jensen, Kristan and O'Bannon, Andy and Tsatis, Efstratios and Wrase, Timm",
    title = "{On Holographic Defect Entropy}",
    eprint = "1403.6475",
    archivePrefix = "arXiv",
    primaryClass = "hep-th",
    reportNumber = "IMPERIAL-TP-2014-JE-01, YITP-SB-14-08, OUTP-14-03P, SU-ITP-14-05",
    doi = "10.1007/JHEP05(2014)084",
    journal = "JHEP",
    volume = "05",
    pages = "084",
    year = "2014"
}

@article{Andrei:2018die,
    author = "Andrei, N. and others",
    title = "{Boundary and Defect CFT: Open Problems and Applications}",
    eprint = "1810.05697",
    archivePrefix = "arXiv",
    primaryClass = "hep-th",
    doi = "10.1088/1751-8121/abb0fe",
    journal = "J. Phys. A",
    volume = "53",
    number = "45",
    pages = "453002",
    year = "2020"
}

@article{Chalabi:2021jud,
    author = "Chalabi, Adam and Herzog, Christopher P. and O'Bannon, Andy and Robinson, Brandon and Sisti, Jacopo",
    title = "{Weyl anomalies of four dimensional conformal boundaries and defects}",
    eprint = "2111.14713",
    archivePrefix = "arXiv",
    primaryClass = "hep-th",
    reportNumber = "UUITP-58/21",
    doi = "10.1007/JHEP02(2022)166",
    journal = "JHEP",
    volume = "02",
    pages = "166",
    year = "2022"
}

@article{Bianchi:2018zpb,
    author = "Bianchi, Lorenzo and Lemos, Madalena and Meineri, Marco",
    title = "{Line Defects and Radiation in $\mathcal{N}=2$ Conformal Theories}",
    eprint = "1805.04111",
    archivePrefix = "arXiv",
    primaryClass = "hep-th",
    reportNumber = "DESY-18-071",
    doi = "10.1103/PhysRevLett.121.141601",
    journal = "Phys. Rev. Lett.",
    volume = "121",
    number = "14",
    pages = "141601",
    year = "2018"
}

@article{Bianchi:2019sxz,
    author = "Bianchi, Lorenzo and Lemos, Madalena",
    title = "{Superconformal surfaces in four dimensions}",
    eprint = "1911.05082",
    archivePrefix = "arXiv",
    primaryClass = "hep-th",
    reportNumber = "CERN-TH-2019-190",
    doi = "10.1007/JHEP06(2020)056",
    journal = "JHEP",
    volume = "06",
    pages = "056",
    year = "2020"
}

@article{DHoker:2007mci,
    author = "D'Hoker, Eric and Estes, John and Gutperle, Michael",
    title = "{Gravity duals of half-BPS Wilson loops}",
    eprint = "0705.1004",
    archivePrefix = "arXiv",
    primaryClass = "hep-th",
    reportNumber = "UCLA-07-TEP-11",
    doi = "10.1088/1126-6708/2007/06/063",
    journal = "JHEP",
    volume = "06",
    pages = "063",
    year = "2007"
}

@article{Chiodaroli:2009yw,
    author = "Chiodaroli, Marco and Gutperle, Michael and Krym, Darya",
    title = "{Half-BPS Solutions locally asymptotic to AdS(3) x S**3 and interface conformal field theories}",
    eprint = "0910.0466",
    archivePrefix = "arXiv",
    primaryClass = "hep-th",
    reportNumber = "KUL-TF-09-21",
    doi = "10.1007/JHEP02(2010)066",
    journal = "JHEP",
    volume = "02",
    pages = "066",
    year = "2010"
}

@article{Chiodaroli:2011fn,
    author = "Chiodaroli, Marco and D'Hoker, Eric and Gutperle, Michael",
    title = "{Simple Holographic Duals to Boundary CFTs}",
    eprint = "1111.6912",
    archivePrefix = "arXiv",
    primaryClass = "hep-th",
    reportNumber = "IGC-11-11-2",
    doi = "10.1007/JHEP02(2012)005",
    journal = "JHEP",
    volume = "02",
    pages = "005",
    year = "2012"
}

@article{Bobev:2013yra,
    author = "Bobev, Nikolay and Pilch, Krzysztof and Warner, Nicholas P.",
    title = "{Supersymmetric Janus Solutions in Four Dimensions}",
    eprint = "1311.4883",
    archivePrefix = "arXiv",
    primaryClass = "hep-th",
    reportNumber = "IPHT-T13-259",
    doi = "10.1007/JHEP06(2014)058",
    journal = "JHEP",
    volume = "06",
    pages = "058",
    year = "2014"
}

@article{Dibitetto:2017klx,
    author = "Dibitetto, Giuseppe and Petri, Nicol{\`o}",
    title = "{6d surface defects from massive type IIA}",
    eprint = "1707.06154",
    archivePrefix = "arXiv",
    primaryClass = "hep-th",
    reportNumber = "UUITP-23-17, UUITP-23/17",
    doi = "10.1007/JHEP01(2018)039",
    journal = "JHEP",
    volume = "01",
    pages = "039",
    year = "2018"
}

@article{Dibitetto:2017tve,
    author = "Dibitetto, Giuseppe and Petri, Nicol{\`o}",
    title = "{BPS objects in D = 7 supergravity and their M-theory origin}",
    eprint = "1707.06152",
    archivePrefix = "arXiv",
    primaryClass = "hep-th",
    reportNumber = "UUITP-22-17",
    doi = "10.1007/JHEP12(2017)041",
    journal = "JHEP",
    volume = "12",
    pages = "041",
    year = "2017"
}

@article{Dibitetto:2020bsh,
    author = "Dibitetto, Giuseppe and Petri, Nicol{\`o}",
    title = "{AdS$_{3}$ from M-branes at conical singularities}",
    eprint = "2010.12323",
    archivePrefix = "arXiv",
    primaryClass = "hep-th",
    doi = "10.1007/JHEP01(2021)129",
    journal = "JHEP",
    volume = "01",
    pages = "129",
    year = "2021"
}

@article{Dibitetto:2019nyz,
    author = "Dibitetto, Giuseppe and Lozano, Yolanda and Petri, Nicol{\`o} and Ramirez, Anayeli",
    title = "{Holographic description of M-branes via AdS$_{2}$}",
    eprint = "1912.09932",
    archivePrefix = "arXiv",
    primaryClass = "hep-th",
    doi = "10.1007/JHEP04(2020)037",
    journal = "JHEP",
    volume = "04",
    pages = "037",
    year = "2020"
}

@article{Gutperle:2017nwo,
    author = "Gutperle, Michael and Kaidi, Justin and Raj, Himanshu",
    title = "{Janus solutions in six-dimensional gauged supergravity}",
    eprint = "1709.09204",
    archivePrefix = "arXiv",
    primaryClass = "hep-th",
    doi = "10.1007/JHEP12(2017)018",
    journal = "JHEP",
    volume = "12",
    pages = "018",
    year = "2017"
}

@article{Dibitetto:2018iar,
    author = "Dibitetto, Giuseppe and Petri, Nicol{\`o}",
    title = "{Surface defects in the D4 $-$ D8 brane system}",
    eprint = "1807.07768",
    archivePrefix = "arXiv",
    primaryClass = "hep-th",
    reportNumber = "UUITP-32/18",
    doi = "10.1007/JHEP01(2019)193",
    journal = "JHEP",
    volume = "01",
    pages = "193",
    year = "2019"
}

@article{Dibitetto:2018gtk,
    author = "Dibitetto, Giuseppe and Petri, Nicol{\`o}",
    title = "{AdS$_{2}$ solutions and their massive IIA origin}",
    eprint = "1811.11572",
    archivePrefix = "arXiv",
    primaryClass = "hep-th",
    doi = "10.1007/JHEP05(2019)107",
    journal = "JHEP",
    volume = "05",
    pages = "107",
    year = "2019"
}

@article{Gutperle:2018fea,
    author = "Gutperle, Michael and Vicino, Matteo",
    title = "{Conformal defect solutions in $N=2,D=4$ gauged supergravity}",
    eprint = "1811.04166",
    archivePrefix = "arXiv",
    primaryClass = "hep-th",
    doi = "10.1016/j.nuclphysb.2019.03.012",
    journal = "Nucl. Phys. B",
    volume = "942",
    pages = "149--163",
    year = "2019"
}

@article{Chen:2019qib,
    author = "Chen, Kevin and Gutperle, Michael",
    title = "{Holographic line defects in F(4) gauged supergravity}",
    eprint = "1909.11127",
    archivePrefix = "arXiv",
    primaryClass = "hep-th",
    doi = "10.1103/PhysRevD.100.126015",
    journal = "Phys. Rev. D",
    volume = "100",
    number = "12",
    pages = "126015",
    year = "2019"
}

@article{Gutperle:2019dqf,
    author = "Gutperle, Michael and Vicino, Matteo",
    title = "{Holographic Surface Defects in $D=5$, $N=4$ Gauged Supergravity}",
    eprint = "1911.02185",
    archivePrefix = "arXiv",
    primaryClass = "hep-th",
    doi = "10.1103/PhysRevD.101.066016",
    journal = "Phys. Rev. D",
    volume = "101",
    number = "6",
    pages = "066016",
    year = "2020"
}

@article{Chen:2020mtv,
    author = "Chen, Kevin and Gutperle, Michael and Vicino, Matteo",
    title = "{Holographic Line Defects in $D=4$, $N=2$ Gauged Supergravity}",
    eprint = "2005.03046",
    archivePrefix = "arXiv",
    primaryClass = "hep-th",
    doi = "10.1103/PhysRevD.102.026025",
    journal = "Phys. Rev. D",
    volume = "102",
    number = "2",
    pages = "026025",
    year = "2020"
}

@article{Faedo:2020nol,
    author = "Faedo, Federico and Lozano, Yolanda and Petri, Nicolo",
    title = "{Searching for surface defect CFTs within AdS$_3$}",
    eprint = "2007.16167",
    archivePrefix = "arXiv",
    primaryClass = "hep-th",
    doi = "10.1007/JHEP11(2020)052",
    journal = "JHEP",
    volume = "11",
    pages = "052",
    year = "2020"
}

@article{Faedo:2020lyw,
    author = "Faedo, Federico and Lozano, Yolanda and Petri, Nicol{\`o}",
    title = "{New $\mathcal{N}=(0,4)$ AdS$_3$ near-horizons in Type IIB}",
    eprint = "2012.07148",
    archivePrefix = "arXiv",
    primaryClass = "hep-th",
    doi = "10.1007/JHEP04(2021)028",
    journal = "JHEP",
    volume = "04",
    pages = "028",
    year = "2021"
}

@article{Chen:2020efh,
    author = "Chen, Kevin and Gutperle, Michael",
    title = "{Janus solutions in three-dimensional $ \mathcal{N} $ = 8 gauged supergravity}",
    eprint = "2011.10154",
    archivePrefix = "arXiv",
    primaryClass = "hep-th",
    doi = "10.1007/JHEP05(2021)008",
    journal = "JHEP",
    volume = "05",
    pages = "008",
    year = "2021"
}

@article{Lozano:2021fkk,
    author = "Lozano, Yolanda and Petri, Nicol{\`o} and Risco, Cristian",
    title = "{New AdS$_{2}$ supergravity duals of 4d SCFTs with defects}",
    eprint = "2107.12277",
    archivePrefix = "arXiv",
    primaryClass = "hep-th",
    doi = "10.1007/JHEP10(2021)217",
    journal = "JHEP",
    volume = "10",
    pages = "217",
    year = "2021"
}

@article{Lozano:2022ouq,
    author = "Lozano, Yolanda and Macpherson, Niall T. and Petri, Nicol{\`o} and Risco, Cristian",
    title = "{New AdS$_{3}$/CFT$_{2}$ pairs in massive IIA with (0, 4) and (4, 4) supersymmetries}",
    eprint = "2206.13541",
    archivePrefix = "arXiv",
    primaryClass = "hep-th",
    doi = "10.1007/JHEP09(2022)130",
    journal = "JHEP",
    volume = "09",
    pages = "130",
    year = "2022"
}

@article{Anabalon:2022fti,
    author = "Anabal{\'o}n, Andr{\'e}s and Chamorro-Burgos, Miguel and Guarino, Adolfo",
    title = "{Janus and Hades in M-theory}",
    eprint = "2207.09287",
    archivePrefix = "arXiv",
    primaryClass = "hep-th",
    doi = "10.1007/JHEP11(2022)150",
    journal = "JHEP",
    volume = "11",
    pages = "150",
    year = "2022"
}

@article{Lozano:2022swp,
    author = "Lozano, Yolanda and Petri, Nicol{\`o} and Risco, Cristian",
    title = "{AdS2 near-horizons, defects, and string dualities}",
    eprint = "2212.11095",
    archivePrefix = "arXiv",
    primaryClass = "hep-th",
    doi = "10.1103/PhysRevD.107.106012",
    journal = "Phys. Rev. D",
    volume = "107",
    number = "10",
    pages = "106012",
    year = "2023"
}

@article{Lozano:2022vsv,
    author = "Lozano, Yolanda and Petri, Nicol{\`o} and Risco, Cristian",
    title = "{Line defects as brane boxes in Gaiotto-Maldacena geometries}",
    eprint = "2212.10398",
    archivePrefix = "arXiv",
    primaryClass = "hep-th",
    doi = "10.1007/JHEP02(2023)193",
    journal = "JHEP",
    volume = "02",
    pages = "193",
    year = "2023"
}

@article{Lozano:2024idt,
    author = "Lozano, Yolanda and Macpherson, Niall T. and Petri, Nicol{\`o} and Ram{\'\i}rez, Anayeli",
    title = "{Holographic $ \frac{1}{2} $-BPS surface defects in ABJM}",
    eprint = "2404.17469",
    archivePrefix = "arXiv",
    primaryClass = "hep-th",
    doi = "10.1007/JHEP08(2024)044",
    journal = "JHEP",
    volume = "08",
    pages = "044",
    year = "2024"
}

@article{Conti:2024qgx,
    author = "Conti, Andrea and Dibitetto, Giuseppe and Lozano, Yolanda and Petri, Nicol{\`o} and Ram{\'\i}rez, Anayeli",
    title = "{Deconstruction and surface defects in 6d CFTs}",
    eprint = "2407.21627",
    archivePrefix = "arXiv",
    primaryClass = "hep-th",
    doi = "10.1007/JHEP11(2024)131",
    journal = "JHEP",
    volume = "11",
    pages = "131",
    year = "2024"
}

@article{Gutperle:2024yiz,
    author = "Gutperle, Michael and Hultgreen-Mena, Charlie",
    title = "{Janus and RG-interfaces in minimal 3d gauged supergravity}",
    eprint = "2412.16749",
    archivePrefix = "arXiv",
    primaryClass = "hep-th",
    doi = "10.1142/S0217751X25501131",
    journal = "Int. J. Mod. Phys. A",
    volume = "40",
    number = "26",
    pages = "2550113",
    year = "2025"
}

@article{Lozano:2019emq,
    author = "Lozano, Yolanda and Macpherson, Niall T. and Nunez, Carlos and Ramirez, Anayeli",
    title = "{AdS$_3$ solutions in Massive IIA with small $\mathcal{N}=(4,0)$ supersymmetry}",
    eprint = "1908.09851",
    archivePrefix = "arXiv",
    primaryClass = "hep-th",
    doi = "10.1007/JHEP01(2020)129",
    journal = "JHEP",
    volume = "01",
    pages = "129",
    year = "2020"
}

@article{Lozano:2019jza,
    author = "Lozano, Yolanda and Macpherson, Niall T. and Nunez, Carlos and Ramirez, Anayeli",
    title = "{1/4 BPS solutions and the AdS$_3$/CFT$_2$ correspondence}",
    eprint = "1909.09636",
    archivePrefix = "arXiv",
    primaryClass = "hep-th",
    doi = "10.1103/PhysRevD.101.026014",
    journal = "Phys. Rev. D",
    volume = "101",
    number = "2",
    pages = "026014",
    year = "2020"
}

@article{Couzens:2021veb,
    author = "Couzens, Christopher and Lozano, Yolanda and Petri, Nicol{\`o} and Vandoren, Stefan",
    title = "{N=(0,4) black string chains}",
    eprint = "2109.10413",
    archivePrefix = "arXiv",
    primaryClass = "hep-th",
    doi = "10.1103/PhysRevD.105.086015",
    journal = "Phys. Rev. D",
    volume = "105",
    number = "8",
    pages = "086015",
    year = "2022"
}

@article{Lozano:2020txg,
    author = "Lozano, Yolanda and Nunez, Carlos and Ramirez, Anayeli and Speziali, Stefano",
    title = "{New AdS$_{2}$ backgrounds and $ \mathcal{N} $ = 4 conformal quantum mechanics}",
    eprint = "2011.00005",
    archivePrefix = "arXiv",
    primaryClass = "hep-th",
    doi = "10.1007/JHEP03(2021)277",
    journal = "JHEP",
    volume = "03",
    pages = "277",
    year = "2021"
}

@article{Lozano:2020sae,
    author = "Lozano, Yolanda and Nunez, Carlos and Ramirez, Anayeli and Speziali, Stefano",
    title = "{AdS$_{2}$ duals to ADHM quivers with Wilson lines}",
    eprint = "2011.13932",
    archivePrefix = "arXiv",
    primaryClass = "hep-th",
    doi = "10.1007/JHEP03(2021)145",
    journal = "JHEP",
    volume = "03",
    pages = "145",
    year = "2021"
}

@article{Faedo:2025kjf,
    author = "Faedo, Federico and Petri, Nicol{\`o} and Segati, Alessia",
    title = "{Defects in 4d SCFTs from supergravity and holographic renormalization}",
    eprint = "2501.17923",
    archivePrefix = "arXiv",
    primaryClass = "hep-th",
    doi = "10.1007/JHEP05(2025)070",
    journal = "JHEP",
    volume = "05",
    pages = "070",
    year = "2025"
}

@article{Arav:2024exg,
    author = "Arav, Igal and Gauntlett, Jerome P. and Jiao, Yusheng and Roberts, Matthew M. and Rosen, Christopher",
    title = "{Superconformal monodromy defects in $ \mathcal{N} $=4 SYM and LS theory}",
    eprint = "2405.06014",
    archivePrefix = "arXiv",
    primaryClass = "hep-th",
    reportNumber = "APCTP Pre2024-005,CCTP-2024-9, ITCP-2024/9, APCTP Pre2024-005",
    doi = "10.1007/JHEP08(2024)177",
    journal = "JHEP",
    volume = "08",
    pages = "177",
    year = "2024"
}

@article{Arav:2024wyg,
    author = "Arav, Igal and Gauntlett, Jerome P. and Jiao, Yusheng and Roberts, Matthew M. and Rosen, Christopher",
    title = "{Superconformal monodromy defects in ABJM and mABJM theory}",
    eprint = "2408.11088",
    archivePrefix = "arXiv",
    primaryClass = "hep-th",
    reportNumber = "APCTP Pre2024-004; CCTP-2024-10; ITCP-2024/10",
    doi = "10.1007/JHEP11(2024)008",
    journal = "JHEP",
    volume = "11",
    pages = "008",
    year = "2024"
}

@article{Conti:2025wwf,
    author = "Conti, Andrea and Lozano, Yolanda and Rogdakis, Filippos and Rosen, Christopher",
    title = "{Defect entanglement entropy for superconformal monodromy defects}",
    eprint = "2511.22695",
    archivePrefix = "arXiv",
    primaryClass = "hep-th",
    doi = "10.1007/JHEP05(2026)036",
    journal = "JHEP",
    volume = "05",
    pages = "036",
    year = "2026"
}

@article{Conti:2025wyj,
    author = "Conti, Andrea and Lozano, Yolanda and Rosen, Christopher",
    title = "{Monodromy defects in massive Type IIA}",
    eprint = "2512.10006",
    archivePrefix = "arXiv",
    primaryClass = "hep-th",
    doi = "10.1007/JHEP04(2026)173",
    journal = "JHEP",
    volume = "04",
    pages = "173",
    year = "2026"
}

@article{Couzens:2026qne,
    author = {Couzens, Christopher and L{\"u}scher, Alice and Sparks, James},
    title = "{IIB an equivariantly localized puncture}",
    eprint = "2601.07598",
    archivePrefix = "arXiv",
    primaryClass = "hep-th",
    month = "1",
    year = "2026"
}

@article{Papadimitriou:2004rz,
    author = "Papadimitriou, Ioannis and Skenderis, Kostas",
    title = "{Correlation functions in holographic RG flows}",
    eprint = "hep-th/0407071",
    archivePrefix = "arXiv",
    reportNumber = "ITFA-2004-23",
    doi = "10.1088/1126-6708/2004/10/075",
    journal = "JHEP",
    volume = "10",
    pages = "075",
    year = "2004"
}

@article{Henningson:1998gx,
    author = "Henningson, M. and Skenderis, K.",
    title = "{The Holographic Weyl anomaly}",
    eprint = "hep-th/9806087",
    archivePrefix = "arXiv",
    reportNumber = "CERN-TH-98-188, KUL-TF-98-21",
    doi = "10.1088/1126-6708/1998/07/023",
    journal = "JHEP",
    volume = "07",
    pages = "023",
    year = "1998"
}

@article{Bianchi:2026sax,
    author = "Bianchi, Lorenzo and de Sabbata, Elia and Meineri, Marco",
    title = "{Conformal defects and Goldstone bosons in Anti-de Sitter space}",
    eprint = "2605.13947",
    archivePrefix = "arXiv",
    primaryClass = "hep-th",
    month = "5",
    year = "2026"
}

@article{Bianchi:2001kw,
    author = "Bianchi, Massimo and Freedman, Daniel Z. and Skenderis, Kostas",
    title = "{Holographic renormalization}",
    eprint = "hep-th/0112119",
    archivePrefix = "arXiv",
    reportNumber = "MIT-CTP-3166, PUTP-1999, DAMTP-2001-63, ROM2F-2001-30",
    doi = "10.1016/S0550-3213(02)00179-7",
    journal = "Nucl. Phys. B",
    volume = "631",
    pages = "159--194",
    year = "2002"
}

@article{deHaro:2000vlm,
    author = "de Haro, Sebastian and Solodukhin, Sergey N. and Skenderis, Kostas",
    title = "{Holographic reconstruction of space-time and renormalization in the AdS / CFT correspondence}",
    eprint = "hep-th/0002230",
    archivePrefix = "arXiv",
    reportNumber = "SPIN-2000-05, ITP-UU-00-03, PUTP-1921",
    doi = "10.1007/s002200100381",
    journal = "Commun. Math. Phys.",
    volume = "217",
    pages = "595--622",
    year = "2001"
}

@article{Romans:1985tw,
    author = "Romans, L. J.",
    editor = "Salam, A. and Sezgin, E.",
    title = "{The F(4) Gauged Supergravity in Six-dimensions}",
    reportNumber = "NSF-ITP-85-137",
    doi = "10.1016/0550-3213(86)90517-1",
    journal = "Nucl. Phys. B",
    volume = "269",
    pages = "691",
    year = "1986"
}

@article{Conti:2024rwd,
    author = "Conti, Andrea and Dibitetto, Giuseppe and Lozano, Yolanda and Petri, Nicol{\`o} and Ram{\'\i}rez, Anayeli",
    title = "{Half-BPS Janus solutions in AdS$_{7}$}",
    eprint = "2407.21619",
    archivePrefix = "arXiv",
    primaryClass = "hep-th",
    doi = "10.1007/JHEP12(2024)198",
    journal = "JHEP",
    volume = "12",
    pages = "198",
    year = "2024"
}

@article{Alday:2014bta,
    author = "Alday, Luis F. and Fluder, Martin and Gregory, Carolina M. and Richmond, Paul and Sparks, James",
    title = "{Supersymmetric gauge theories on squashed five-spheres and their gravity duals}",
    eprint = "1405.7194",
    archivePrefix = "arXiv",
    primaryClass = "hep-th",
    doi = "10.1007/JHEP09(2014)067",
    journal = "JHEP",
    volume = "09",
    pages = "067",
    year = "2014"
}

@article{Klebanov:1999tb,
    author = "Klebanov, Igor R. and Witten, Edward",
    title = "{AdS / CFT correspondence and symmetry breaking}",
    eprint = "hep-th/9905104",
    archivePrefix = "arXiv",
    reportNumber = "PUPT-1863, IASSNS-HEP-99-49",
    doi = "10.1016/S0550-3213(99)00387-9",
    journal = "Nucl. Phys. B",
    volume = "556",
    pages = "89--114",
    year = "1999"
}

@article{Ferrara:1998gv,
    author = "Ferrara, S. and Kehagias, A. and Partouche, H. and Zaffaroni, A.",
    title = "{AdS(6) interpretation of 5-D superconformal field theories}",
    eprint = "hep-th/9804006",
    archivePrefix = "arXiv",
    reportNumber = "CERN-TH-98-111",
    doi = "10.1016/S0370-2693(98)00560-7",
    journal = "Phys. Lett. B",
    volume = "431",
    pages = "57--62",
    year = "1998"
}

@article{Ferrara:1998ur,
    author = "Ferrara, Sergio and Zaffaroni, Alberto",
    title = "{N=1, N=2 4-D superconformal field theories and supergravity in AdS(5)}",
    eprint = "hep-th/9803060",
    archivePrefix = "arXiv",
    reportNumber = "CERN-TH-98-68",
    doi = "10.1016/S0370-2693(98)00567-X",
    journal = "Phys. Lett. B",
    volume = "431",
    pages = "49--56",
    year = "1998"
}

@article{Romans:1985ps,
    author = "Romans, L. J.",
    title = "{Gauged $N=4$ Supergravities in Five-dimensions and Their Magnetovac Backgrounds}",
    reportNumber = "NSF-ITP-85-113",
    doi = "10.1016/0550-3213(86)90398-6",
    journal = "Nucl. Phys. B",
    volume = "267",
    pages = "433--447",
    year = "1986"
}

@article{Lu:1999bw,
    author = "Lu, Hong and Pope, C. N. and Tran, Tuan A.",
    title = "{Five-dimensional N=4, SU(2) x U(1) gauged supergravity from type IIB}",
    eprint = "hep-th/9909203",
    archivePrefix = "arXiv",
    reportNumber = "CTP-TAMU-41-99, UPR-862-T",
    doi = "10.1016/S0370-2693(00)00073-3",
    journal = "Phys. Lett. B",
    volume = "475",
    pages = "261--268",
    year = "2000"
}

@article{Gauntlett:2007sm,
    author = "Gauntlett, Jerome P. and Varela, Oscar",
    title = "{D=5 SU(2) x U(1) Gauged Supergravity from D=11 Supergravity}",
    eprint = "0712.3560",
    archivePrefix = "arXiv",
    primaryClass = "hep-th",
    reportNumber = "IMPERIAL-TP-2007-JG-04",
    doi = "10.1088/1126-6708/2008/02/083",
    journal = "JHEP",
    volume = "02",
    pages = "083",
    year = "2008"
}

@article{Arav:2020obl,
    author = "Arav, Igal and Cheung, K. C. Matthew and Gauntlett, Jerome P. and Roberts, Matthew M. and Rosen, Christopher",
    title = "{Spatially modulated and supersymmetric mass deformations of $ \mathcal{N} $ = 4 SYM}",
    eprint = "2007.15095",
    archivePrefix = "arXiv",
    primaryClass = "hep-th",
    reportNumber = "Imperial/TP/2020/JG/03; ICCUB-20-XXX",
    doi = "10.1007/JHEP11(2020)156",
    journal = "JHEP",
    volume = "11",
    pages = "156",
    year = "2020"
}

@article{Gurarie:1993xq,
    author = "Gurarie, V.",
    title = "{Logarithmic operators in conformal field theory}",
    eprint = "hep-th/9303160",
    archivePrefix = "arXiv",
    reportNumber = "PUPT-1391",
    doi = "10.1016/0550-3213(93)90528-W",
    journal = "Nucl. Phys. B",
    volume = "410",
    pages = "535--549",
    year = "1993"
}

@article{Gaberdiel:2001tr,
    author = "Gaberdiel, Matthias R",
    editor = "Flohr, M. A. I. and Rouhani, S.",
    title = "{An Algebraic approach to logarithmic conformal field theory}",
    eprint = "hep-th/0111260",
    archivePrefix = "arXiv",
    reportNumber = "KCL-MTH-01-46",
    doi = "10.1142/S0217751X03016860",
    journal = "Int. J. Mod. Phys. A",
    volume = "18",
    pages = "4593--4638",
    year = "2003"
}

@article{Cuomo:2021rkm,
    author = "Cuomo, Gabriel and Komargodski, Zohar and Raviv-Moshe, Avia",
    title = "{Renormalization Group Flows on Line Defects}",
    eprint = "2108.01117",
    archivePrefix = "arXiv",
    primaryClass = "hep-th",
    doi = "10.1103/PhysRevLett.128.021603",
    journal = "Phys. Rev. Lett.",
    volume = "128",
    number = "2",
    pages = "021603",
    year = "2022"
}
\end{document}